\documentclass[11pt]{article}
\pdfoutput=1
\usepackage{jcapmod}
\usepackage{shorthand}

\usepackage{mathtools}
\usepackage{booktabs}
\usepackage[english]{babel}
\usepackage{amsmath,amssymb,amsbsy,amstext, amsthm, simplewick, amsfonts}
\usepackage{hyperref}
\usepackage{graphicx}
\usepackage[small]{caption}
\usepackage{siunitx}
\usepackage{upgreek}
\usepackage{framed}
\usepackage{wrapfig}
\usepackage{multirow}
\usepackage{bbm}
\usepackage[svgnames,dvipsnames,x11names]{xcolor}
\usepackage[utf8x]{inputenc}
\usepackage{selinput}
\usepackage{bm}
\usepackage{float}
\usepackage{geometry}
\usepackage{yfonts}
\usepackage{caption}
\usepackage{subcaption}
\usepackage{sidecap}
\usepackage{longtable}
\usepackage{anyfontsize}
\usepackage{dsfont}

\usepackage[framemethod=default]{mdframed}
\newmdenv[skipabove=7pt,
skipbelow=7pt,
rightline=false,
leftline=false,
topline=false,
bottomline=false,
backgroundcolor=gray!10,
linecolor=gray,
innerleftmargin=5pt,
innerrightmargin=5pt,
innertopmargin=5pt,
innerbottommargin=5pt,
leftmargin=0cm,
rightmargin=0cm,
linewidth=4pt]{eBox}
\newmdenv[skipabove=7pt,
skipbelow=7pt,
rightline=false,
leftline=false,
topline=false,
bottomline=false,
backgroundcolor=gray!10,
linecolor=gray,
innerleftmargin=5pt,
innerrightmargin=5pt,
innertopmargin=-5pt,
innerbottommargin=5pt,
leftmargin=0cm,
rightmargin=0cm,
linewidth=4pt]{eBox2}

\definecolor{blue3}{RGB}{31, 119, 180}
\definecolor{red3}{RGB}{	214, 39, 40}
\definecolor{orange3}{RGB}{255, 127, 14}
\definecolor{green3}{RGB}{44, 160, 44}
\definecolor{repBlue}{RGB}{31, 119, 180}
\definecolor{repRed}{RGB}{	214, 39, 40}
\definecolor{repGreen}{RGB}{44, 160, 44}

\renewcommand{\(}{\left(}
\renewcommand{\)}{\right)}
\renewcommand{\[}{\left[}
\renewcommand{\]}{\right]}

\def\be{\begin{equation}}
\def\ee{\end{equation}}
\newcommand{\bea}{\begin{eqnarray}}
\newcommand{\eea}{\end{eqnarray}}
\def\fnl{f_{\rm NL}}
\def\cG{\mathcal{K}}
\def\vp{\varphi}

\usepackage{colortbl}
\definecolor{lightgreen}{cmyk}{0.2, 0, 0.2, 0.2}
\definecolor{lightgray}{cmyk}{0.1,0.2,0,0.1}
\definecolor{lightgray2}{cmyk}{0.1,0.1,0,0.1}

\makeatletter
\newlength{\apb@width}
\newcommand{\autoparbox}[2][c]{\settowidth{\apb@width}{#2}\parbox[#1]{\apb@width}{#2}}

\makeatother


\def\beq{\begin{equation}}
\def\eeq{\end{equation}}

\newcommand{\mpl}{M_{\rm Pl}}

\allowdisplaybreaks[1]
\begin{document}


\newgeometry{top=2cm, bottom=2cm, left=2.9cm, right=2.9cm}

\begin{titlepage}
\setcounter{page}{1} \baselineskip=15.5pt 
\thispagestyle{empty}

\begin{center}
{\fontsize{20}{18} \bf Bootstrapping multi-field inflation:}\\ [12pt]
{\fontsize{17}{18} \it Non-Gaussianities from light scalars revisited}\\[15pt]
\end{center}

\vskip 20pt

\begin{center}
\noindent
{\fontsize{12}{18}\selectfont Dong-Gang Wang$^1$, 
Guilherme L.~Pimentel$^2$ and Ana Ach\'ucarro$^{3,4}$ }
\end{center}

\vskip 20pt

\begin{center}
  \vskip8pt
 {$^1$ \fontsize{10.6}{18}\it Department of Applied Mathematics and Theoretical Physics, University of Cambridge,\\
Wilberforce Road, Cambridge, CB3 0WA, UK}

  \vskip8pt
 {$^2$ \fontsize{10.6}{18}\it Scuola Normale Superiore and INFN, Piazza dei Cavalieri 7, 56126, Pisa, Italy
}

  \vskip8pt
 {$^3$ \fontsize{10.6}{18}\it Lorentz Institute for Theoretical Physics, Leiden University, Leiden, 2333 CA, The Netherlands}
  
  \vskip8pt
 {$^4$ \fontsize{10.6}{18}\it
Department of Physics, University of the Basque Country, UPV/EHU, 48080, Bilbao, Spain}
\end{center}

%
%

\vspace{0.4cm}
 \begin{center}{\bf Abstract} 
 \end{center}
 \noindent
Primordial non-Gaussianities from multi-field inflation are a leading target for cosmological observations, because of the possible large correlations generated between long and short distances. 
These signatures are captured by the local shape of the scalar bispectrum. In this paper, we revisit the nonlinearities of the conversion process from additional light scalars into curvature perturbations during inflation. 
We provide analytic templates for correlation functions valid at any kinematical configuration, using  the cosmological bootstrap as a main computational tool. Our results include the possibility of large breaking of boost symmetry, in the form of small speeds of sound for both the inflaton and the mediators. 
We consider correlators coming from the tree-level exchange of a massless scalar field.
By introducing a late-time cutoff, we identify that the symmetry constraints on the correlators are modified. This leads to anomalous conformal Ward identities, and consequently the bootstrap differential equations acquire a source term that depends on this cutoff.  
The solutions to the differential equations are scalar seed functions that incorporate these late-time growth effects. 
 Applying weight-shifting operators to auxiliary ``seed" functions, we obtain a systematic classification of shapes of non-Gaussianity coming from massless exchange.  
For theories with de Sitter symmetry, we compare the resulting shapes with the ones obtained via the $\delta N$ formalism, identifying missing contributions away from the squeezed limit. For boost-breaking scenarios, we derive a novel class of shape functions with phenomenologically distinct features. 
Specifically, the new shape provides a simple extension of equilateral non-Gaussianity: the signal peaks at a geometric configuration controlled by the ratio of the sound speeds of the mediator and the inflaton.

\noindent

\end{titlepage}

\newpage

\restoregeometry
\setcounter{tocdepth}{3}
\setcounter{page}{1}
\tableofcontents

\newpage
\section{Introduction}

Are there additional light scalar degrees of freedom beyond the primordial curvature perturbation?
This intriguing question is particularly important for inflationary cosmology, as it is closely related to the dynamics of the primordial Universe and provides great opportunities to test fundamental physics at extremely high energies \cite{Meerburg:2019qqi, Achucarro:2022qrl}.
Theories of inflation with multiple scalar fields have been extensively investigated for many years. They have distinctive signatures, due to correlations generated by the extra particles with masses much smaller than the Hubble scale \cite{Lyth:2005fi, Seery:2005gb, Bassett:2005xm,Rigopoulos:2005ae,Vernizzi:2006ve,Seery:2006js, Byrnes:2006vq, Byrnes:2009qy, Byrnes:2010em, Wands:2010af,Senatore:2010wk,Peterson:2010mv, Gong:2016qmq}.  
Specifically, the light scalars can be converted into the curvature perturbation after horizon crossing, and the nonlinearity of this process 
leads to non-Gaussian statistics in primordial fluctuations coupling long and short distances. 
In the scalar bispectrum of the curvature perturbation $\zeta$, this corresponds to the well-known {\it local shape} \cite{Komatsu:2001rj}
\be \label{local}
\langle \zeta_{{\bf k}_1} \zeta_{{\bf k}_2} \zeta_{{\bf k}_3}  \rangle'  =  
\frac{6}{5} \fnl \,P_\zeta^2\, S_{\rm local}(k_1,k_2,k_3)~, ~~~~~~{\rm with}~~ S_{\rm local}(k_1,k_2,k_3) = \frac{k_1^3+k_2^3+k_3^3}{k_1^3 k_2^3 k_3^3}
\ee
 where $P_\zeta \equiv  k^3 \langle \zeta_{\bf k} \zeta_{\bf -k}  \rangle' $ is the primordial power spectrum of $\zeta$.\footnote{The prime on  correlators means that we have stripped the momentum-conserving $\delta$-function $(2\pi)^3 \delta(\sum_n{\bf k}_n)$.} As the smoking gun of additional light scalars beyond the inflaton, a detection of local non-Gaussianity would rule out (almost) all models of single field inflation.\footnote{See non-attractor inflation as a counterexample which has one scalar field but two degrees of freedom in the background evolution. In other words, this class  of models are not ``single-clock,"  and thus local non-Gaussianity can be generated \cite{Namjoo:2012aa, Chen:2013eea, Chen:2013aj, Cai:2017bxr}.}
That, together with its very distinctive observational imprint, makes the local form of the bispectrum a major target of observations probing primordial non-Gaussianity.
The latest CMB data from the Planck satellite gives the current limit on the size parameter $\fnl= -0.9\pm 5.1$ \cite{Akrami:2019izv}. In many upcoming surveys, of both galaxies and the CMB,
we expect the local shape to be further constrained, and potentially detected \cite{Achucarro:2022qrl}.

\vskip4pt
Meanwhile, on the theory frontier, there have been significant improvements on our understanding of cosmological correlators. 
This partly comes from the ``cosmological bootstrap'' program  \cite{Arkani-Hamed:2018kmz,Baumann:2019oyu,Baumann:2020dch, Arkani-Hamed:2017fdk,Arkani-Hamed:2018bjr,Benincasa:2018ssx, Sleight:2019mgd, Sleight:2019hfp,Sleight:2020obc, Pajer:2020wnj,Pajer:2020wxk, Jazayeri:2021fvk,Bonifacio:2021azc,Cabass:2021fnw,Hillman:2021bnk, Pimentel:2022fsc, Jazayeri:2022kjy, Cabass:2022jda, Goodhew:2020hob, Cespedes:2020xqq, Melville:2021lst, Goodhew:2021oqg, Baumann:2021fxj, Meltzer:2021zin,Hogervorst:2021uvp,DiPietro:2021sjt, Baumann:2022jpr, Goodhew:2022ayb, Qin:2022fbv, Bonifacio:2022vwa, Salcedo:2022aal}, which allows us to derive theoretically accurate predictions based on fundamental principles, such as symmetries, unitarity and locality, while being relatively model agnostic. The bootstrap approach provides a comprehensive classification of the inflationary correlators based on minimal assumptions. Two parallel lines of development follow different symmetry assumptions: first, the idea of bootstrap was implemented for theories that respect all de Sitter (dS) isometries (the {\it dS bootstrap}) \cite{Arkani-Hamed:2018kmz,Baumann:2019oyu,Baumann:2020dch}. Later, a broader class of theories with broken dS boost symmetry were considered in the {\it boostless bootstrap}, where large signals and richer phenomenology are naturally expected \cite{Pajer:2020wxk, Jazayeri:2021fvk, Bonifacio:2021azc,Cabass:2021fnw,Hillman:2021bnk,Pimentel:2022fsc, Jazayeri:2022kjy, Cabass:2022jda}. Beyond reproducing the known results in the literature, many new non-Gaussian signals were bootstrapped from this novel formalism.

\vskip4pt
Our theoretical prior is that there are two broad classes of primordial non-Gaussianties: one from self-interactions of the inflaton, leading to equilateral-type correlations; another from the presence of new species of particles, which mediate long distance correlations during inflation.
A systematic study of these shapes---dubbed {\it cosmological collider physics}---has provided a remarkable avenue for testing new physics in the extremely high energy environment of the primordial Universe \cite{Chen:2009zp, Baumann:2011nk, Assassi:2012zq, Chen:2012ge, Pi:2012gf, Noumi:2012vr, Baumann:2012bc, Assassi:2013gxa, Gong:2013sma, Arkani-Hamed:2015bza, Lee:2016vti, Flauger:2016idt, Chen:2016uwp, Chen:2016hrz, Kehagias:2017cym, Kumar:2017ecc, An:2017hlx, An:2017rwo, Baumann:2017jvh, Kumar:2018jxz, Bordin:2018pca, Goon:2018fyu, Anninos:2019nib, Kim:2019wjo, Alexander:2019vtb, Hook:2019zxa, Kumar:2019ebj, Liu:2019fag, Wang:2019gbi, Wang:2019gok, Maru:2021ezc, Lu:2021wxu, Wang:2021qez, Tong:2021wai, Cui:2021iie, Tong:2022cdz, Reece:2022soh, Pimentel:2022fsc, Chen:2022vzh, Qin:2022lva, Jazayeri:2022kjy, Xianyu:2022jwk, Niu:2022quw}. For example, now we understand how to generically extract spectroscopic information (masses, spins and couplings) of mediator particles from the shapes of non-Gaussianity.
One notable case is when the intermediate states are massless scalars: they can be the source of massless isocurvature perturbations.
In the {\it exactly massless} limit of multi-field inflation, the curvature perturbation can be continuously sourced during inflation, and the action for the inflationary perturbations acquires an extra “scaling” symmetry \cite{Achucarro:2016fby, Achucarro:2019lgo} (see also Ref. \cite{Achucarro:2021pdh}).\footnote{As a proof of concept, a class of multi-field models was recently constructed with exact background solutions  and  neutrally
stable attractor behaviour \cite{Achucarro:2019pux}. Unlike many other scenarios, the isocurvatue perturbations here remain massless for the whole duration of inflation. This model serves as a benchmark example for the analysis presented in the current work.
}

\vskip4pt
These recent developments encourage us to re-examine the cosmological correlators mediated by additional very light scalars, given the importance of these shapes to observations. 
In this work, we perform a systematic analysis of cosmological correlators from multi-field inflation, using the bootstrap as a main tool. 
This paper complements our previous work, where we derived a large set of massive exchange correlators with broken boost symmetry \cite{Pimentel:2022fsc}. Here, instead, our focus is to bootstrap {\it massless exchanges}. An important difference in this case is the appearance of the well-known {\it infrared (IR) divergences} for interacting massless scalars in dS space \cite{Ford:1984hs, Antoniadis:1985pj, Tsamis:1994ca, Tsamis:1996qm, Seery:2010kh, Giddings:2010nc,Giddings:2010ui,Anninos:2014lwa,Hu:2018nxy, Gorbenko:2019rza, Mirbabayi:2019qtx, Cohen:2020php,Cohen:2021fzf, Green:2022ovz, Cohen:2022clv}, which are typically addressed within the framework of stochastic inflation \cite{Vilenkin:1983xq, Starobinsky:1986fx}. 
We will remain in the perturbative regime, and compute correlators at tree-level, while carefully accounting for the IR effects. An important technical step will be how to incorporate IR effects within the bootstrap. We show that they introduce new terms in the ``boundary" (late-time) differential equations. 
We consider both the dS-invariant and boost-breaking scenarios.
We will also compare our results with the literature on primordial non-Gaussianity within multi-field inflation. When there is overlap, we find agreement. Nonetheless, we find new shapes of non-Gaussianity, with new phenomenology of potential interest for future observations.

\subsection{Summary of Results}

Our main results can be summarized in three points:

\begin{itemize}
\item {\bf IR divergences in the cosmological bootstrap.} We incorporate IR divergences in the cosmological bootstrap. 
Within the validity of perturbation theory, the tree-level IR divergent terms are regularized by an explicit  late-time cutoff $\eta_0$ that is related to the end of inflation. 
Technically, the resulting boundary correlators satisfy {\it anomalous} conformal Ward identities.
In particular, for exchange diagrams with an intermediate massless state, the IR cutoff $\eta_0$ modifies the boundary differential equations with new source terms. As a result, the correlators contain $\eta_0$-dependent terms which must be accounted for when computing the full shape.
We also perform the bootstrap analysis using the wavefunction method, where the massless-exchange wavefunction coefficients remain IR-finite and the $\eta_0$-dependent divergent terms are found in the disconnected part.
As an important outcome, we derive the three-point and four-point ``seed functions" of massless exchanges for both dS-invariant and boost-breaking theories, from which more general shapes can be computed using differential (weight-shifting) operators. 

\item {\bf Classification of massless-exchange correlators.} The possible correlators of the inflaton $\phi$ from the single exchange of a massless scalar $\s$ fall in three categories:
\begin{itemize}
\item {\it Correlators with (approximate) dS symmetry}: two typical couplings here are $\dot\phi\s$ and $(\partial_\mu\phi)^2\s$.
As the simplest setup, the scalar bispectrum  contains IR-divergent terms,  
and the shape function has a mild logarithmic deviation from the local ansatz \eqref{local}:
\begin{small}
\begin{align} \label{true-local0}
S(k_1,k_2,k_3)  \propto &  \frac{ 1}{k_1^3k_2^3k_3^3} \Big[ \left(\gamma_E-3 - \log(-k_t\eta_0)\right)(k_1^3+k_2^3+k_3^3) +k_t e_2 -4 e_3   \\
&  +(k_2^3+k_3^3) \log(-2k_1\eta_0)
+   (k_1^3+k_3^3) \log(-2k_2\eta_0)
+   (k_1^3+k_2^3) \log(-2k_3\eta_0) \Big] .\nn
\end{align}\end{small}This result is derived and analyzed in \eqref{true-local} in Section \ref{sec:dS4pt3pt}.
The trispectrum is IR-finite, with the standard $\tau_{\rm NL}$ and $g_{\rm NL}$-type local forms. 
\item {\it Correlators from $\dot\phi\s$ and boost-breaking cubic interactions, with arbitrary $c_s$}: 
the bispectra here are also IR-divergent, with various new shapes that resemble the local shape in the squeezed limit. One example is given by
\be \label{ir-bispec0}
S(k_1,k_2,k_3)  \propto \frac{1}{k_1k_2k_3^3} \frac{c_sk_{12}+2k_3}{(c_sk_{12}+k_3)^2}\log\[-(1+c_s)k_3\eta_0\] + 2~{\rm perms}+...~,
\ee
where we have only kept the IR-divergent terms for demonstration.
See \eqref{dotphi2sigma} in Section \ref{sec:cs-IR} for more details.
For both the bispectra and trispectra, their sizes can become potentially large, and we identify richer analytical structure in their shape functions away from the squeezed limit.

\item {\it Multi-speed non-Gaussianity:} this is a new class of bispectrum shapes from higher-derivative quadratic interactions (e.g. $\dot\phi\dot\sigma$,  $\dot\phi^2\partial_i^2\s$) and multiple reduced sound speeds. As an example, a simple template with three sound speeds parameter $c_{1,2,3}$ is given by
\be \label{multics0}
 {S}^{{\rm multi}-c_s} (k_1,k_2,k_3)= \frac{1}{k_1k_2k_3(c_1 k_1+ c_2 k_2 +c_3 k_3)^3} +  5~{\rm perms}~.
\ee
There is no logarithmic divergences, and the IR-finite shapes are of the equilateral-type in terms of rational polynomials. However, as multiple sound horizon crossings are involved,
the peaks of shapes are shifted by the sound speed ratios. These simple results with possibly large sizes  provide new signatures of light degree of freedom during inflation, with rich phenomenology. See Section \ref{sec:multispeed} for more discussions.
\end{itemize}

\item {\bf Comparison with multi-field $\delta N$ analysis.} We compare the shape function \eqref{true-local0} from the bootstrap and the one computed within the $\delta N$ formalism in a benchmark example. 
Explicitly, we consider a simple two-field model, which can be analyzed by both the dS bootstrap and the $\delta N$ method. 
For the squeezed limit of the scalar bispectrum, we find agreement in these two results. Nevertheless, there is a {\it mismatch} away from the squeezed limit.
As the $\delta N$ formalism mainly focuses on the conversion on super-horizon scales, the bootstrap approach provides a more accurate shape function by also including sub-horizon field interactions.

\end{itemize}

\subsection{Outline and Reading Guide}

\paragraph{Outline} The rest of the paper is organized as follows. In Section \ref{sec:pandora}, we briefly review some key aspects of multi-field inflation, and present a model-independent reformulation of the conversion mechanism based on field interactions. In addition, we also list the goals and assumptions of the subsequent bootstrap analysis.
In Section \ref{sec:ir}, we perform a detailed analysis of IR divergences in the cosmological bootstrap with the presence of massless scalars. We study how an explicit IR cutoff modifies the boundary differential equations for correlators, and derive the scalar seed functions of massless exchanges in both the dS and boostless bootstrap. A complementary analysis using the wavefunction of the universe is presented in Appendix \ref{app:wave}.
In Section \ref{sec:nonG} we use weight-shifting operators to bootstrap a complete set of inflaton correlators from massless exchanges. We also discuss the phenomenology of the shapes of primordial non-Gaussianity.
In Section \ref{sec:revisit}, we compare the bootstrap results with the ones from the previous literature on multi-field inflation. Our conclusions are summarized in Section \ref{sec:concl}.

\paragraph{Reading Guide}
As these results are of interest for physicists with different research backgrounds, we strived to make the paper self-contained. Therefore, it may be helpful to provide a brief reading guide. 

\vskip4pt
Theoretical cosmologists familiar with the bootstrap may skip ahead to Section \ref{sec:ir}, which incorporates IR effects from massless exchanges into the boundary differential equations. Alternatively, they can read Appendix \ref{app:wave} if their preference is the wavefunctional perspective. Then Section \ref{sec:nonG} presents the classification of the correlators with massless external fields, which includes new phenomenology in boost-breaking scenarios. Sections \ref{sec:pandora} and \ref{sec:revisit} show how the bootstrap results are related to  previous analyses of multi-field inflation.

\vskip4pt
For experts who are more familiar with multi-field inflation, we recommend beginning with Section \ref{sec:pandora} to get familiar with our basic assumptions and notations. 
On a first reading, Section \ref{sec:ir} can be skipped,  while the reader may directly turn to Section \ref{sec:dS4pt3pt} for the dS bootstrap results for the primordial bispectrum \eqref{true-local} and trispectrum \eqref{trispec-dS}. 
Next, the comparison of these two results with the previous literature is demonstrated within a simple example in Section \ref{sec:revisit}. 
After that, we recommend reading Section \ref{sec:boostless}, where we discuss the new phenomenology associated to boost-breaking scenarios.

\paragraph{Notation and Conventions}
Throughout the paper, the metric signature is $−+++$. We use natural units  $\hbar = c ≡ 1$ and the reduced Planck mass $\mpl^2=1/8\pi G$. We use Greek letters for spacetime indices, $\mu = 0, 1, 2, 3$, Latin
letters for spatial indices, $i = 1, 2, 3$, and $a,b,c,...$ for internal field space indices. The background fields are denoted by $\Phi^a$, while fluctuations $\phi$ and $\s$ corresponds to the inflaton (adiabatic modes) and additional light scalars (isocurvature modes) respectively. 
The momentum of the $n$-th external leg of a correlator is denoted by ${\bf k}_n$ and its magnitude is $k_n ≡ |{\bf k}_n|$.
We use $k_t\equiv k_1+k_2+k_3$ as the total energy of three-point functions. In four-point functions,  the total energy is denoted by $k_T\equiv k_1+k_2+k_3+k_4$, 
and we mainly focus on the $s$-channel exchange with $s=|{\bf k}_1+{\bf k}_2|$.
Functions with a hat, such as $\hat{I}$, $\hat{F}$ and $\hat{\cG}$, are dimensionless by definition.

\newpage
\section{Disassembling the Pandora's Box of Multi-Field Inflation}
\label{sec:pandora}

In this section, we give a lightning review for some key aspects of multi-field inflation, and identify the universal features of the non-linear conversion process. This streamlines our analysis in the following sections, allowing us to say a few general things about multi-field inflation, despite the large freedom in terms of model building.

\vskip4pt
Light scalars with masses much smaller than the Hubble scale 
are ubiquitous in UV completions of inflation \cite{Baumann:2014nda}.
For instance, they could be moduli fields
arising from string compactifications, or they appear as pseudo-Nambu-Goldstone bosons from the breaking of a global symmetry.
Thus  the inflaton field may not be the only light scalar degree of freedom during inflation. 
When there are additional light fields, both the background dynamics and the behaviour of perturbations can become significantly different from the scenarios with only the inflaton. 
In general, the background evolution involving multiple scalars traces a complicated trajectory in field space, which in turn generates many interactions among these light scalars (see Ref.~\cite{Turzynski:2014tza, Renaux-Petel:2015mga, Achucarro:2016fby, Brown:2017osf, Achucarro:2017ing, Christodoulidis:2018qdw, Garcia-Saenz:2018ifx, Achucarro:2019pux, Achucarro:2019mea,  Bjorkmo:2019qno, Romano:2020kmj} for recent examples).
As a result, predictions of multi-field inflation are expected to be model-dependent, and the vast range of possible scenarios is like Pandora's Box, lacking some unifying theme.

\vskip4pt
However, we can still look for generic features of curvature perturbations when additional light fields are present, and try to extrapolate to more general lessons about multi-field inflation.
A key feature of multi-field inflation is the conversion from isocurvature perturbations to the adiabatic ones \cite{Gordon:2000hv}. They lead to the super-horizon evolution of curvature perturbations and change their statistics.
Based on the time when this conversion happens, multi-field models can be broadly classified as follows:

\begin{itemize}
\item {\it Conversion after inflation}: In this class of models, the additional light fields are spectators during inflation. One way to achieve this is to consider a two-field system with canonically normalized kinetic terms and a hierarchy between the inflaton mass and the extra field mass, such that the extra field rolls much slower than the inflaton and the field space trajectory can be approximated as a geodesic. As a result, the extra fields do not contribute to the curvature perturbations during inflation and the single field results remain unaffected. However, there can be some nontrivial effects at the end of inflation or in the post-inflation eras. Famous examples include the curvaton scenario \cite{Enqvist:2001zp, Lyth:2001nq, Moroi:2001ct} and modulated reheating \cite{Dvali:2003em}. In the former, after inflation the energy density of the curvaton field dominates over the energy density of the inflaton, and the curvaton fluctuations source the nonlinear evolution of curvature perturbations on the super-horizon scales. As has been extensively discussed in the literature, this process generates $\mathcal{O}(1)$ local non-Gaussianity.

\item {\it Conversion during inflation}: When multiple fields are actively involved in the inflationary background dynamics, the resulting trajectory can deviate from geodesic motion in field space, and the curvature perturbation suffers significant backreaction from these other fields during inflation. 
Depending on the choice of field basis, there are basically two approaches to describe this class of scenarios:
\begin{itemize}
\item {\it The ``multi-inflaton" analysis.} Since multiple scalars are dynamical in this scenario, one natural choice is to consider their evolution in some convenient field basis
\be
\Phi_1(t), ~\Phi_2(t), ~\Phi_3(t), ~...
\ee
For simplicity, in this approach the choice of field basis normally results in multi-field models with canonical kinetic terms and sum-separable/product-separable potentials, such that the background dynamics can be approximately solved. 
One simple but typical example of this class of models is double inflation, where we have two canonically normalized fields with a bowl-shaped potential, such as
\be \label{double}
V(\Phi_1,\Phi_2)= \frac{1}{2} m_1^2 \Phi_1^2 + \frac{1}{2} m_2^2 \Phi_2^2~.
\ee
This model has been well-studied in the literature \cite{Starobinsky:1985ibc, Langlois:1999dw}. In general, the inflaton rolls down along a bent trajectory.
The perturbations are usually analyzed using the $\delta N$ formalism (see Section \ref{sec:deltaN} for a brief review). In this scenario, the primordial non-Gaussianities are typically small, because field interactions are slow-roll suppressed. 
 In most cases, the conversion from isocurvature to adiabatic perturbations is not significant, and the models behave more like single-field inflation.

\item {\it The covariant formalism.} This approach begins with an adiabatic/isocurvature basis for the two types of perturbations \cite{Gordon:2000hv, GrootNibbelink:2000vx, GrootNibbelink:2001qt, Achucarro:2010da, Gong:2011uw,Achucarro:2012sm,Achucarro:2012yr}. 
The inflaton trajectory in the internal field space $\Phi^a(t)$ picks a tangential vector along the trajectory $T^a$, with the orthogonal directions parametrized by normal vectors $N^a$. 
Field fluctuations along $T^a$ are associated with adiabatic perturbations, while the others correspond to isocurvature.
The two types of perturbations are coupled if the inflation trajectory deviates from geodesic motion dictated by the metric of the field manifold, with all couplings having a geometrical interpretation.

\end{itemize}
\end{itemize}

To summarize, if the background expansion history is known in specific models, the $\delta N$ formalism provides a simple description for the nonlinear evolution of perturbations on super-horizon scales. This approach can also be applied for the conversion in post-inflation stages. 
Meanwhile, the covariant formalism may seem quite complicated for the analysis of specific models, as detailed information about the inflaton trajectory is needed. 
Previously this approach was mainly used in  studies of inflation models with curved field spaces and/or sharp-turn trajectories. 
However, one of its advantages is that field interactions  between the adiabatic and isocurvature perturbations take   constrained forms.
In the following we will take a closer look at the covariant formalism, and try to learn some generic lessons for the bootstrap analysis.

\begin{figure}
\centering
\includegraphics[scale=0.28]{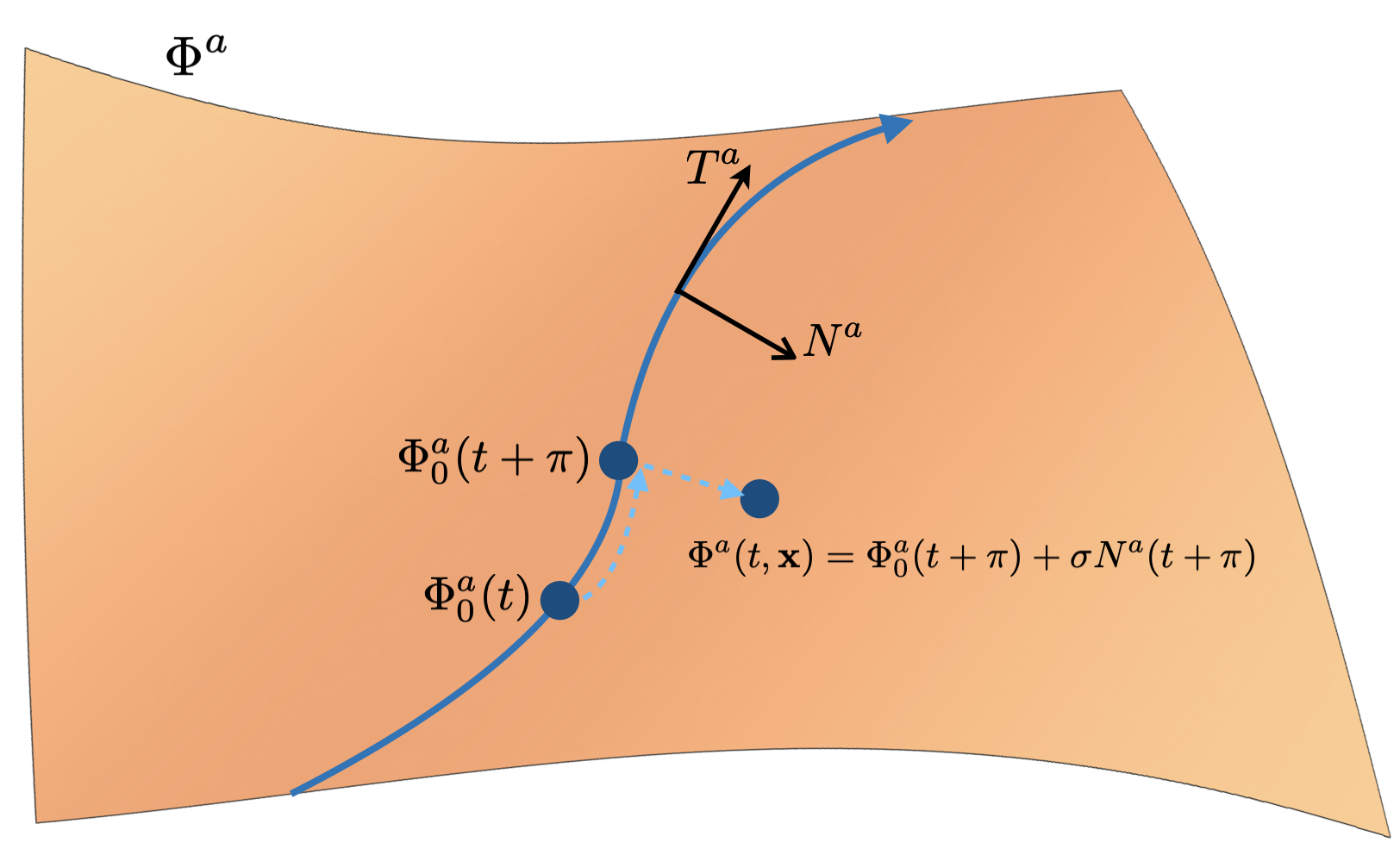}
\caption{An inflaton trajectory $\Phi_0^a(t)$ in a curved field manifold with multiple scalars $\Phi^a$. Adiabatic and isocurvatue modes are defined by the decomposition of field fluctuations along the tangent and normal vectors at each point of the trajectory. At the linear order $\s N^a$ is a vector living on the tangent space of one particular point on the manifold.}
\label{fig:traj}
\end{figure}

\subsection{Conversion from Interactions}
\label{sec:convert}
Let's look at a simple model to illustrate some of the points made above. Consider a theory with a set of light scalars $\Phi^a$ in a curved manifold with field space metric $G_{ab}$. A generic Lagrangian with two-derivative kinetic terms takes the form
\be \label{scalarL}
 \mathcal{L} = -\frac{1}{2} G_{ab}(\Phi) \partial_\mu\Phi^a \partial^\mu\Phi^b - V(\Phi) ,
\ee 
where both $G_{ab}$ and the potential are functions of the field coordinates. 
In multi-field inflation, the background trajectory is given by $\Phi_0^a(t)$ as shown in Figure \ref{fig:traj}. For the two-field case, the tangent and normal vectors of the trajectory are
\be
T^a \equiv \frac{\dot\Phi_0^a}{\dot\Phi_0}~,~~~~~~ N_a\equiv \sqrt{{\rm det}G} \epsilon_{ab} T^b~, 
\ee
where $\dot\Phi_0 = G_{ab} \dot\Phi_0^a \dot\Phi_0^b$ and $\epsilon_{ab}$ is the anti-symmetric tensor.
One important background parameter here is the turning rate $\Omega=-T^a D_t N_a$, with $D_t$ being the covariant derivative with respect to cosmic time $t$. The size of $\Omega$ tells us how much the inflationary trajectory  deviates from geodesic motion in field space.
For perturbations, we decompose them in the following way 
\be \label{pert}
\Phi^a(t,{\bf x}) \equiv \Phi^a_0(t+\pi) + \s N^a(t+\pi) = \Phi_0^a(t)+ \phi(t,{\bf x}) T^a + \s(t,{\bf x}) N^a + ...~
\ee
where $\cdots$ represent higher order contributions.\footnote{See Ref. \cite{Gong:2011uw} for a systematic study on the higher order perturbations  via a geometrical approach.} We see that $\phi = \dot\Phi_0 \pi$ is the canonically normalized fluctuations along the trajectory and $\s$ is the isocurvature modes.
Next, we substitute \eqref{pert} in the Lagrangian \eqref{scalarL} and identify the couplings between the perturbations. For the purpose of highlighting multi-field effects, we work in flat gauge and take the decoupling limit, where gravitational interactions vanish.
The curvature and isocurvature modes remain orthogonal along the background motion, which is a constraining feature. It implies that the general form of the quadratic and cubic interacting Lagrangian is fixed to be  \cite{Gong:2016qmq, Garcia-Saenz:2019njm} 
\be \label{int3}
 \mathcal{L}^{\rm int} = -\lambda \dot\phi \sigma  -g  (\partial_\mu\phi)^2 \sigma + \alpha  \dot\phi \sigma^2 + \beta \sigma^3
\ee
up to (small) contributions from the potential term.\footnote{The inflaton mass and self-interactions are suppressed by slow-roll parameters.
The masses of extra scalars receive contributions not only from the Hessian of the potential, but also from the turning  and field space curvature. For light fields, we assume the mass is much smaller than the Hubble scale, and the self-interactions are negligible.}
While we are only left with a few possibilities for interactions, $\dot\phi\s$ and $(\partial_\mu\phi)^2 \sigma$ are the two most important ones.\footnote{The other two cubic vertices may become important for models with highly curved field manifolds, since the couplings $\alpha$ and $\beta$ are related to the field space geometry \cite{Gong:2016qmq, Garcia-Saenz:2019njm}. However, they are not necessarily associated with the conversion mechanism of multi-field inflation. Therefore, we don't focus on those couplings in this paper.} 
First, the $\dot\phi\s$ linear mixing is responsible for the  conversion process in multi-field inflation. On super-horizon scales, in terms of the curvature perturbation $\zeta\equiv \frac{H}{\dot\Phi_0 } \phi $ and the isocurvature $\mathcal{S} \equiv \frac{H}{\dot\Phi_0 } \s $, the equation of motion approximately reduces to
\be \label{source}
\dot{\zeta} \simeq \lambda \hskip 2pt \mathcal{S}~,
\ee
which basically describes the growth of curvature perturbations sourced by the isocurvature modes. 
In addition, the couplings $\lambda$ and $g$ are both proportional to the turning rate $\Omega$:
\be
\lambda = 2 \Omega~,~~~~
g =- \frac{\Omega}{\dot\Phi_0} ~,
\ee
which means that these two interaction operators are expected to be nonzero as long as the inflaton trajectory is not geodesic.
Therefore, a complete treatment of the conversion process should include not only the linear mixing $\dot\phi\s$, but also the cubic coupling $(\partial_\mu\phi)^2 \sigma$, regardless of specific models. {This is not explicit in the ``multi-inflaton'' analysis for  models like double inflation \eqref{double} where the scalar fields are thought to be non-interacting. However, as multiple fields are rolling, these scalars actually  become  coupled as long as the trajectory deviates from field space geodesics.}

\vskip4pt
Generally speaking, the correlation of the quadratic and cubic interactions can be  seen as a consequence of the spontaneously broken boost symmetry during inflation.
To show this, let's take a look at the effective field theory of inflation \cite{Cheung:2007st} with an additional scalar $\s$. In this framework, the adiabatic degree of freedom is contained in the metric fluctuations, such as $\delta g^{00}\equiv 1+g^{00} $. Then at lowest derivatives, the mixing with the $\s$ field is given by\footnote{Notice that for the EFT with multiple scalars, here we adopted a different strategy from the construction in \cite{Senatore:2010wk}. We are particularly interested in the interaction operators responsible for the conversion, while Ref. \cite{Senatore:2010wk} parametrized the conversion effects via a $\delta N$-like field redefinition, and focused on other interactions among the Goldstone $\pi$ and extra light scalars. More comments are left to the end of Section \ref{sec:revisit}.}
\be
\mathcal{L}_{\rm int} \propto \delta g^{00} \s ~\longrightarrow ~ \big[-2\dot\pi + (\partial_\mu\pi)^2 \big] \s  ~,
\ee
where in the second step we have introduced the Goldstone boson $\pi$ from the breaking of the time-translation symmetry and taken the decoupling limit.
The particular form of $\delta g^{00}$ is fixed by the nonlinearly realized boost symmetry.
By using the field rescaling $\phi = \dot\Phi_0 \pi$, we see that these two interactions are the same with the first two terms in \eqref{int3}. Thus they have the unique origin from the same EFT operator, and the couplings are related to each other.\footnote{This is similar with the situation in the single field EFT  \cite{Cheung:2007st}, that a reduced sound speed is correlated with the enhanced cubic interaction $\dot\pi(\partial_\mu\pi)^2$, as they are both uniquely generated by the EFT operator $(\delta g^{00})^2$. Recently, the nonlinearly realized boost symmetry has been analyzed in the context of soft theorems in Ref. \cite{Hui:2022dnm}.}
As long as we have the conversion caused by the $\dot\phi\s$ linear mixing, the corresponding cubic interaction is also expected.
It is easy to check that this conclusion remains valid if we include higher-derivative operators in the EFT, though a systematic construction needs to be done for the EFT description of internal field manifold of inflation.

\subsection{Towards the Bootstrap of Multi-Field Inflation}
With this knowledge, now we move forward to  bootstrapping multi-field inflation. Our goal is to derive general results of cosmological correlators due to the presence of light scalar fields during inflation. 
To be specific, there are two novelties we aim to achieve via the bootstrap approach.

\vskip4pt
The first one is related to the comparison with previous studies.
Instead of using the $\delta N$ formalism and the separate universe approximation, here we would like to perform the first-principle computation of the primordial bispectra and trispectra based on field interactions.   
Particularly, we will focus on the conversion process from the isocurvature to adiabatic perturbations, and present a full description by using the mixed propagator from the $\dot\phi\s$ quadratic interaction.

\vskip4pt
Next, in addition to the simplest version of the conversion, here we will also systematically investigate all the possible boost-breaking interactions between curvature and isocurvature modes, and take into account the reduced sound speeds for these two perturbations.
In this most general setup, by releasing the power of the bootstrap, we will be able to go beyond the standard $\delta N$ analysis, and find a complete classification of cosmological correlators from additional light fields. New phenomenologies will be identified.

\vskip4pt
As a first step, it is important to draw some boundaries in the Pandora's box, and specify which regions can we derive definite answers by using the bootstrap method.
For this purpose, the fences are built as follows:

\begin{itemize}
\item First, we focus on the situations where the conversion happens during inflation. This simplifies the bootstrap analysis, as we will be allowed to exploit some of the de Sitter symmetries. For scenarios with post-inflation conversion, such as curvaton, we expect similar physics would apply, while it becomes more complicated to perform concrete computation based on field interactions. 

\item Second, we are interested in theories with (approximately) constant couplings for perturbations during inflation, such that the dS dilation symmetry is still respected, and perturbations are nearly scale-invariant.  For multi-field inflation, this requirement not only tells us all the model parameters should be constant, but also gives constraints on time dependence caused by the field space trajectories. In other words, we do not consider sharp-turn trajectories, but focus on the ones with (approximately) constant turning rate.

\item Third, we restrict ourselves to the perturbative regime. This first means that the dimensionless couplings of two fields are required to be smaller than unity. Furthermore, as logarithmic IR divergences are generally expected for massless scalar interactions in dS, a stronger condition is needed such  that in the regularized correlators, the IR-divergent term multiplied by the coupling is smaller than one. For instance, the $\lambda \dot\phi\s$ linear mixing leads to $ (\lambda/H)\log(-k\eta_0) \simeq  (\lambda/H) N_* <1$, where $\eta_0$ is the end of inflation, and $N_*$ is the number of e-folds from the horizon-exit time of the $k$ mode. We can trust the perturbative computation in this weakly coupled regime, but may need to consider the stochastic effects if we go beyond. In the literature, the condition $\lambda\ll H$ is normally known as the ``slow-turn approximation". 
\end{itemize}

The three conditions above define the {\it Elpis} region in the Pandora's box of multi-field inflation, where {\it hope} remains for a model-independent description. 
In fact, the conditions are satisfied by a majority of multi-field models, as we shall see through one particular example in Section \ref{sec:ssoi}. Meanwhile the {Elpis} region also contains more general theories of multiple interacting scalars, such as the ones with higher derivative couplings. 
With these preparations, we bootstrap the inflationary correlators with the presence of additional light scalars in the following two sections.

\section{IR Divergences in the Cosmological Bootstrap}
\label{sec:ir}

The cosmological bootstrap is based on the assumption that cosmological correlators become constant (or vanishing) at the future boundary of de Sitter space $\eta\rightarrow0$ (i.e. the end of inflation).  
However, we may encounter circumstances where the correlators keep growing before the end of inflation. This secular behaviour is a consequence of the well-known IR divergence of quantum field theory in de Sitter spacetime. 
In particular, cosmological correlators involving massless scalars typically contain logarithmic terms, which may become divergent in the late-time limit $\eta\rightarrow0$. 
At the practical level, as inflation must end, a nonzero conformal time $\eta_0$ is expected to provide an explicit late-time cutoff to regularize the singular correlators.

\vskip4pt
In this section, we present a systematic investigation of the IR-divergent correlators using the bootstrap approach.
After a brief review of the cosmological bootstrap, we begin with the examination of the IR divergence of the $\langle \vp\vp\phi\rangle$ correlator from contact interaction in Section \ref{sec:contact}, and show that how the explicit IR cutoff $\eta_0$ leads to the anomalous conformal Ward identities in the boundary perspective. 
After this warmup exercise, we move to consider the cases with exchange diagrams, for both the four-point function in de Sitter bootstrap in Section \ref{sec:4pt-ex} and the three-point function with a mixed propagator in Section \ref{sec:3pt-ex}. We find analytic expressions for these  ``seed" functions. They serve as building blocks for bootstrapping non-Gaussianities of multi-field inflation. In Section \ref{sec:wave}, we investigate the IR divergence in the wavefunctional approach, leaving further details to Appendix \ref{app:wave}.

\subsection{Recap of Bootstrap}
\label{sec:recap}

Let's first give a brief review of the cosmological bootstrap and explain our notations.
We fix the spacetime background to be the de Sitter (dS) Universe which serves as a good approximation for cosmic inflation
\be \label{dS}
ds^2=a(\eta)^2(-d\eta^2+d{\bf x}^2)~,~~~~~ a(\eta)=-\frac{1}{H\eta}~, ~~~~{\rm with}~ -\infty <\eta < \eta_0 ~.
\ee
where $H$ is the Hubble scale, $\eta$ is the conformal time, and $\eta_0 $ corresponds to the end of inflation.
The late-time limit  $\eta_0 \rightarrow 0$ can also be seen as the future boundary of dS.
Our goal is to find correlation functions of quantum fields on this boundary.
The standard approach to compute primordial correlators is the in-in (or Schwinger-Keldysh) formalism, where we need to track the field interactions in the bulk of dS (during inflation) and then derive observables on the boundary (at the end of inflation). 
The starting point of this bulk perspective computation is the propagation of free fields in dS.
We are mainly interested in the massless scalar $\phi$ with $m^2=0$ and the conformally coupled scalar $\vp$ with $m^2=2H^2$. Their mode functions in Fourier space are given by
\bea \label{phik}
\phi_{k}(\eta)  &=&  \frac{H}{\sqrt{2k^3}}(1+i k\eta)e^{-ik\eta}~,
\\ 
\label{vpk}
\vp_{k}(\eta) &=& \frac{iH\eta}{\sqrt{2k}}e^{-ik\eta}~.
\eea
For the bulk computation of their correlators, as the two fields correspond to the external lines of  Feynman diagrams, we introduce their 
 bulk-to-boundary propagators
\be \label{Kpm}
K_+(k,\eta) =  \phi_{k}(\eta_0) \phi^*_{k}(\eta) , ~~~~~~~~
K_-(k,\eta) =  \phi^*_{k}(\eta_0) \phi_{k}(\eta) ,
\ee
\be \label{Kvppm}
K_+^\vp(k,\eta) =  \vp_{k}(\eta_0) \vp^*_{k}(\eta) , ~~~~~~~~
K_-^\vp(k,\eta) =  \vp^*_{k}(\eta_0) \vp_{k}(\eta) ,
\ee
which describe the propagation of free fields from some bulk time $\eta$ to the late-time boundary at $\eta_0$. 
The massless scalar $\phi$ is related to inflaton fluctuations. There can also be additional massless fields $\s$ which will mix with $\phi$ through exchange diagrams. 
Their bulk-to-bulk propagators are given by
\bea \label{b2b}
G_{++}^\sigma(k,\eta, \eta') &=& \sigma_{k}(\eta) \sigma^*_{k}(\eta') \theta(\eta - \eta') +
\sigma^*_{k}(\eta) \sigma_{k}(\eta') \theta(\eta' - \eta) \nn\\ 
G_{+-}^\sigma(k,\eta, \eta') &=&  
\sigma^*_{k}(\eta) \sigma_{k}(\eta')  \nn\\ 
G_{-+}^\sigma(k,\eta, \eta') &=&  
\sigma_{k}(\eta) \sigma^*_{k}(\eta') \nn\\ 
G_{--}^\sigma(k,\eta, \eta') &=& \sigma_{k}(\eta) \sigma^*_{k}(\eta') \theta(\eta' - \eta) +
\sigma^*_{k}(\eta) \sigma_{k}(\eta') \theta(\eta - \eta') 
\eea
which describe the propagation of the $\s$ field between time $\eta$ and $\eta'$.
These propagators satisfy the following differential equation
\be \label{eom}
\mathcal{O}_\eta G_{\pm\pm}^\sigma(k,\eta, \eta') = \mp i H^2 \eta^2\eta'^2 \delta(\eta-\eta')  , ~~~~
\mathcal{O}_\eta G_{\pm\mp}^\sigma(k,\eta, \eta') = 0 ,
\ee
where $\mathcal{O}_\eta \equiv \eta^2 \partial^2_\eta - 2\eta \partial_\eta + k^2\eta^2$.
With these propagators, we can apply the Feynman rules to write down the in-in integrals over bulk time to compute boundary correlators. See Ref.~\cite{Maldacena:2002vr, Weinberg:2005vy, Chen:2010xka,Chen:2017ryl} for more details.

\vskip4pt
The idea of the cosmological bootstrap is that we can derive the self-consistent results of boundary correlators directly from basic principles, such as symmetries, unitarity and locality, without referencing to specific bulk evolutions. This ``boundary perspective" can be realized in various guises.
Here we mainly follow the symmetry-guided approach developed in \cite{Arkani-Hamed:2018kmz}.

\vskip4pt
As a maximally symmetric spacetime, the dS space \eqref{dS} has four types of isometries with the following Killing vectors
\begin{equation}
\begin{aligned}
P_i&= \partial_i\,, &  \qquad\quad D & = -\eta \partial_\eta - x^i\partial_i\,,\\
J_{ij} &= x_i\partial_j - x_j\partial_i\,, & K_i &=  2x_i \eta\partial_\eta +\left(2x^jx_i+(\eta^2- x^2) \delta^j_i\right)\partial_j\,.
\end{aligned}
\label{eq:dssymms}
\end{equation} 
While the spatial translation $P_i$ and rotation $J_{ij}$ act in the same way as in Minkowski spacetime, the dS dilation $D$ and the dS boosts $K_i$ require special attention.
In particular, the latter act as special conformal transformations (SCTs) on the late-time boundary. 
For field theories that respect all the dS isometries, their boundary correlators must be invariant under all these transformations. 
On the late-time boundary $\eta\rightarrow 0$, a general scalar has the scaling behaviour
\be \label{scaling}
\lim_{\eta\rightarrow 0} \s({\bf k},\eta)= O^+({\bf k})\eta^{\Delta^+} + O^-({\bf k})\eta^{\Delta^-}
\ee
where the scaling dimensions are determined by the scalar field mass
\be
\Delta^\pm = \frac{3}{2}\pm i \mu~,~~~~~~  \mu = \sqrt{\frac{m^2}{H^2}-\frac{9}{4}}~.
\ee
An important observation is that the $O^\pm$ operators satisfy the transformation rules of the three-dimensional conformal group. Therefore they can be seen as primary operators with weights $\Delta_\pm$ in conformal field theory (CFT), and the structure of their correlators $\langle O^n\rangle$ is strongly constrained by the conformal symmetry. In the boundary perspective, these CFT correlators are the object of interest which we would like to bootstrap.

\vskip4pt
Before spelling out the conformal symmetry constraints on correlators, let us clarify the notations first. 
Following the standard convention, we shall mainly focus on the correlation functions of $O^-$ operators in the rest of the paper, and drop the superscript for convenience. The result of the dual operator $O^+$ can be obtained via a rescaling $O^+=k^{\Delta^+-\Delta^-}O^-$.
Also, for a general scalar, we set the conformal dimension $\Delta  = \Delta^-$, and use $\s_\Delta$ to denote the bulk field and $O_\Delta= O^-$ for the boundary CFT operator.
As we are mainly interested in the massless exchange in this work, we shall drop the subscript $\Delta$ and simply use $\s$ for the internal massless scalar. 
For the two external fields in \eqref{phik} and \eqref{vpk}, the massless scalar $\phi$ corresponds to $\Delta =3$  and the conformally coupled scalar has $\Delta=2$.
For light fields  with $m<3H/2$, the $\Delta_+$ fall-off dominates at the late time $\eta_0$, and thus the $O^+$ operator contributes to the $\s_\Delta$-correlators with $\s_\Delta ({\bf k},\eta_0)=\eta_0^{\Delta^+} O^+({\bf k}) =\eta_0^{3-\Delta} k^{3-2\Delta} O_\Delta({\bf k}) $.
Explicitly, the $n$-point cosmological correlator of these light scalars $\s_{\Delta}$ is related to the boundary CFT correlator through
\be \label{cc-cft}
\langle \s_{\Delta_1}({\bf k}_1)\s_{\Delta_2}({\bf k}_2)...\s_{\Delta_n} ({\bf k}_n)\rangle'
= \eta_0^{3n-\Delta_t}  k_1^{3-2\Delta_1}
...k_n^{3-2\Delta_n}
\langle O_{\Delta_1}({\bf k}_1)O_{\Delta_2}({\bf k}_2)...O_{\Delta_n}({\bf k}_n) \rangle'~,
\ee
where $\Delta_t=\Delta_1+\Delta_2+...+\Delta_n$ and the prime means that we have stripped the momentum-conservation $\delta$-function in the correlators. In the end, we are interested in computing inflaton correlators with $\Delta_t=3n$ such that the decaying prefactor of $\eta_0$ vanishes. But we shall also consider correlators with conformally coupled fields ($\Delta=2$) in the intermediate steps of the bootstrap.

\vskip4pt
Now let's look at how the dS dilation and SCTs act on the boundary CFT operators. In Fourier space, we have
\bea \label{dilak}
D O_\Delta &=& \(  3-\Delta+k^j\partial_{k^j} \) O_\Delta
\\[4pt] \label{sctk}
K_i O_\Delta &=& \[ 2(\Delta-3)\partial_{k^i} -2k^j\partial_{k^j}\partial_{k^i} +k_i\partial_{k^j}\partial_{k_j} \] O_\Delta~.
\eea
As a result, the boundary correlators satisfy the {\it conformal Ward identities} associated with the two symmetries above
\bea
 \label{cwid}
 \left[ \hskip 2pt-3+ \sum_{a=1}^n D_{a}  \right] \langle {O}_{\Delta_1} \cdots {O}_{\Delta_a} \cdots {O}_{\Delta_n} \rangle' &=&0\, ,   \\[4pt]
 \sum_{a=1}^n K_{a}^i\, \langle {O}_{\Delta_1} \cdots {O}_{\Delta_a} \cdots {O}_{\Delta_n} \rangle' &=& 0 \label{cwis}
  \, ,
\eea
where $D_{a}$ and $K_{a}^i$ are differential operators given in \eqref{dilak} and \eqref{sctk}, with ${\bf k} \rightarrow {\bf k}_a$ and $\Delta\rightarrow \Delta_a$.
While the dilation Ward identity simply require the correlators to be scale-invariant, the special conformal Ward identities provide a set of boundary differential equations that determine the functional form of the $n$-point function.
Solving these differential equations, one can directly bootstrap boundary correlators with full generalities.

\vskip4pt
There is one subtlety in the above analysis.
On the boundary, as the time-dependence of $\s_\Delta$ has been separated into the scaling behaviours in \eqref{scaling} and $O_\Delta$ operators are constant, it is typical to assume that the CFT correlators $\langle O_{\Delta_1}...O_{\Delta_n} \rangle'$  are also time-independent.
In cosmology this provides good description for many circumstances, as correlators are expected to be frozen on super-horizon scales before the end of inflation.
However, the assumption of time independence breaks down for correlators that become singular at the late-time limit $\eta_0\rightarrow 0 $. This circumstance is typically associated with IR divergences in dS
when massless fields are involved.
From the perspective of the boundary CFT, it corresponds to the situation where  $\langle O_{\Delta_1}...O_{\Delta_n} \rangle'$ diverge and the renormalization leads to conformal anomalies.
 In the rest of this section, we shall look at these particular cases and demonstrate how to bootstrap the singular correlators on the boundary.

\subsection{Contact Three-Point Function $\langle \vp\vp\phi \rangle$}
\label{sec:contact}

Now we consider modifications of the cosmological bootstrap due to the presence of IR divergences. 
As a warmup, we first study the contact three-point function $\langle \vp\vp\phi \rangle$. 
To characterize the differences from the IR-finite cases, we examine this simple example from both bulk and boundary perspectives.

\begin{figure} [h]
   \centering
            \includegraphics[width=.28\textwidth]{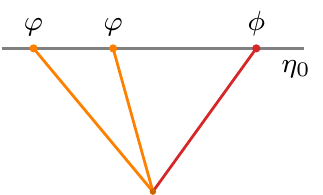}  
   \caption{As a warm up, we look into the IR behaviour of the contact interaction with two conformally coupled scalars $\vp$ (with $m_\vp^2=2H^2$) and one massless field $\phi$.}
  \label{fig:contact}
\end{figure}

\vskip4pt
Let's first take a look at the bulk computation. By assuming a contact interaction $\vp^2\phi$, we can easily compute the three-point correlator 
\bea \label{vvp}
\langle \vp_{k_1}\vp_{k_2} \phi_{k_3} \rangle' &=& i\int_{-\infty}^{\eta_0} d\eta a(\eta)^4 \[K_+^\vp(k_1,\eta)K_+^\vp(k_2,\eta)K_+ (k_3,\eta)-c.c. \] +{\rm perm.}\nn\\
&=& \frac{H^2\eta_0^2}{8k_1k_2k_3^3}{{I}}_{\vp\vp\phi}\(k_{12}, k_3, \eta_0 \) + {\rm perm.} ~,
\eea
where the bulk integral is given by
\be \label{Ivvp}
 {I}_{\vp\vp\phi} \equiv {i} \int_{-\infty}^{\eta_0} \frac{d\eta}{\eta^2} \[ e^{i k_t \eta }(1-ik_3\eta) - c.c. \] = 2k_3 -2k_{12}\[\gamma_E -1 +\log(-k_t\eta_0) \] ~,
\ee
with $k_{12}=k_1+k_2$  and $k_t=k_{12}+k_3$.
Note that we have explicitly introduced the end of inflation  $\eta_0$ as the upper limit of the integration, and taken $|\eta_0|\ll 1$ in the final result.\footnote{\label{eta000}
There is one subtlety about the $\langle \vp\vp  \phi \rangle$ correlator: in principle, the ${I}_{\vp\vp\phi}$ integral should be given by
\be 
 {I}_{\vp\vp\phi} \equiv {i} \int_{-\infty}^{\eta_0} \frac{d\eta}{\eta^2} \[ e^{i k_t \eta +i k_{12}\eta_0 }(1-ik_3\eta) - c.c. \] = 2k_3 -2k_{12}\[\gamma_E +\log(-k_t\eta_0) \] ~,
\ee
where $e^{\pm ik_{12}\eta_0}$ from boundary mode functions in \eqref{Kvppm} change the coefficient of the $k_{12}$ term. 
As this difference is irrelevant when we consider inflaton correlators with derivative interactions, for simplicity ${I}_{\vp\vp\phi}$ is defined as the one without these $e^{\pm ik_{12}\eta_0}$ terms. 
We would like to thank Enrico Pajer for pointing this out.}
The correlator $\langle \vp\vp  \phi \rangle$ is actually vanishing in the late-time limit because of the $\eta_0^2$ prefactor from $\vp$-propagators. To highlight the logarithmically divergent term, we focus on the CFT correlator $\langle O_\vp O_\vp O_\phi \rangle' \propto I_{\vp\vp\phi} $.
Without explicitly solving the integral, its IR divergence can be identified by noticing that
\be \label{tdt}
\eta_0\partial_{\eta_0} I_{\vp\vp\phi} = -2k_{12}~.
\ee
Meanwhile, we notice that the late-time cutoff $\eta_0$ introduces a new scale in the correlator, which explicitly breaks the dilation constraint of the conformal group. 
Indeed we find the conformal Ward identity in \eqref{cwid} is violated
\be
\[~-3+\sum_{a=1}^{3}D_a\]\langle O_\vp O_\vp O_\phi \rangle' ~\propto ~ \[~-1+ \sum_{a=1}^{3} k_a^i \partial_{k_a^i}\] I_{\vp\vp\phi} = -2 k_{12}~.
\ee
In the CFT language, this corresponds to the {\it anomalous} conformal Ward identity of dilation when the renormalized correlators contain logarithms \cite{Bzowski:2013sza, Bzowski:2015pba, Bzowski:2018fql, Bzowski:2019kwd, Cespedes:2020xqq, Bzowski:2022rlz}. Instead of focusing on the conformal boundary, we may also restore the time dependence and check the constraint equation on equal-time correlators \cite{Cespedes:2020xqq}  from the dS dilation in \eqref{eq:dssymms}:
\be \label{cwidt}
\[-\eta_0\partial_{\eta_0}-3+\sum_{a=1}^{3}D_a\]\langle O_\vp O_\vp O_\phi \rangle' = 0~.
\ee
Thus the conformal anomaly is precisely cancelled by the $\eta_0\partial_{\eta_0} $ term in \eqref{tdt}, and the dS dilation isometry is not broken.

\vskip4pt
Next, let's consider the boundary perspective. A similar three-point function has been analyzed in \cite{Arkani-Hamed:2015bza, Arkani-Hamed:2018kmz}, with the massless field $\phi$ being replaced by a general scalar $\s_\Delta$.
There, from the symmetry constraints, the boundary CFT correlator can be expressed as 
\be
\langle O_\vp ({\bf k}_1) O_\vp ({\bf k}_2) O_\Delta ({\bf k}_3) \rangle' = k_3^{\Delta-2} {\hat{I}}_{\vp\vp\Delta} (u)~, ~~~~~~{\rm with} ~~ u\equiv k_3/k_{12}~.
\ee
Furthermore, it has been shown that the conformal Ward identities of dilation and SCTs in \eqref{cwid} and \eqref{cwis} lead to the homogeneous differential equation\footnote{Recall that the conformal weight $\Delta$ is related to the mass of the $\s_\Delta$ field via $ (\Delta-1) (\Delta-2)=2-m^2/H^2 $.} \cite{Arkani-Hamed:2018kmz}
\be \label{eq:IvvD}
\Big[\Delta_{{u}} -(\Delta-1) (\Delta-2) \Big]  \hat{I}_{\vp\vp\Delta}  = 0~,
\ee
with the differential operator of $u$ defined by
\be \label{deltau}
\Delta_u \equiv u^2(1-u^2)\partial_u^2 -2u^3\partial_u~.
\ee
This equation can be solved by noticing that the correlator should be regular at the folded limit $k_{12}=k_3$ as a consequence of the Bunch-Davies vacuum. Thus, using the absence of singularity at $u\rightarrow1$ as a boundary condition, we find the hypergeometric solution
\be
\hat{I}_{\vp\vp\Delta}  \propto {}_2F_1\[2-\Delta,\Delta-1; 1; \frac{u-1}{2u}\]~.
\ee
If we want to generate the result with a massless scalar by choosing $\Delta = 3$ here,  we  find $\hat{I}_{\vp\vp\Delta} \propto u^{-1}$, which differs from the bulk computation in \eqref{Ivvp}.  
This mismatch is expected, since the $\langle O_\vp O_\vp O_\phi \rangle'$ correlator does not satisfy the conformal Ward identities as we have shown. Therefore, one can no longer use the constraint equations in \eqref{cwid} and \eqref{cwis} to derive the boundary differential equation in \eqref{eq:IvvD}.

\vskip4pt
Does this signal the breakdown of the boundary approach when we have IR-divergent correlators due to the presence of massless scalars? Or could there be another way to derive the boundary differential equation when there are IR divergences?
The major problem here is that fixing the boundary at $\eta_0$ explicitly breaks the dilation symmetry. Therefore, we may apply a simple trick to bypass this issue by introducing a rescaled cutoff $x_0=k_3\eta_0$. Then the upper limit of the bulk integral in \eqref{Ivvp} becomes $x_0/k_3$. Now we do not solve the integral in \eqref{Ivvp} explicitly, and notice that the bulk-to-boundary propagator of the massless scalar satisfies
\be
\(k^2\partial_k^2 -2k\partial_k + k^2\eta^2 \)\[ e^{i k \eta}(1-ik\eta ) \] =0 ~.
\ee
Using this differential equation, we find that $I_{\vp\vp\phi}$ satisfies
\be
\(k_3^2\partial_{k_3}^2 -2k_3\partial_{k_3} - k_3^2\partial_{k_{12}}^2 \) I_{\vp\vp\phi} = -6k_{12}~.
\ee
The source term is generated when the $k_3$-derivatives hit on the upper-limit of the integral $x_0/k_3$. Next, we consider the dimensionless function $\hat{I}_{\vp\vp\phi}=I_{\vp\vp\phi}/k_3$ which depends on the ratio $u\equiv k_3/k_{12}$ only.  We find the differential equation
\begin{eBox}
\be \label{eq:Ivvp}
\(\Delta_{{u}} -2 \) \hat{I}_{\vp\vp\phi} (u) = -\frac{6}{u}~,
\ee\end{eBox}
with $\Delta_u$ being the differential operator introduced in \eqref{deltau}. This inhomogeneous boundary equation with a nontrivial source provides the modified version of \eqref{eq:IvvD} for $\Delta=3$. The appearance of this source term is due to the fact that the massless scalar becomes constant on super-horizon scales. If we perform the same derivation for $\hat{I}_{\vp\vp\Delta}$ with general massive scalar $\s_\Delta$, we find a decaying source term with a positive power of $x_0$.
Therefore, by taking the $x_0\rightarrow0$ limit, the source term vanishes, and we reproduce the boundary equation \eqref{eq:IvvD}. 
From the CFT point of view, we suspect that \eqref{eq:Ivvp} may be seen as a consequence of the anomalous special conformal Ward identities.

\vskip4pt
Solving \eqref{eq:Ivvp}, we find the general solution
\be
\hat{I}_{\vp\vp\phi} (u) = c_1\frac{1}{u} + c_2 \[ 1+\frac{1}{2u} \log\( \frac{1-u}{1+u}\)\] -\frac{1}{u} \log \( \frac{1-u^2}{u^2}\)~,
\ee
with two free constants. Again, one boundary condition is given by the absence of the folded singularity at $u\rightarrow1$, which fixes $c_2=2$. The other constant $c_1$ is related to the cutoff scale which is arbitrary. This can be normalized by imposing the soft behaviour  $\lim_{k_3\rightarrow 0}\hat{I}_{\vp\vp\phi}$, 
which gives $c_1=-2[\gamma_E -1  + \log(-x_0)]$ (or alternatively, at least in part, by using the scaling behavior in \eqref{tdt}). In the end we find
\be \label{Ivvphat}
\hat{I}_{\vp\vp\phi} (u,x_0) = 2-\frac{2}{u}\[ \gamma_E -1 + \log\(\frac{1+u}{u}\)+ \log(-x_0) \]~,
\ee
which matches the bulk computation \eqref{Ivvp}, after restoring $x_0=k_3\eta_0$.
As expected, the final result only contains the total-energy pole, while the suspicious logarithmic $k_3$-pole is cancelled.

\vskip4pt
Although this warmup example is simple and can be easily computed from direct bulk integration, there are lessons about how to treat IR divergences in the cosmological bootstrap, and we shall get back to this contact example in the subsequent analysis of massless exchange diagrams. 
We close this section with a few observations:
\begin{itemize}
\item With no need for solving the bulk integral, one simple criterion to tell if a correlator is IR-divergent or not is to use the $\eta_0\partial_{\eta_0}$ operator. Only if 
\be
\lim_{\eta_0\rightarrow0} ~\eta_0\partial_{\eta_0} \langle O^n \rangle \rightarrow0~,
\ee
the correlator is IR-finite, and one can safely take the late-time cutoff to $0$, otherwise one needs to be careful with the singular behaviour of boundary correlators. 
This condition becomes useful when we consider exchange diagrams for which the explicit integration may become difficult.
Meanwhile, as we can see from the time-dependent dS dilation constraint on equal-time correlators \eqref{cwidt}, this criterion also tells us if the dilation conformal Ward identity in \eqref{cwid} remains valid, or becomes the anomalous one.
\item  The  singular behaviour of the boundary correlators is typically associated with massless fields with no derivatives in the interaction vertices, as they become constant on super-horizon scales and keep contributing in the bulk integral. For massive fields which decay after horizon-exit, the  correlators are regular. For massless fields with derivative interactions, the  correlators are given by rational polynomials with no logarithmic terms \cite{Goodhew:2022ayb}.
\item For IR-divergent correlators, the conformal Ward identities become the anomalous ones. As a new scale, the cutoff $\eta_0$ explicitly breaks scale-invariance. One useful trick to ``restore" the dilation symmetry is to consider a dimensionless cutoff by rescaling $x_0=k\eta_0$. As a consequence, the boundary differential equation acquires one extra source term. We will see this behaviour again in the analysis of exchange processes.
\item In the end, we notice that the IR divergence is also present in other contact $n$-point functions with one or more massless scalar fields.
Another well-known example is the $\langle \phi\phi \phi \rangle$ correlator from the $\phi^3$ interaction
\bea \label{3pt-self}
\langle \phi_{k_1}\phi_{k_2} \phi_{k_3} \rangle' &=& i\int_{-\infty}^{\eta_0} d\eta a(\eta)^4 \[K_+ (k_1,\eta)K_+ (k_2,\eta)K_+ (k_3,\eta)-c.c. \]+{\rm perm.} \nn\\
&=&\frac{H^2}{2k_1^3k_2^3k_3^3}\[(k_1^3+k_2^3+k_3^3) \(\gamma_E-1+ \log(-k_t\eta_0)\)+4e_3-e_2 k_t \]~,
\eea
with $e_2=k_1k_2+k_1k_3+k_2k_3$ and $e_3=k_1k_2k_3$.
This is known as the conformal non-Gaussianity from the inflaton self-interaction, and can also be analyzed in the same approach from the boundary perspective \cite{Pajer:2016ieg}.
\end{itemize}

\subsection{Massless Exchange in dS Bootstrap}
\label{sec:4pt-ex}

Now we are ready to investigate the IR divergences in exchange diagrams. In this section, we shall focus on the seed function of dS bootstrap, which is the four-point function of conformally coupled scalars exchanging one additional scalar field, while we leave the analysis of the three-point scalar seed of the boostless bootstrap in Section \ref{sec:3pt-ex}.

\begin{figure} [h]
   \centering
            \includegraphics[width=.4\textwidth]{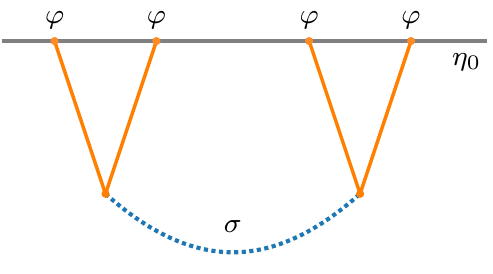}  
            \caption{The four-point scalar seed function of massless exchange.}
  \label{fig:4ptseed}
\end{figure}

\vskip4pt
Let's restrict our discussion to the $s$-channel contribution to the tree-level exchange, where $s\equiv |{\bf k}_1+{\bf k}_2|$ is the Mandelstam-like variable.
In the dS bootstrap, because of the symmetry constraints on kinematics, the boundary four-point correlator of $\vp$ mediated by a general scalar $\s_\Delta$ with mass $m$ can be expressed in the following form
\be
\langle O_\vp ({\bf k}_1) O_\vp ({\bf k}_2) O_\vp ({\bf k}_3)O_\vp ({\bf k}_4) \rangle' = \frac{1}{s} \hat{F}(u,v)~,~~~~{\rm with}~~~~ u\equiv \frac{s}{k_{12}}~,~~v\equiv \frac{s}{k_{34}}~,
\ee
where $\hat{F}$ is the so-called four-point scalar seed function, which depends on two momentum ratios $u$ and $v$ only. It was shown that the conformal Ward identities of SCTs in \eqref{cwis} lead  to a set of differential equations for $\hat{F}$
\bea \label{eq:4pt}
\[\Delta_u + \frac{m^2}{H^2} -2\] \hat{F} &=& \frac{uv}{u+v}~,\nn\\
\[\Delta_v + \frac{m^2}{H^2} -2\] \hat{F} &=& \frac{uv}{u+v}~,
\eea
where $\Delta_u$ is the differential operator given in \eqref{deltau}.
Solving this equation with proper consideration of boundary conditions from singularities, we can derive the full analytical results of the massive exchange.
In this work we are interested in the situation where we take the intermediate scalar mass to zero. 
We would like to examine if this seed function $\hat{F}$ becomes IR-divergent, and if it does, how the differential equation \eqref{eq:4pt} will be modified.

\vskip4pt
We first notice that the boundary four-point function above corresponds to the following bulk computation of the correlator for conformally coupled scalars
\be
 \langle \vp_{k_1}\vp_{k_2} \vp_{k_3} \vp_{k_4} \rangle'
 = \frac{\eta_0^4H^{2}}{2k_1k_2k_3k_4 s} \hat{F}(k_{12},k_{34},s)  \ +\ \text{$t$- and $u$-channels}\, ,
\ee
where the integral form of the seed function $\hat{F}$ is given by
\be \label{Fhat}
\hat{F} = -\frac{s}{2H^2}\int_{-\infty}^{\eta_0} \frac{d\eta}{\eta^2}  \int_{-\infty}^{\eta_0} \frac{d\eta'}{\eta'^2}\left[
\sum_{\pm\pm} (\pm)(\pm)e^{\pm i k_{12}\eta  \pm i k_{34} \eta'}  {G}_{\pm\pm}(s,\eta,\eta')  \right] .
\ee
Again, we have kept the late-time cutoff $\eta_0$ explicit as the upper limit of the integration. In this integral form, we also neglected the $e^{\pm ik_n\eta_0}$'s from  $\vp$-propagators in \eqref{Kvppm}.\footnote{Like we discussed in footnote \ref{eta000} for the $\langle \vp\vp\phi \rangle$ correlator, in principle the $\langle \vp^4 \rangle$ correlator corresponds to the double integral with $e^{\pm ik_{12,34}\eta_0}$'s from the  boundary $\vp_k(\eta_0)$.
The revised integral has similar IR behaviour but more complicated form.
As our goal is to bootstrap inflaton correlators with derivative interactions, the difference becomes negligible 
after the weight-shifting procedure (see Section \ref{sec:nonG}). Thus we shall use $\hat{F}$ as the scalar seed for analysis, but notice that this subtlety may become nontrivial for correlators from non-derivative interactions.}
Here ${G}$ is the bulk-to-bulk propagator introduced in \eqref{b2b}. This is a nested double integral, which becomes more difficult to solve. To trace its IR behaviour, let's take the $\eta_0 \partial_{\eta_0}$ operation on $\hat{F}$
\be \label{deta0F}
\eta_0 \partial_{\eta_0} \hat{F} = -\frac{1}{2s^2} \Big[ k_{12}I_{\vp\vp\phi}(k_{34},s,\eta_0) +k_{34}I_{\vp\vp\phi}(k_{12},s,\eta_0) \Big]~,
\ee
where $I_{\vp\vp\phi}$ is the integral introduced in \eqref{Ivvp}, which contains logarithmic divergence. Therefore the four-point scalar seed of massless exchange has explicit $\eta_0$-dependence, and becomes singular in the $\eta_0\rightarrow0$ limit. As a result, the boundary equation \eqref{eq:4pt} should be modified when $m=0$.

\vskip4pt
Due to the presence of the cutoff scale $\eta_0$, the scalar seed $\hat{F}$ may not be a function of two momentum ratios $u$ and $v$ only.
To remove the explicit $\eta_0$ dependence, we use the dimensionless cutoff $x_0=s\eta_0$, and then rescale $\eta'$ by
$x=s\eta'$.
The $\hat{F}$ function becomes
\be
\hat{F} = -\frac{1}{2s}\int_{-\infty}^{x_0/s} \frac{d\eta}{\eta^2}  \int_{-\infty}^{x_0} \frac{dx}{x^2}\left[
\sum_{\pm\pm} (\pm)(\pm)e^{\pm i k_{12} \eta  \pm i  x /v} \hat{G}_{\pm\pm}(s\eta,x)  \right] ,
\ee
where $\hat{G}(s\eta,s\eta')=s^3G(s,\eta,\eta')/H^2$ is the dimensionless bulk-to-bulk propagator. 
One nontrivial consequence of this rescaling is that the upper limit of the $\eta$ integral now becomes $s$-dependent.
Without solving this nested double integral, we notice that the $G$-propagators satisfy Eq. \eqref{eom}. 
As the dependence of $\hat{G}$ on $\eta$ and $\eta'$ is through the combination $s\eta$ and $s\eta'$, we can trade $\eta$-derivatives with $s$-derivatives. 
 To derive the differential equation for $\hat{F}$ in terms of $u\equiv s/k_{12}$, we first set $v\equiv s/k_{34}$ to be a constant. 
Then we find
\be
\frac{1}{s} \( s^2\partial_s^2-2s\partial_s -s^2 \partial_{k_{12}}^2 \) \(s \hat{F} \)  =\frac{s}{k_T} -\frac{3k_{12}}{s v} \[1-\gamma_E+v + \log\(\frac{v}{1+v}\)-\log(-x_0) \]~.
\ee
The first source term is the standard contact term in the dS bootstrap, which can be seen as a result of collapsing the internal line.
The second source term, which has the form of the $\langle \vp\vp\phi \rangle$ correlator in \eqref{Ivvp}, is generated when the $s$-derivatives act on the upper limit of the $\eta$ integral. Changing variable to $u$, we find the differential equation
\begin{eBox}
\be \label{eq:4pt-ir}
(\Delta_u -2) \hat{F} = \frac{uv}{u+v} - \frac{3}{2u} \hat{I}_{\vp\vp\phi}(v, x_0)~,
\ee
\end{eBox}
with $\hat{I}_{\vp\vp\phi}$ given in \eqref{Ivvphat}.
If we rescale another integration variable $\eta$, we find the second differential equation in terms of $u$, which can also be obtained simply from the permutation symmetry $u \leftrightarrow v$. Comparing with the IR-finite equations \eqref{eq:4pt}, we find 
an additional source term for correlators which become singular on the late-time boundary. This result is in analogy with what we have shown for contact interactions in \eqref{eq:Ivvp}.
Schematically, when an explicit late-time cutoff is present, the $(\Delta_u -2)$ operator reduces the four-point exchange diagram into the contact one, as well as the three-point function $\langle \vp\vp\s \rangle$ by taking the internal line to the boundary.
For the exchange of a general massive scalar $\s_\Delta$, it is easy to apply the same derivation, and in the end we find the second source term is simply given by  $\propto x_0^{\Delta^{\pm}}\hat{I}_{\vp\vp\Delta}$. Thus, in the late-time limit $x_0\rightarrow0$, this term vanishes, and we return to the equations in \eqref{eq:4pt}. 

\vskip4pt
We now wish to find the solution for this modified differential equation of massless exchange. Let's first take a look at the $u$-equation with $v$ being a constant.
Its general solution can be expressed in a closed-form, which we first separate into two parts
\be \label{Fhat2}
\hat{F}  = \hat{F}_{\rm fin} (u,v) + \hat{F}_{\rm div} (u,v,x_0) ~.
\ee
Let's first look at $\hat{F}_{\rm fin}$. This is the IR-finite part of the solution, which satisfies $(\Delta_u -2)\hat{F}_{\rm fin} = uv/(u+v)$. This solution does not depend on the IR cutoff $\eta_0$, and has been derived in \cite{Arkani-Hamed:2018kmz} \footnote{In \cite{Arkani-Hamed:2018kmz, Baumann:2020dch, Bzowski:2022rlz}, there are differences for the last term in the first line because of choices of the boundary condition at $u,v\rightarrow 0$. As this term can be moved to the homogenoues solution, without losing generality here we choose $-\pi^2/6$ which makes the terms in the bracket vanish at $u\rightarrow0$.}
\bea
\label{eq:Ffin}
\hat{F}_{\rm fin} &= &\,-\frac{1}{2uv} \[{\rm Li}_2\left(\frac{u(1-v)}{u+v}\right)+ {\rm Li}_2\left(\frac{v(1-u)}{u+v}\right)+ \log\left(\frac{u(1+v)}{u+v}\right)\log\left(\frac{v(1+u)}{u+v}\right)-\frac{\pi^2}{6}\]\nn \\
&&~+ \frac{1}{v} \log\left(\frac{u(1+v)}{u+v}\right) + \frac{1}{u} \log\left(\frac{v(1+u)}{u+v}\right) -1 \,, 
\eea
where ${\rm Li}_2(x)$ is the dilogarithm. To analyze its analytical structure, we notice that $\hat{F}_{\rm fin}$ has IR-finite logarithmic poles, which can be classified into total-energy pole at $u+v\propto k_T \rightarrow 0$, and partial-energy poles at $u+1 \rightarrow 0$ and $v+1 \rightarrow 0$.

\vskip4pt
The second term  $\hat{F}_{\rm div}$ has been missed in previous considerations. 
It corresponds to the singular piece of the  solution that
has been regularized by the IR cutoff and satisfies $(\Delta_u -2)\hat{F}_{\rm div} = -3 \hat{I}_{\vp\vp\phi}/2u$. 
Solving this equation explicitly, we find $\hat{F}_{\rm div}$ is given by a sum of the particular and  homogeneous solutions 
\bea 
\hat{F}_{\rm div} &=& -\frac{1}{4u}\log\( \frac{1-u^2}{u^2} \) \hat{I}_{\vp\vp\phi}(v, x_0) +c_1\frac{1}{u} + c_2 \[ 1+\frac{1}{2u} \log\( \frac{1-u}{1+u}\)\]\, ,
\eea
with two arbitrary constants $c_1$ and $c_2$.
To impose boundary conditions, we first notice that the absence of the folded singularity at $u=1$ fixes $c_2=\hat{I}_{\vp\vp\phi}(v, x_0)/2$, while $c_1$ can be determined by requiring the solution to be symmetric in $u \leftrightarrow v$
\be
c_1 = -\frac{1}{2}\Big[ \gamma_E- 1 +\log(-x_0)  \Big] \hat{I}_{\vp\vp\phi}(v, x_0)~.
\ee
This completely fixes the IR-divergent solution to be
\bea \label{eq:Fdiv}
\hat{F}_{\rm div} &=& \frac{1}{4} \hat{I}_{\vp\vp\phi}(u, x_0) \hat{I}_{\vp\vp\phi}(v, x_0) \nn\\
 &=& \left[ 1-\frac{1}{u}\Big(\gamma_E - 1 +\log (-E_L\eta_0) \Big)\right]
\left[ 1-\frac{1}{v}\Big(\gamma_E - 1 +\log (-E_R\eta_0) \Big)\right]~,
\eea
where in the second line we have restored $\eta_0$ by $x_0=s\eta_0$, and introduced $E_L \equiv k_{12}+s$ and $E_R \equiv k_{34}+s$.
Thus $\hat{F}_{\rm div}$ has IR-divergent partial-energy poles.
The factorized form suggests that $\hat{F}_{\rm div}$ belongs to the disconnected part of the four-point function that can be written as the product of two three-point functions. We shall confirm this expectation from the wavefunctional approach in Section \ref{sec:wave}. 

\vskip4pt
Combining $\hat{F}_{\rm fin}$ and $\hat{F}_{\rm div}$, we find the complete solution of the four-point scalar seed of massless exchange. The IR-divergent part of the solution is particularly important when we use this seed function to compute non-Gaussianities from multi-field inflation, as we shall see in Section \ref{sec:nonG}.
In the end, we notice that one has $u,v\in [0,1]$ in the dS bootstrap as a consequence of triangle inequality.
But
this is not assumed for deriving the closed-form solution above.
Thus the seed function here can also be applied in the boost-breaking scenarios with reduced sound speeds, where $u$ and $v$ can be any positive number. We will elaborate on this point in Section \ref{sec:trispectr}.

\subsection{Mixed Propagator and Three-Point Scalar Seed}
\label{sec:3pt-ex}

The above analysis on IR divergences has assumed the full dS isometries and then allowed the mild breaking of the dilation symmetry by introducing the late-time cutoff $\eta_0$. 
In cosmology, a broader class of theories correspond to the circumstances where the dS boost symmetry is strongly broken, and thus one can no longer rely on the special conformal Ward identities for deriving differential equations of boundary correlators. These theories typically have reduced sound speeds and large field interactions, which give large signals of primordial non-Gaussianity with immediate observational relevance. 
Recently, systematical  investigations into these cases have been performed in the context of the {\it boostless bootstrap} \cite{Pajer:2020wxk, Jazayeri:2021fvk, Bonifacio:2021azc,Cabass:2021fnw,Hillman:2021bnk,Pimentel:2022fsc, Jazayeri:2022kjy, Cabass:2022jda}. Here we mainly follow the approach in \cite{Pimentel:2022fsc} to examine the IR divergences of  massless exchange in boost-breaking scenarios. 

\begin{figure} [h]
   \centering
             \includegraphics[width=.4\textwidth]{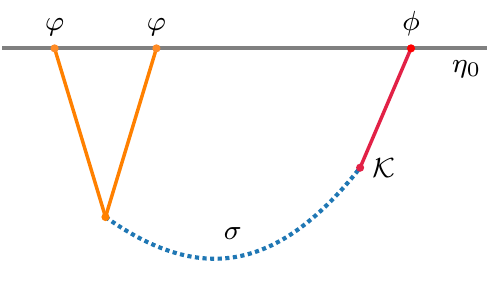}
   \caption{The three-point scalar seed function of massless exchange with one mixed propagator $\cG$.}
  \label{fig:3ptseed}
\end{figure}
 
\vskip4pt
Without dS boost symmetry, the main object of interest is the  exchange bispectrum as shown in Figure \ref{fig:3ptseed}, and thus it is much more convenient to introduce a {\it mixed propagator} between the inflaton field $\phi$ and another massless scalar $\s$.
Consider the transfer vertex $\dot\phi \sigma$, and then a new bulk-to-boundary propagator  is given as \cite{Chen:2017ryl, Pimentel:2022fsc}
\begin{small}
\be
\cG_\pm (k, \eta , \eta_0) = \pm i  \int_{-\infty}^{\eta_0} d\eta' a(\eta')^3 \[ G_{\pm\pm}^\sigma(c_\s k,\eta, \eta') \partial_{\eta'} K_{\pm}(c_s k, \eta')
- G_{\pm\mp}^\sigma(c_\s k,\eta, \eta') \partial_{\eta'} K_{\mp}(c_s k, \eta')
\] ,
\ee
\end{small}which describes the propagation from $\sigma$ at some bulk time $\eta$ to the inflaton $\phi$ at future boundary $\eta_0$. 
Here we also introduced the sound speed of the inflaton field $c_s$ and the one of the additional scalar $c_\s$.
For simplicity, we can remove the $c_\s$-dependence by rescaling $c_\s k \rightarrow k$ and $c_s \rightarrow c_s/c_\s$, after which $c_s$ becomes the {\it ratio} of two sound speeds and thus can be any positive number. We will restore the $c_\s$ parameter when we consider one particular new phenomenology in Section \ref{sec:multispeed}.
While the free bulk-to-boundary propagators satisfy a homogeneous equation of motion, from \eqref{eom},
 this mixed propagator is found to be governed by the following inhomogeneous equation
\be \label{eqmixK}
\mathcal{O}_\eta \cG_\pm (k, \eta,\eta_0) = - \frac{H\eta^2}{2c_sk} e^{\pm ic_sk\eta} .
\ee
As both $\phi$ and $\s$ are massless scalars, the mixed propagator can be easily solved.
For illustration, the analytical expression of $\cG_+ $ is given by
\be \label{mixKp}
\cG_+ (k, \eta, \eta_0) = \frac{H}{4k^3} \[e^{-i k \eta} (1+ik \eta)~{\rm Ei}\big(i(1+c_s)k\eta\big) + e^{i k \eta} (1-ik\eta )~\mathcal{D} -\frac{2}{c_s} e^{ic_sk \eta}  \] ~,
\ee
with ${\rm Ei}(x)$ being the exponential integral and 
\begin{align}
    \mathcal{D}= 
        \begin{dcases}
             \gamma_E-2+ \frac{i\pi}{2}+\log \( \frac{2k^2\eta_0^2}{-k\eta}\) ~,\quad & c_s = 1\\
            2 \gamma_E-2 + {i\pi} +2\log \( {-k\eta_0} \) 
            + \log\(1-c_s^2\)
            - {\rm Ei}\big(i(-1+c_s)k\eta\big) ~,\quad & c_s \neq 1
        \end{dcases} \qquad .
\end{align}
The result of $\cG_- $ is the complex conjugation of $\cG_+ $.
At first sight, the mixed propagator seems to have a non-Bunch-Davies state, with a mixture of positive- and negative-frequency modes. However, the negative-frequency mode only becomes comparable to the positive frequency component at late times. In the early-time limit, we find
\begin{align} \label{earlycG}
 \lim_{\eta\rightarrow-\infty} \cG_+ (k, \eta, \eta_0) \rightarrow 
        \begin{dcases}
             \frac{iH\eta}{4k^2}  e^{ik\eta} \log(-k \eta) ~,\quad & c_s = 1\\
            -\frac{iH\eta}{4k^2}  e^{ik\eta} \log \[(1-c_s^2)k^2 \eta_0^2\] ~,\quad & c_s \neq 1
        \end{dcases} \qquad .
\end{align}
Thus we still have the adiabatic vacuum deep inside the horzion, but there could be a deformation from the standard Bunch-Davies state because of the linear mixing.
Meanwhile, we can see that the mixed propagator explicitly depends on the IR cutoff $\eta_0$.
For perturbations outside of the horizon $-k\eta\ll 1$, we find the $\eta$-dependence drops out with
\bea \label{tau0}
\lim_{-k\eta\ll 1} \cG_+ (k, \eta, \eta_0 )  =  \frac{ H}{2c_sk^3} \Big[ (\gamma_E-1)c_s-1+c_s\log\(-(1+c_s) k\eta_0\) \Big] ~,
\eea
which diverges when $\eta_0\rightarrow 0$.
This secular behaviour of the mixed propagator basically captures the super-horizon conversion effect in multi-field inflation, where the isocurvature mode keeps sourcing the growth of the curvature perturbation.
As this super-horizon evolution is widely believed to be responsible for the generation of local non-Gaussianity, later we will see in Section \ref{sec:revisit} that indeed this extensively studied shape is  closely related to the IR-divergent behaviour of the mixed propagator.

\vskip4pt
In the following we will mainly use the dimensionless mixed propagator $\hat\cG_+  \equiv c_s k^3 \cG_+  /  H $ and rescale the IR cutoff $\eta_0 = x_0/k$. As a result, $\hat\cG_+$ depends on $k$ and $\eta$ only through the combination $k\eta$.
Therefore, we can trade $\eta$-derivatives with $k$-derivatives on $\hat \cG_+$, and the differential equation \eqref{eqmixK} is equivalent to
\be \label{eq:Khat}
\( k^2\partial_{k}^2 - 2k\partial_k + k^2\eta^2 \)\hat\cG_\pm (k \eta, x_0)  =
-\frac{1}{2}  k^2 \eta^2  e^{\pm ic_sk\eta}  .
\ee
The late-time limit becomes a function of $x_0$ only
\bea \label{Kx0}
\lim_{\eta\rightarrow 0}\hat\cG_\pm (k \eta,x_0 )= \hat\cG  (x_0 )  =  \frac{1}{2} \Big[ (\gamma_E-1)c_s-1+c_s\log\(-(1+c_s) x_0\) \Big] ~.
\eea

Next, we consider the single-exchange three-point correlator with a mixed propagator. The starting point of the bootstrap approach is the bispectrum
 $\langle \vp  \vp  \phi  \rangle$ with two conformally coupled scalars and an inflaton, exchaning one additional massless scalar $\s$. 
At the practical level, one advantage of the mixed propagator is that the exchange correlator can be expressed in a ``contact-like" form in the bulk computation.
For the cubic vertex $\vp^2\s$, the three-point function becomes
\bea \label{ccphi}
\langle \vp_{{\bf k}_1} \vp_{{\bf k}_2} \phi_{{\bf k}_3} \rangle'  &=& i \int_{-\infty}^{\eta_0} d\eta a(\eta)^4 \[ K^\vp_+(c_sk_1,\eta) K^\vp_+(c_sk_2,\eta) \cG
_+(k_3,\eta,\eta_0) -c.c.\] + {\rm perms.}\nn\\
&=&  -\frac{H\eta_0^2}{4c_s^3k_1k_2k_3^2} \mathcal{\hat I}  (k_{12}, k_3, x_0) + {\rm perms.}~,
\eea
where we have set $\vp$ has the same sound speed with the inflaton.
Then the {\it three-point scalar seed} is given by the following integral
\bea \label{singleM}
\mathcal{\hat I} (k_{12}, k_3, x_0) 
& \equiv & -\frac{i}{k_3} \int_{-\infty}^{x_0/k_3} \frac{d\eta}{\eta^2} \[
e^{i c_s k_{12}\eta}
\hat\cG_+ (k_3\eta;x_0) -
e^{-i c_s k_{12}\eta}
\hat\cG_- (k_3\eta;x_0)\] .
\eea
Notice here the upper limit of the integral is taken to be $x_0/k_3$, as we have already used $x_0=k_3\eta_0$ as the rescaled IR cutoff in the mixed propagator.
This integral is still IR-divergent when the upper limit goes to zero, and it is rather complicated to compute, as the explicit expression of $\cG$ contains exponential integral and logarithms. Instead, we will find its differential equation and solve it from the boundary approach.
First, we see that $\mathcal{\hat I}$ is dimensionless by definition, and depends on $k_{12}$ and $k_3$ through the momentum ratio
\be
w\equiv \frac{k_3}{c_s k_{12}}~,
\ee
which can take any positive value $w<c_s^{-1}$ as $c_s$ is an arbitrary sound speed ratio.
Using the differential equation of $\hat{\cG}$ in \eqref{eq:Khat}, and following the same approach for the four-point scalar seed, we find the boundary equation in terms of $w$
\be \label{seedeq}
\( \Delta_w -2 \)   \mathcal{\hat I} (w,x_0)
=   \frac{w}{1+c_s w} + \frac{6}{w} \hat\cG  (x_0 ) ~.
\ee
where $\Delta_w$ is the operator \eqref{deltau} with $u\rightarrow w$.
The second source term is a consequence of the nontrivial upper limit of the $\mathcal{\hat I}$ integral \eqref{singleM}. It is interesting to notice that this equation can be reproduced from \eqref{eq:4pt} by replacing $u\rightarrow w$ and $v\rightarrow 1/c_s$. This is due to the fact that $\partial_\eta\phi$ has the same mode function with a conformally coupled scalar $\vp$, and the four-point scalar seed \eqref{Fhat} matches with $ \mathcal{\hat I}$ by taking $k_4\rightarrow 0$. The explicit connection between the three-point and four-point seed functions  has been analyzed in \cite{Pimentel:2022fsc}.

\vskip4pt
With this observation in mind, it is straightforward to obtain the solution of \eqref{seedeq}. The IR-finite part of the solution which satisfies $( \Delta_w -2 )\mathcal{\hat I}_{\rm fin} ={w}/({1+c_s w}) $ is simply given by the ${\hat F}_{\rm fin}$ solution in \eqref{eq:Ffin} via
\be \label{eq:Ifin}
\mathcal{\hat I}_{\rm fin} (w) = {\hat F}_{\rm fin} (w,c_s^{-1})~.
\ee
Meanwhile, the IR-divergent part regularized by $x_0$ satisfies $( \Delta_w -2 )\mathcal{\hat I}_{\rm div} =6 \hat\cG  (x_0 )/w $, and can be solved as
\be
\mathcal{\hat I}_{\rm div} (w,x_0) = \frac{1}{w}\log\( \frac{1-w^2}{w^2}\) \hat\cG  (x_0 ) + c_1\frac{1}{w} + c_2 \[ 1+\frac{1}{2w} \log\( \frac{1-w}{1+w}\)\]\, .
\ee
Again, we need to impose boundary conditions to fix the two free coefficients. 
For non-unity $c_s$, $w=1$ does not correspond to the folded configuration, but still no physical singularity is allowed in this limit, which fixes $c_2= -2\hat\cG  (x_0 )$. The other boundary condition can be obtained by taking the soft limit $k_3\rightarrow 0 $, where the mixed propagator is given by the late-time limit \eqref{Kx0}, and the seed function \eqref{singleM} factorizes into
$\lim_{w\rightarrow 0} \mathcal{\hat I} = {2} \left[ \gamma_E- 1 +\log({-x_0}/{w})  \right]\hat\cG  (x_0 )/{w}$.
This leads to $c_1=2 \(\gamma_E- 1 +\log({-x_0})\right)\hat\cG  (x_0 ) $, and thus we find
\be \label{eq:Idiv}
\mathcal{\hat I}_{\rm div} (w,x_0) = \frac{2}{w} \hat\cG  (x_0 ) \left[\gamma_E- 1-w+\log\(\frac{1+w}{w}\) +\log({-x_0})\right] .
\ee
The full analytical solution is then given by $\mathcal{\hat I}=\mathcal{\hat I}_{\rm fin} + \mathcal{\hat I}_{\rm div}$, which will be used as our main building block for bootstrapping multi-field non-Gaussianities in Section \ref{sec:nonG}.
As an illustration, let's take a look at the three-point scalar seed with $c_s=1$, which has the following simple form
\bea \label{eq:3ptseed}
\mathcal{\hat I}(k_{12},k_3,\eta_0)    &=& \frac{k_{12}}{k_3} \Big[ \gamma_E- 1 +\log({-k_t\eta_0}) \Big] \Big[ \gamma_E- 2 -\frac{k_3}{k_{12}} +\log({-2 k_3\eta_0}) \Big] \nn\\
&& +\frac{k_{12}}{2k_3} \[\frac{\pi^2}{6}- {\rm Li}_2\(\frac{k_{12}-k_3}{k_t}\)\]~.
\eea
Here we have restored $\eta_0$, $k_{12}$ and $k_3$ explicitly. This result shows that the $\eta_0$-dependent logarithm arise in two ways: it comes with $2k_3$ and also with $k_t$. While the $k_3$ term is associated with the mixed propagator, the $k_T$-type IR divergence is a consequence of the cubic interaction vertex\footnote{For cases with a general sound speed ratio, this term is given by $\log(-E_L\eta_0)$, with $E_L=c_s k_{12}+k_3$. When $c_s=1$, the partial-energy $E_L$-pole concides with the logarithmic $k_t$-pole.}, as we observed in the contact example in Section \ref{sec:contact}.
In the exchange bispectrum, the IR-divergent term is a product of these two $\eta_0$-dependent logarithms, like in the exchange four-point function. We shall also confirm that the IR-divergence is given by the disconnected part in the wavefunctional approach in Section \ref{sec:wave}.

\subsection{Wavefunction Approach}
\label{sec:wave}

The recent development of cosmological bootstrap shows that the wavefunction of the Universe provides a convenient approach for the analysis of boundary correlators.
We leave the detailed discussion   in Appendix \ref{app:wave}, while here we mainly present the results for dS-invariant theories, and demonstrate the behaviour of IR divergences using the wavefunction method.

\vskip4pt
The primary object of interest here is the wavefunction coefficients $\psi_n$ in the Fourier space at the late-time boundary of the dS spacetime. 
In perturbation theory, the bulk computation of $\psi_n$ can also be performed in a diagrammatic fashion, where similarly we introduce bulk-to-boundary propagator $K_\Psi(k,\eta)$ and the bulk-to-bulk propagator $G_\Psi(k,\eta,\eta')$.
Explicitly, the bulk-to-boundary propagators of massless and conformally coupled scalars  are expressed as
\be \label{KPsi}
K_\Psi(k,\eta) = (1-ik\eta)e^{ik\eta}~,~~~~
K_\Psi^{\varphi}(k,\eta) = \frac{\eta}{\eta_0} e^{ik\eta}~,
\ee
which become 1 in the late-time limit $\eta=\eta_0\rightarrow0$. They are similar with the ones in the in-in formalism, but with different normalizations. Meanwhile, the bulk-to-bulk propagator has a boundary condition that it becomes 0 in the late-time limit, and thus takes a different form. For instance, the one for massless scalars is given by 
\bea \label{GPsi}
G^\s_\Psi(k,\eta,\eta') &=& \frac{H^2}{2k^3} \[(1+ik\eta)(1-ik\eta') e^{ik(\eta'-\eta)}\theta(\eta-\eta') + (1-ik\eta)(1+ik\eta')e^{ik(\eta-\eta')}\theta(\eta'-\eta)\right.\nn\\
&&~~~~~~~~\left. -(1-ik\eta)(1-ik\eta')e^{ik(\eta'+\eta)} \]~.
\eea
As we see, the presence of the last term ensures that $G^\s_\Psi$  vanishes when we take $\eta$ or $\eta'$ to 0.
As a result, in the wavefunction approach, the bulk-to-bulk propagator of massless fields decays after the perturbation mode exits the horizon, which essentially differs from the behaviour of $G$ in the in-in formalism. As we have mentioned for several times, the super-horizon freezing of the massless scalar plays an important role for the appearance of the IR-divergences. Next, we will show that, in the wavefunction approach the $\psi_n$ from massless exchanges remain IR-finite due to the decaying behaviour of  $G^\s_\Psi$ outside of the horzion.

\vskip4pt
Here let's consider the two diagrams we analyzed in the dS bootstrap: the contact cubic interaction $\varphi^2\phi$ and the  four-point seed function from the $s$-channel massless exchange. Their corresponding wavefunction coefficients are given by
\bea
\psi_3^{\vp\vp\phi} &=& -2i\int_{-\infty}^{\eta_0} d\eta a(\eta)^{4} K_\Psi^\vp (k_1,\eta) K_\Psi^\vp (k_2,\eta)   {K}_\Psi(k_3,\eta)\nn\\
\psi_4 &=& 4\int_{-\infty}^{\eta_0} d\eta  d\eta' a(\eta)^{4} a(\eta')^{4} K_\Psi^\vp (k_1,\eta) K_\Psi^\vp (k_2,\eta)  G_\Psi^\s(s,\eta,\eta')  K_\Psi^\vp (k_3,\eta') K_\Psi^\vp (k_4,\eta') ~.
\eea
We leave the derivation of their solutions in Appendix \ref{app:wave}, and here let's take a look at the final results 
\be
\psi_3^{\vp\vp\phi}\propto -\frac{2i}{\eta_0}+i\pi + {I}_{\vp\vp\phi} (k_{12},k_3,\eta_0)~,~~~~~~
\psi_4 \propto \frac{1}{s} \hat{F}_{\rm fin} (u,v)~,
\ee
with ${I}_{\vp\vp\phi}$ and $\hat{F}_{\rm fin}$ given in \eqref{Ivvp} and \eqref{eq:Ffin} respectively.
The contact three-point function $\psi_3^{\vp\vp\phi}$ remains divergent at the late-time boundary, as the bulk-to-boundary propagator of $\phi$ becomes contant outside of the horizon and thus keeps contributing to the bulk integral.
Meanwhile, because of the decay of $G_\Psi^\s$ on super-horizon scales, we see that $\psi_4$ is independent of the IR cutoff $\eta_0$, and corresponds to the IR-finite part of the four-point scalar seed.

\vskip4pt
The wavefunction coefficients are not physical observables, and correlation functions can be computed via simple
algebraic relations of their real parts.
For the $\vp\vp\phi$ contact bispectrum, the real part of $\psi_3^{\vp\vp\phi}$ simply gives us the result in \eqref{vvp}, while the unphysical $1/\eta_0$ divergence drops out as it is purely imaginary. 
For the exchange four-point function, in addition to ${\rm Re} \psi_4$, there is also contribution to the corresponding correlator from the disconnected part, which is proportional to a product of two three-point functions ${\rm Re} \psi_3^{\vp\vp\s}$. Combining these two parts, we find the correlator becomes
\be
 \langle \vp_{k_1}\vp_{k_2} \vp_{k_3} \vp_{k_4} \rangle'
\propto \[ \frac{1}{s} \hat{F}_{\rm fin} (u,v) + \frac{1}{4s^3} {I}_{\vp\vp\phi} (k_{12},s,\eta_0){I}_{\vp\vp\phi} (k_{34},s,\eta_0) \] +  \text{$t$- and $u$-channels}\, ,
\ee
which precisely agrees with the result of the four-point seed function \eqref{Fhat2} with the disconneted part given by the IR-divergent term in \eqref{eq:Fdiv}.
The detailed deviation with proper consideration of various prefactors is left in Appendix \ref{app:wave}. 
The agreement between two different approaches  provides a useful consistency check for our analysis of IR divergences in massless exchange. Similarly, one can perform the computation for the boost-breaking three-point scalar seed with a mixed propagator using the the wavefunction approach. This is presented in Appendix \ref{app:wave} as well, and the final result matches what we found in Section \ref{sec:3pt-ex}.

\

As a concluding remark of this section, we note that in our analysis of IR divergences of cosmological bootstrap, we have looked into three objects: the cosmological correlation functions at the end of inflation, the boundary CFT correlators and the wavefunction coefficients.
In many circumstances, such as for contact diagrams and massive exchanges, their distinctions are not so important, and they may be simply related with each other by using normalization factors, such as \eqref{cc-cft}. However, when we have IR divergences, the distinction among these objects become nontrivial. 
In particular, we have seen that for the massless exchange, the constraints on CFT correlators from conformal Ward identities lead to the bootstrap equations \eqref{eq:4pt} with the IR-finite term only, which agrees with the result for the corresponding wavefunction coefficient $\psi_4$. Meanwhile, the physical observable -- the cosmological correlator $\langle \vp^4 \rangle$ is IR-divergent, as we need to include the disconnected part which becomes singular at the late-time limit. 
In this sense, the CFT correlators on the boundary are  associated with the wavefunction coefficients, instead of the cosmological correlators. This can be explained by the fact that it is more natural to see the appearance of the conformal group as a result of dS isometries in the late-time wavefunction. 
As we will show in the next section, the IR-divergent terms are particularly important for predictions on inflationary correlators, thus one needs to be careful when bootstrapping these observables of primordial non-Gaussianty by exploiting conformal symmetry or using the wavefunction method.

\section{Inflationary Massless-Exchange Correlators}
\label{sec:nonG}

For the IR-divergent correlators in massless exchanges, in the previous section we have presented the four-point scalar seed of the dS bootstrap and the three-point seed function for the boostless bootstrap, which both contain conformally coupled scalars as external fields.
Meanwhile, for inflationary predictions of the primordial curvature perturbation, we are interested in results with all external lines being the inflaton fluctuations (a nearly massless field). To derive inflationary bispectra and trispectra from the scalar seeds, we will apply the {\it weight-shifting operators} as the major tool. These are differential operators which map the conformally coupled scalar $\vp$ to the massless inflaton $\phi$.
By using this approach, we generate a complete set of inflationary predictions from the single exchange of a massless scalar in both dS-invariant and boost-breaking theories, many of which are of immediate interest for ongoing and upcoming observations.

\vskip4pt

In Section \ref{sec:dS4pt3pt}, we derive the inflaton four-point and the three-point correlation functions from massless exchange in theories where the full dS isometries are (approximately) respected. 
In Section \ref{sec:boostless}, we look into all the possible massless exchange correlators from boost-breaking theories with nontrivial sound speeds. In particular, we consider the bispectra with IR-divergent terms in Section \ref{sec:cs-IR}, and then we identify a new class of non-Gaussianity shapes in IR-finite correlators in Section \ref{sec:multispeed}.

\paragraph{Weight-shifting operators} Before moving to the inflationary correlators, let's first give a brief review of the weight-shifting operators. We shall mainly follow the approach in \cite{Pimentel:2022fsc} which is based on the bulk intuition and generalizes to theories with broken boost symmetries.
See \cite{Arkani-Hamed:2018kmz, Baumann:2019oyu} for the symmetry-based derivation of the weight-shifting operators in the dS bootstrap from a purely boundary perspective.

\vskip4pt
Our goal here is to raise the conformal weight of the external fields from $\Delta=2$ (conformally coupled scalar) to $\Delta=3$ (massless scalar). 
To achieve this, we 
mainly use the observation that  the massless bulk-to-boundary propagator in \eqref{Kpm} can be generated from the one of the conformally coupled scalar by some differential operators.
For simplicity, let's strip the overall normalization factors with $H$ and $k$, and look at the $\eta$-dependent part of these two propagators
\be
 \phi_k  \rightarrow (1-ic_sk\eta)e^{ic_sk\eta}~,~~~~~~~~ \vp_k  \rightarrow \eta e^{ic_sk\eta}~.
\ee
Next, we are interested in generating the $\phi\phi\s$-type cubic interactions from the $\varphi^2\s$ vertex used in the scalar seeds. 
For the inflaton coupling, here  we mainly focus on the boost-breaking ones with lowest derivatives
$\dot\phi^2\s$, $(\partial_i\phi)^2 \s$, and their dS-invariant combination $(\partial_\mu\phi)^2 \s$.
From the EFT point of view,
they are normally expected to provide the leading  vertices for inflationary predictions.
It is convenient to look at the products of field operators $\partial \phi_{k_1} \partial \phi_{k_2}$ and $\vp_{k_1}\vp_{k_2}$.
Then in the bulk computation of the particular cubic interactions, we find the two products can be connected by
\begin{align}
 \label{dphi2}
\dot\phi^2\s ~ :& ~~~ \eta\partial_\eta \phi_{k_1} \eta\partial_\eta \phi_{k_2}   = k_1 k_2~ \mathcal{W}^{\dot\phi^2\sigma}_{12}    \big[\vp_{k_1}\vp_{k_2} \big]   \\
\label{diphi2}(\partial_i\phi)^2 \s ~:& ~~~ (-{\bf k}_1\cdot{\bf k}_2)\eta^2 \phi_{k_1}   \phi_{k_2}   =k_1 k_2 ~\mathcal{W}^{(\partial_i\phi)^2\sigma}_{12}   \big[ \vp_{k_1}\vp_{k_2}\big]~.
\end{align}
The two $\mathcal{W}_{12}$'s are the weight-shifting operators for corresponding cubic interactions: 
\bea \label{ws1}
\mathcal{W}^{\dot\phi^2\sigma}_{12} &=& -c_s^2 k_1k_2 \partial_{k_{12}}^2~, \\
\label{ws2}
\mathcal{W}^{(\partial_i\phi)^2\sigma}_{12} &=& -\frac{1}{2k_1k_2} (s^2 - k_1^2 - k_2^2) {\(1-k_1\partial_{k_1}\)\(1-k_2\partial_{k_2}\)}~,
\eea
where we have used $s^2=({\bf k}_1 + {\bf k}_2)^2$ to rewrite ${\bf k}_1\cdot {\bf k}_2 = (s^2 - k_1^2 - k_2^2)/2$. For the exchange bispectrum, we simply have $s=k_3$. By setting $c_s=1$ and combining the two interactions above, we also find
the  weight-shifting operator in the dS bootstrap
\be \label{ws-dS}
\mathcal{W}_{12}^{\rm dS} = - \mathcal{W}^{\dot\phi^2\sigma}_{12}  + \mathcal{W}^{(\partial_i\phi)^2\sigma}_{12}= \frac{1}{2} \(k_{12}^2-s^2\) \partial^2_{k_{12}} - \frac{1}{2 k_1k_2}\( s^2 - k_1^2 -k_2^2 \) \(1-k_{12}\partial_{k_{12}}\)~,
\ee
which maps the $\vp^2\s$ vertex to the dS-invariant one $(\partial_\mu\phi)^2\s$.
Using the same approach,
we are able to derive the weight-shifting operators for all the $\phi\phi\s$-type boost-breaking cubic interactions with any number of time and spatial derivatives. The most general form is presented in \cite{Pimentel:2022fsc}.

\begin{figure} [t!]
   \centering
            \includegraphics[width=.4\textwidth]{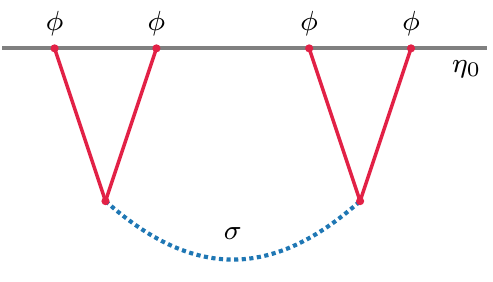} \hspace{20pt}
             \includegraphics[width=.4\textwidth]{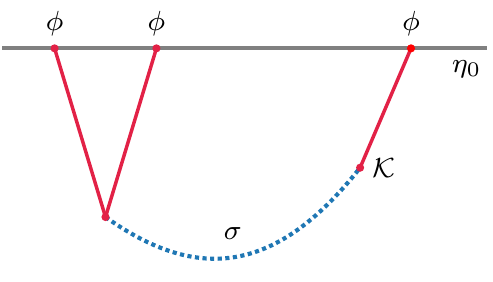}
   \caption{The inflaton four-point and three-point correlators mediated by one additional massless scalar.}
  \label{fig:inflaton}
\end{figure}

\vskip4pt
Now we consider how to derive the inflaton correlators from the scalar seed functions. In the bulk computation with $\phi\phi\s$-type cubic couplings, it is easy to see that  we can apply the relations connecting two field products, such as \eqref{dphi2} and \eqref{diphi2},  and then take the $\mathcal{W}$
operators outside of bulk integrals. By doing so, we find the maps from the scalar seed functions to the corresponding inflaton three-point and four-point correlators
\bea \label{phi3b}  
\langle \phi_{{\bf k}_1} \phi_{{\bf k}_2} \phi_{{\bf k}_3} \rangle' &=&  \frac{-H^3}{4 c_s^7 k_1^2k_2^2k_3^2}\ \mathcal{W}_{12}\ 
\mathcal{\hat I} + {\rm perms.}~,\\
 \label{phi4t}
\langle \phi_{{\bf k}_1} \phi_{{\bf k}_2} \phi_{{\bf k}_3} \phi_{{\bf k}_4} \rangle' &=&  \frac{H^6}{2 c_s^{12} k_1^2k_2^2k_3^2k_4^2 s}\ \mathcal{W}_{12}\ \mathcal{W}_{34}\ 
 {\hat F} \ +\ \text{$t$- and $u$-channels}\,~,
\eea
where $\mathcal{W}_{34}$ is the weight-shifting operator associated with the momenta $k_3$ and $k_4$.
As we already have the analytical results for $\hat{\mathcal{I}}$ and $\hat{F}$, we can simply use the weight-shifting operators to derive inflaton correlators on the boundary, with no need to solve the complicated bulk integrals case by case.  
In the following, we shall apply this  approach to bootstrap inflationary bispectra and trispectra  from the single-exchange of a massless scalar.

\subsection{(Almost) dS-Invariant Correlators}
\label{sec:dS4pt3pt}

Let's first consider the cosmological correlators from (approximately) dS-invariant theories. 
For cosmic inflation, the full dS symmetries restrict us to to slow-roll models where the boost isometry can only be mildly broken by the time dependence of the inflaton field, and thus one always finds small level of non-Gaussianities. 
In single-field inflation, the resulting signal is due to graviton exchange and slow-roll suppressed, which is widely known as the gravitational floor \cite{Maldacena:2002vr}.
When an additional light scalar is present and coupled to the inflaton, the famous conlcusion is that {\it local} non-Gaussianity is generated. We show that
the dS-invariant massless scalar exchange in our analysis  provides the minimal amount of non-Gaussianities from multi-field inflation.
In particular, here we derive the inflationary four-point and three-point functions from the first-principle computation using the bootstrap method. 
Then, in Section \ref{sec:revisit} we will compare these  results with the local non-Gaussianity from the approximated computation in multi-field inflation models.

\paragraph{Scalar Trispectrum}  The inflaton four-point function from massless exchange can be generated by two exactly dS-invariant cubic vertices $g(\partial_\mu\phi)^2\s$. Here we keep the coupling constant $g$ for the later convenience.
Applying the $\mathcal{W}^{\rm dS}$ operator on the four-point seed function, we can derive the scalar trispectrum from \eqref{phi4t}.
First, let's take a look at the contribution from the IR-finite part of the scalar seed in \eqref{eq:Ffin} 
\be
\frac{1}{s} \mathcal{W}^{\rm dS}_{12}\ \mathcal{W}^{\rm dS}_{34}\ 
 {\hat F}_{\rm fin} + (\text{$t$- and $u$-channels}) = \frac{1}{4k_1k_2k_3k_4} \( k_1^3+k_2^3+k_3^3+k_4^3 - s^3 -t^3 -u^3 \)~,
\ee
with $s=|{\bf k}_1+{\bf k}_2|$, $t=|{\bf k}_1+{\bf k}_3|$ and $u=|{\bf k}_1+{\bf k}_4|$.\footnote{For the rest of the paper, we use $u$ as one of the Mandelstam variables, no longer as the ratio $s/k_{12}$.}
We see that the weight-shifting operator annihilates the logarithmic and dilogarithmic functions in ${\hat F}_{\rm fin}$, and change the expression into simple polynomials of the momenta.
Similarly, the $s$-channel contribution from the IR-divergent part of the scalar seed \eqref{eq:Fdiv} is given by
\be
\frac{1}{s} \mathcal{W}^{\rm dS}_{12}\ \mathcal{W}^{\rm dS}_{34}\ 
 {\hat F}_{\rm div} = \frac{1}{4k_1k_2k_3k_4} \[ \frac{(k_1^3+k_2^3)(k_3^3+k_4^3)}{s^3} -k_1^3-k_2^3-k_3^3-k_4^3 + s^3  \]~,
\ee
where the singular $\eta_0$-dependence in ${\hat F}_{\rm div} $ is completely removed, and we find an IR-finite  polynomials of the external and internal fields energies. These two results are in agreement with the proof in  Ref.~\cite{Goodhew:2022ayb}  that only rational functions are allowed for interactions of massless scalars with at least two derivatives. Combining the  two contributions above, we find the final inflaton four-point function
\begin{eBox}
\be \label{trispec-dS}
\langle \phi_{{\bf k}_1} \phi_{{\bf k}_2} \phi_{{\bf k}_3} \phi_{{\bf k}_4} \rangle'= \frac{g^2H^6}{8 }\[ T_{\rm local1} - 2 T_{\rm local2}  \]~,
\ee
\end{eBox}
with the two shape functions
\begin{align} \label{tlocal1}
T_{\rm local1} &= \frac{1}{{k_1^3 k_2^3 k_3^3 k_4^3} } \[\frac{(k_1^3+k_2^3)(k_3^3+k_4^3)}{s^3} +\frac{(k_1^3+k_3^3)(k_2^3+k_4^3)}{t^3}    +\frac{(k_1^3+k_4^3)(k_2^3+k_4^3)}{u^3} \] ~,\\
T_{\rm local2} &= \frac{k_1^3+k_2^3+k_3^3+k_4^3}{k_1^3 k_2^3 k_3^3 k_4^3}~. \label{tlocal2}
\end{align}
This massless-exchange trispectrum has the standard local shape.
Recall that  the primordial trispectrum has two size parameters $g_{\rm NL}$ and $\tau_{\rm NL}$ for the corresponding local ansatzs
\be \label{tri-local}
\langle \zeta_{{\bf k}_1} \zeta_{{\bf k}_2} \zeta_{{\bf k}_3} \zeta_{{\bf k}_4} \rangle'  =   \[ {\tau_{\rm NL}} T_{\rm local1} 
   + \frac{54}{25} g_{\rm NL} T_{\rm local2} \] P_\zeta^3~.
\ee  
In our result \eqref{trispec-dS}, we find the particular combination of these two shapes makes the trispectrum vanish in the soft limit $k_1\rightarrow 0$. This is a consequence of the  cubic coupling $(\partial_\mu\phi)^2\s$ where the external field $\phi$ has a shift symmetry .

\vskip4pt
 In addition, we notice that the trispectrum above has no total-energy singularities.\footnote{As a contrast, the non-derivative quartic $\phi^4$ contact interaction gives a trispectrum of the following form
\beq
(k_1 k_2 k_3 k_4)^3 \langle \phi \phi \phi \phi \rangle \propto \frac{3 E_4+k_T^2 E_2-4k_T E_3}{k_T}-(k_1^3+k_2^3+k_3^3+k_4^3)\log[-k_T\eta_0]
\eeq
with $k_T\equiv \sum_i k_i$, $E_2\equiv \sum_{i<j} k_i k_j$, $E_3 \equiv\sum_{i<j<k}k_i k_j k_k$, $E_4\equiv k_1k_2k_3k_4$. The contact interactions with derivatives lead to rational polynomials with higher-order $k_T$-poles, which are systematically classified in Ref.~\cite{Bonifacio:2021azc}. In addition to the $k_T$-poles, partial energy poles are also expected for exchange diagrams at $E_L=k_1+k_2+s\rightarrow0$, $E_R=k_3+k_4+s\rightarrow0$.}
One intuitive way to understand why this happens is the following: we start from the cubic vertex  $(\partial_\mu \phi)^2 \sigma$. By doing integration by parts, we obtain $(\square \phi )\s$ with $\square =\partial_\mu\partial^\mu$, and another term with the derivative hitting $\sigma$. As $\phi$ is massless, its equation of motion gives $\square \phi=0$, and thus only the $\phi\partial_\mu \phi \partial^\mu\sigma$ term remains. If we integrate by parts again, we get $\square \sigma$, which has $\square \sigma = m_\s^2\s$ on shell and again vanishes if $\sigma$ is massless.
More manifestly, using integration by parts and on-shell condition we have
\be \label{IbP}
\int d^4 x \sqrt{-g} (\partial_\mu \phi)^2 \sigma = -\int d^4 x \sqrt{-g} \phi \partial_\mu \phi  \partial^\mu\sigma
= \frac{m^2_\s}{2}\int d^4 x \sqrt{-g}  \phi^2 \sigma \rightarrow 0
\ee
We are cavalier with boundary terms here, as their job is to ensure that the final shape vanishes in the soft limit. The upshot is that the cubic vertex can be reduced to $\phi^2 \sigma$ plus boundary terms that ensure shift symmetry. As the non-derivative interaction  breaks shift symmetry, one is likely to be  cornered into the case where the coefficient of $\phi^2 \sigma$ being zero.
Therefore, the trispectrum \eqref{trispec-dS} has no total- or partial-energy poles, can be seen as the consequence of a local field redefinition, for instance
$\phi \rightarrow \phi+ g \phi^2 - g^2 \phi^3$.
In Section \ref{sec:ssoi}, we will compare this bootstrap result  with the one from the $\delta N$ analysis in  one particular model of multi-field inflation.

\paragraph{Scalar Bispectrum} To have the exchange three-point function, one needs to consider the mild breaking of the dS symmetry by taking one of the inflaton legs to the background $\Phi(t)$. For the cubic vertex we considered above, this is simply achieved by
\be \label{cubic+mix}
g (\partial_\mu \Phi)^2 \s \rightarrow  -2 g \dot\Phi \dot{\phi} \s + g (\partial_\mu \phi)^2 \s,
\ee 
which leads to the linear mixing between the inflaton and $\s$ with the coupling $\lambda=-2 g \dot\Phi$.
Thus, by using the mixed propagator from $\lambda \dot{\phi} \s $, the massless exchange bispectrum of the inflaton can be derived from the three-point scalar seed in \eqref{eq:3ptseed}.
By using the weight-shifting operator \eqref{ws-dS}, the IR-finite part of the scalar seed leads to
\be
\mathcal{W}^{\rm dS}_{12} ~\hat{\mathcal{I}}_{\rm fin} \({k_3}/{k_{12}}\) = -\frac{k_3}{2k_1k_2k_t}(k_1^2+k_1k_2+k_2^2-k_3^2)-\frac{k_3^2}{2k_1k_2} \log(2k_3/k_t)~.
\ee
This contribution corresponds to the massless-exchange bispectrum Eq.(6.10) in \cite{Arkani-Hamed:2018kmz}.
Although this result contains logartithmic functions of momenta, it is IR-finite with no dependence on the late-time cutoff $\eta_0$. 
Meanwhile, to find the complete result, we also need to include the IR-divergent part of the seed function, which gives
\be \label{phi3bir}
\mathcal{W}^{\rm dS}_{12} ~\hat{\mathcal{I}}_{\rm div} \( {k_3}/{k_{12}}, k_3 \eta_0\) = -\frac{k_1^3+k_2^3-k_3^3}{2k_1k_2k_3}\big[\gamma_E-2+ \log(-2k_3\eta_0)\big]~.
\ee
This contribution, which becomes large and dominates over the IR-finite term in the late-time limit $\eta_0 \rightarrow 0$, was missed in Ref.~\cite{Arkani-Hamed:2018kmz}.
Combining these two parts and adding permutations, we use \eqref{phi3b} to find the full expression of the inflaton bispectrum from massless exchange
\begin{eBox}
\begin{align} \label{true-local}
\langle \phi_{{\bf k}_1} \phi_{{\bf k}_2} \phi_{{\bf k}_3} \rangle'   = &  \frac{ g\lambda H^3}{4k_1^3k_2^3k_3^3} \Big[ \left(\gamma_E-3 - \log(-k_t\eta_0)\right)(k_1^3+k_2^3+k_3^3) +k_t e_2 -4 e_3   \nn\\
&  +(k_2^3+k_3^3) \log(-2k_1\eta_0)
+   (k_1^3+k_3^3) \log(-2k_2\eta_0)
+   (k_1^3+k_2^3) \log(-2k_3\eta_0) \Big] .
\end{align}
\end{eBox}
As one key result of the paper, this is the bispectrum shape that corresponds to the {\it local non-Gaussianity} from additional light fields during inflation.
While we leave the detailed discussion and comparison in Section \ref{sec:revisit}, the connection with the multi-field analysis can be understood in the following way.
Recall that the well-known explanation for the generation of local non-Gaussianity is the nonlinearities of the super-horizon conversion process that transfers the isocurvature perturbations to the curvature ones. 
In the above computation based on field interactions, the couplings in \eqref{cubic+mix} provide the minimal  interactions between the inflaton and the light scalar. In particular,  the $\dot\phi\s$ mixed propagator captures the conversion effect from the the additional light field (isocurvature modes) to the curvature pertubation.
As the linear mixing $\dot\phi\s$ is always accompanied by the cubic coupling $(\partial_\mu\phi)^2\s$, the first-principle computation here provides the full consideration of the nonlinearities from the conversion process.

\vskip4pt

Although the bispectrum shape in \eqref{true-local} is not exactly the same with the local ansatz in \eqref{local}, there are similarities. We first notice that there are  two types of logarithmic IR divergences: $\log(-k_t\eta_0)$ and $\log(-2k_a\eta_0)$ with $a=1,2,3$. 
The first line in \eqref{true-local} with $\log(-k_t\eta_0)$ is the same with the bispectrum shape \eqref{3pt-self} from the $\phi^3$ contact interaction, which is not explicitly associated with the conversion effect.
The appearance of the logarithmic $k_t$-pole indicates that this contribution comes from a cubic vertex. 
Meanwhile, the second line with $\log(-2k_a\eta_0)$ terms is the super-horizon contribution of the mixed propagator.
It can be generated by a field redefinition with time-dependent coefficients.
As we shall show in Section \ref{sec:revisit},  the $\delta N$ formula provides a particular form of this field redefinition.
 In other words, the full bispectrum of massless exchange contains two parts: one has the same form as  the shape from a   contact interaction of massless scalars; another is due to the  super-horizon conversion process.
 Both of these contributions have shape functions that are similar to the local ansatz.
 
\begin{figure}[t!]
   \centering
      \includegraphics[height =5.35cm]{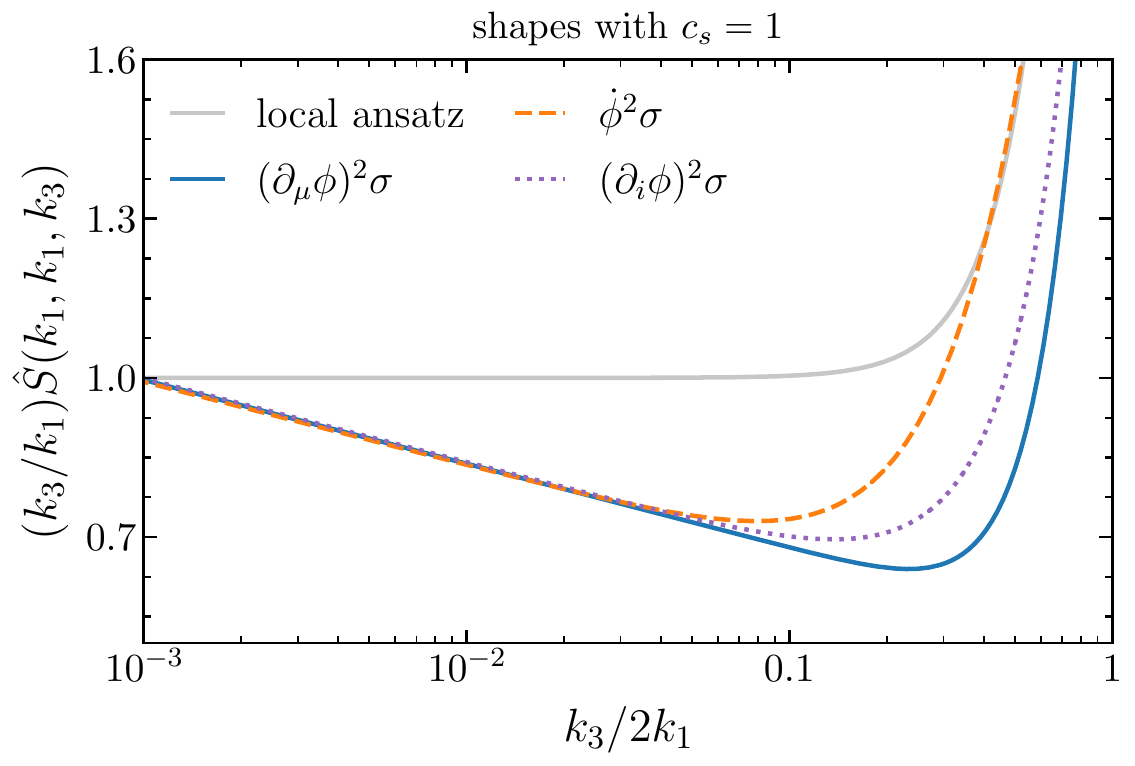}  \hspace{0.1cm}
\includegraphics[height =5.35cm]{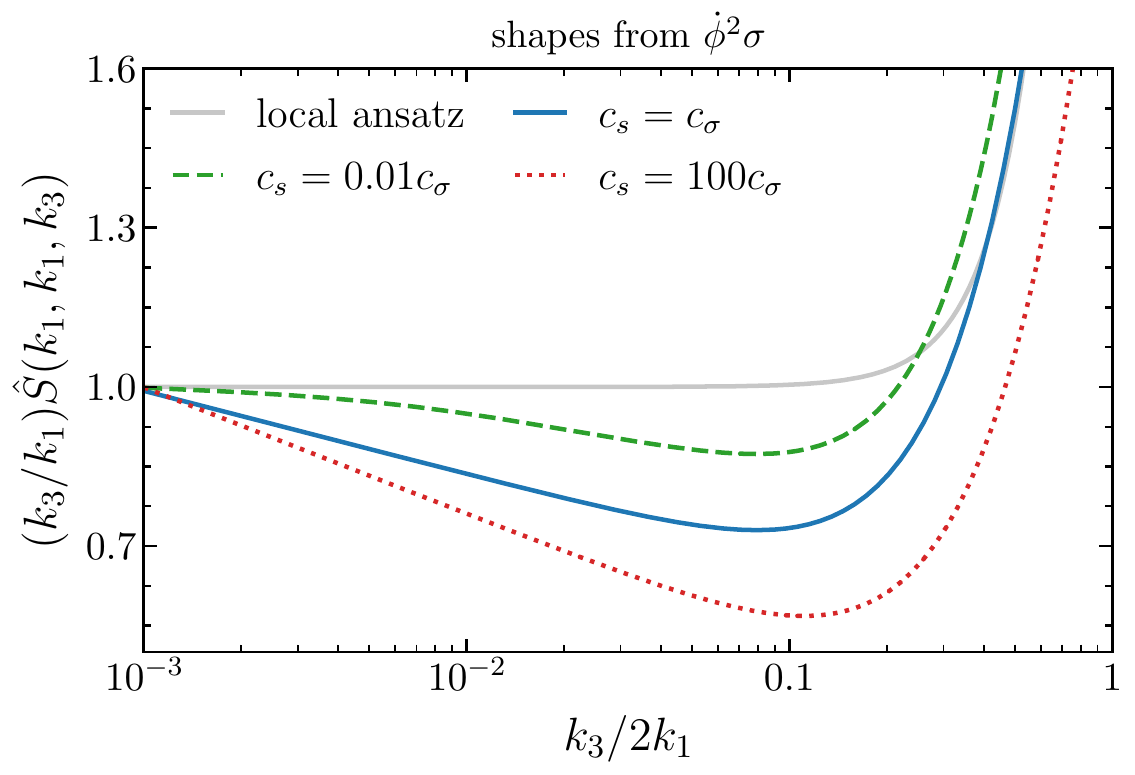} 
      \caption{The bootstrap results of the bispectrum shapes (with $k_1=k_2$) demonstrate mild logarithmic deviations from the local ansatz. {\it Left panel:} the shape functions from the dS-invariant cubic vertex and two boost-breaking ones with $c_s=1$. {\it Right panel:} boost-breaking bispectrum shapes with various sound speed ratios. In this figure we have reintroduced the $\s$ sound speed by $c_s\rightarrow c_s/c_\s$. We choose $\eta_0=-10^{-3}$ for demonstration. The dimensionless shape functions $\hat{S}=(k_1k_2k_3)^2S$ are normalized to be 1 at $k_3/k_{12}=10^{-3}$.} 
      \label{fig:local}
\end{figure}

\vskip4pt
Next, let's look at the squeezed limit of the bispectrum, where the effect of the exchanged massless scalar is mostly manifested
\be \label{soft-dSbi}
\lim_{k_3\rightarrow 0} \langle \phi_{{\bf k}_1} \phi_{{\bf k}_2} \phi_{{\bf k}_3} \rangle'   =    \frac{ g\lambda H^3}{4k_1^3 k_3^3}  \[\gamma_E-2 + \log(-2k_3\eta_0) \]~.
\ee
 This soft behaviour is contributed from the IR-divergent part of the seed function as shown by \eqref{phi3bir}, and it encodes the super-horizon form of the mixed propagator in \eqref{Kx0}. Here we find a logarithmic deviation from the local ansatz, which is shown in the left panel of Figure \ref{fig:local}. 
For perturbation modes within the range of the observational test, the logarithmic deviation corresponds to the number of e-folds $N_l$ from the horizon-exit of the long wavelength mode $k_l=k_3$ to the adiabatic limit when perturbation freezes.  Thus  the mild logarithmic dependence can be approximately taken as a constant number $0<N_l < 60$, and the squeezed bispectrum returns to the standard result of the local ansatz.
 
\paragraph{Correction to the power spectrum} With  the mild breaking of the dS symmetry, we can also put two external legs of the inflaton to the background in the four-point exchange diagram, which leads to corrections to the two-point correlator.
This contribution comes from two $\lambda\dot\phi\s$ linear mixing vertices, and can be computed by taking $k_1=0$ and $k_2=k_3=k$ in the three-point scalar seed \eqref{eq:3ptseed}
\be \label{2pt-corr}
\delta \langle \phi_{{\bf k}} \phi_{-{\bf k}} \rangle' = \frac{\lambda^2}{2k^3} \hat{\mathcal{I}}(1, k\eta_0)  =  {\tilde\lambda^2}  \[ \Big( \log(-2k\eta_0) + \gamma_E-2 \Big)^2-1 + \frac{\pi^2}{12}  \]   
\mathcal{P}_\phi (k)~,
\ee
with $\tilde\lambda= \lambda/H  $ and $ \mathcal{P}_\phi (k) =\frac{H^2}{2k^3}  $ being the two-point function of a free massless scalar. We also see the appearance of the IR-divergent term, which dominates the correction to the power spectrum. 
This result agrees with the in-in computation presented in Ref.~\cite{Achucarro:2016fby}.
Here the logarithmic function can be rewritten in terms of the number of e-folds $N_k$ counting from the horizon-exit of the $k$-mode. Then the size of this correction is approximately given by ${\tilde\lambda^2}  N_k^2$.
For massive exchange, this correction is order of ${\tilde\lambda^2}$, and $ \tilde\lambda<1 $ is sufficient to have perturbative control.
For massless exchange,  we need ${\tilde\lambda }  N_k <1$  to ensure perturbativity, and thus  this  coupling  with additional light scalars is further constrained.
As a consequence, we find a suppressed signal for the massless-exchange bispectrum in dS-invariant scenarios.
This also explains the observation from many multi-field examples that it is usually difficult to generate large local $\fnl$ in models with nearly scale-invariant perturbations.

\subsection{Boost-Breaking Correlators}
\label{sec:boostless}

To achieve larger levels of non-Gaussianity, we need to consider the strong breaking of the dS boost isometry. In these scenarios, the field interactions can become significantly enhanced, and perturbations have small sound speeds.
Now we provide a systematic classification of the inflationary massless-exchange correlators in the boost-breaking theories. 
By using the mixed propagator and three-point scalar seed derived in Section \ref{sec:3pt-ex}, we mainly focus on the scalar bispectra, while similar analysis applies for the inflationary four-point function. 
In the following we show that there are two classes of massless-exchange bispectra: one has IR-divergent terms caused by the $\dot\phi\s$ linear mixing, and others are rational functions of momenta due to the higher derivative quadratic interactions.
We comment on the comparison with previous studies of multi-field inflation in Section \ref{sec:revisit}.

\subsubsection{IR-Divergent Bispectra}
\label{sec:cs-IR}

For this class of correlators, we consider that the quadratic interaction is still given by $\lambda\dot\phi\s$, while the cubic vertices can take any boost-breaking form.
Our starting point is the three-point scalar seed $\mathcal{\hat I}$ introduced in \eqref{singleM} with an arbitrary sound speed ratio $c_s$. 
Because of the sizable cubic couplings and reduced sound speeds, the bispectrum signal can become large and potentially detectable.
Here we mainly focus on the  two leading cubic vertices given by the ones with lowest derivatives from
the EFT of inflation
\be
g_{\rm I} \dot\phi^2 \s~,~~~~~~ g_{\rm II}  a^{-2} {(\partial_i\phi)^2} \s~,
\ee
where the  $g_{\rm II}$ coupling is correlated with the quadratic coupling by $\lambda =-g_{\rm II} ( 8|\dot{H}|)^{1/2}\mpl  /c_s$ due to the nonlinearly realized spacetime symmetry, though the $g_{\rm I}$ coupling is free.

\vskip4pt
By applying the weight-shifting operator \eqref{ws1}, we find the inflaton bispectrum from the $g_{\rm I}\dot\phi^2\s$ and $\lambda \dot\phi\s$ couplings
\bea \label{dotphi2sigma}
\langle \phi_{\bf k_1} \phi_{\bf k_2} \phi_{\bf k_3} \rangle' &= &  \frac{g_{\rm I}\lambda H^3}{4c_s^2k_1 k_2 k_3^3 E_L^2 (c_sk_{12}-k_3)^2} \Bigg[   \frac{(c_s k_{12}-k_3)(1-c_s)k_3^3}{c_s^2k_t} -\frac{1}{c_s}k_3^2E_{L}        \nn\\ [4pt]
&&  + 2 k_3^3 \Big(\gamma_E-1   + \log(-c_sk_t\eta_0)   \Big)
+ 2k_{12}\big(c_s^2 k_{12}^2 -3k_3^2\big)\hat{\cG}(k_3\eta_0)
\Bigg] + {\rm perm.}
\eea
where $E_L=c_sk_{12}+k_3$ is the total energy entering the cubic vertex, and $\hat{\cG}(k_3\eta_0)$ is the late-time limit of the mixed propagator given in \eqref{Kx0}.
Unlike the dS-invariant case, the correlator above cannot be seen as a combination of contributions from field redefinition and cubic contact interactions. Thus this massless-exchange result corresponds to a new class of bispectrum shapes uncategorised in previous classifications. The dominant part of the shape function is the IR-divergent terms with $\eta_0$, which is the result shown in \eqref{ir-bispec0}.
In Figure \ref{fig:local}, we see the deviations from the standard local ansatz.

Let's take a closer look at the singularity structure of this three-point function.
First, there seems to be a pole at $k_3=c_sk_{12}$, however it is easy to check the bispectrum remains regular in this limit, as required by the absence of the $w=1$ singularity in the scalar seed function.
For $c_s\neq 1$, the correlator has total and partial-energy poles
\bea
\lim_{k_t\rightarrow 0} (k_1  k_2  k_3)^3\langle \phi_{\bf k_1} \phi_{\bf k_2} \phi_{\bf k_3} \rangle' &\rightarrow & \frac{g_{\rm I}\lambda H^3}{4c_s^4(c_s^2-1)}~ \frac{e_2^2}{k_t}
\\
\lim_{E_L\rightarrow 0} (k_1  k_2  k_3)^3 \langle \phi_{\bf k_1} \phi_{\bf k_2} \phi_{\bf k_3} \rangle' &\rightarrow & \frac{g_{\rm I}\lambda H^3}{4c_s^2  }~ \frac{k_1  k_2 e_3}{E_L^2} \[ \gamma_E-1+\log\(-k_3\eta_0\)+\frac{1}{2}\log \({c_s^2-1} \) \]~,
\eea
while for $c_s=1$ these two physical singularities coincide with each other and we find
\be
\lim_{k_t\rightarrow 0} (k_1  k_2  k_3)^3 \langle \phi_{\bf k_1} \phi_{\bf k_2} \phi_{\bf k_3} \rangle' \rightarrow \frac{g_{\rm I}\lambda H^3}{4   } ~\frac{e_2 e_3}{k_t^2} \log(k_t) ~.
\ee
At last, the soft limit of the bispectrum, which also corresponds to the $E_R=2k_3\rightarrow 0$ partial-energy pole, is given by
\be
\lim_{k_3\rightarrow 0} \langle \phi_{\bf k_1} \phi_{\bf k_2} \phi_{\bf k_3} \rangle' = \frac{g_{\rm I}\lambda H^3}{4c_s^4k_1^3k_3^3   } ~\hat{\cG}(k_3\eta_0)~.
\ee
This soft behaviour also encodes the late-time limit of the mixed propagator, similar with what we found for the dS-invariant bispectrum in \eqref{soft-dSbi}, except for the appearance of the sound speed ratio $c_s$.
For the massive exchange of cosmological colliders, the effect of the sound speed is to shift the phase of the squeezed limit signal, which can lead to new types of non-Gaussianity such as the {\it equilateral collider shapes} \cite{Pimentel:2022fsc}. In the massless exchange here, the squeezed bispectrum has a nearly constant scaling, and thus changing the sound speed does not lead to large modifications to the shape function, as shown in Figure \ref{fig:local}.

\vskip4pt

The inflaton bispectrum from the $g_{\rm II}(\partial_i\phi)^2\sigma$ interaction can be obtained by using the weight-shifting operator \eqref{ws2} in \eqref{phi3b}
\begin{small}
\be  \label{diphi2sigma}
\langle \phi_{\bf k_1} \phi_{\bf k_2} \phi_{\bf k_3} \rangle' = \frac{g_{\rm II}\lambda H^3(k_1^2+k_2^2-k_3^2)}{8c_s^7k_1^3 k_2^3 k_3^3 E_L^2 } \Bigg[ \hat{\cG}(k_3\eta_0) {\rm Poly}_{\rm I} + \log\(\frac{(1+c_s)k_3}{c_sk_t}  \){\rm Poly}_{\rm II}+ {\rm Poly}_{\rm III}
\Bigg] + {\rm perm.} ,
\ee 
\end{small} 
with the three polynomial functions of the momenta given by
\begin{small}
\bea
 {\rm Poly}_{\rm I} &=&  2c_s^2(k_1^2+k_2^2+k_1k_2) (E_L+k_3) + {2k_3^2(2c_sk_{12} +k_3)}  \nn\\
 {\rm Poly}_{\rm II} &=& c_s k_3^3 \[ \frac{E_L}{c_sk_{12}-k_3} +\frac{2c_s^2 k_1 k_2}{(c_sk_{12}-k_3)^2} \]  \nn\\
 {\rm Poly}_{\rm III} &=& k_3^2 \[ E_L + \frac{c_s^2k_1k_2(k_t+k_3)-c_sk_1k_2k_3}{k_t(c_sk_{12}-k_3)} \] ~.
\eea
\end{small}Again, it is easy to check that the shape function is regular at $k_3=c_s k_{12}$. The singularity structure and soft-limit behaviour are similar with the ones in the $\dot\phi^2\s$ bispectrum.
From these two examples we see that the massless-exchange bispectra become more complicated in the boost-breaking scenarios. 
However, as the main contribution  to the squeezed bispectra still comes from the IR-divergent terms in the $\dot\phi\s$ mixed propagator, the shape functions remain close to the dS-invariant one, as shown in Figure \ref{fig:local}. 
Thus here we find no significant deviations from the local ansatz either. Nevertheless, the size of the non-Gaussian signal can become potentially large.
While in literature it was found to be difficult to generate large local non-Gaussianity in multi-field inflation with canonical scalars \cite{Byrnes:2010em}, 
our results here show that this could be achieved  in boost-breaking scenarios.

\subsubsection{Boostless Trispectra}
\label{sec:trispectr}

Another interesting example is  the inflationary trispectrum from massless exchange with two boost-breaking $\phi\phi\s$ interactions. 
Similarly, we can start from the seed function $\hat{F}$ and apply the weight-shifting procedure to derive the results.
Before moving to the computation, one extra thing to notice is that the four-point scalar seed presented in Section \ref{sec:4pt-ex} is for fields with sound speeds being unit.
To extend this seed function to the one with nontrivial sound speeds ${\hat F}^{\rm BB}$, we introduce a new set of momentum-ratio variables $\tilde{u}\equiv s/c_sk_{12}$ and $\tilde{v}\equiv s/c_sk_{34}$. Then we find the differential equation of ${\hat F}^{\rm BB}(\tilde{u},\tilde{v},x_0)$ in terms of these new variables has the same form with \eqref{eq:4pt-ir}, which also leads to the IR-finite and -divergent solutions \eqref{eq:Ffin} and \eqref{eq:Fdiv}.
One major difference with the previous case is the range of the two new variables $\tilde{u},\tilde{v} \in [0, c_s^{-1}]$.\footnote{Recall that in our notation $c_s$ is the sound speed ratio between $\phi$ and $\s$, which can take any positive value.
Thus $\tilde{u}$ and $\tilde{v}$ may become larger than $1$ for small $c_s$, differing from the range $u,v\in [0,1]$ in the dS-invariant case.} However, as we noted earlier at the end of Section \ref{sec:4pt-ex}, no assumptions for the range of $u$ and $v$ are needed to derive the closed-form solutions for the seed function.\footnote{This is different from the massive-exchange case, where extra work needs to be done to extend the series solution to the $\tilde{u},\tilde{v}>1$ regime, as recently discussed in \cite{Jazayeri:2022kjy}.}
Therefore, the ${\hat F}$ solution can be easily extended to the situation with arbitrary sound speeds.
Explicitly, the boost-breaking version of the four-point scalar seed is  given by
\be
{\hat F}^{\rm BB}   = {\hat F}_{\rm fin} \(\frac{s}{c_s k_{12}}, \frac{s}{c_s k_{34}}\) + {\hat F}_{\rm div} \(\frac{s}{c_s k_{12}}, \frac{s}{c_s k_{34}}, s\eta_0\)~.
\ee
We consider the trispectrum from two $\dot\phi^2\s$ cubic vertices for demonstration.
Substituting  ${\hat F}^{\rm BB} $ into \eqref{phi4t} and using the weight-shifiting operator \eqref{ws1}, we find
\be
\langle \phi_{{\bf k}_1} \phi_{{\bf k}_2} \phi_{{\bf k}_3} \phi_{{\bf k}_4} \rangle' =  \frac{g_{\rm I}^2H^6}{c_s^{7} k_1 k_2 k_3 k_4}  
\[ \frac{c_s^3(E_L+s)(E_R+s)}{2 s^3E_L^2E_R^2} -  \frac{E_L  E_R + c_s k_T s}{k_T^3E_L^2E_R^2}  \]   +  \text{$t$- and $u$-channels}~,
\ee
with $E_L=c_sk_{12}+s$ and  $E_R=c_sk_{34}+s$.
Again, the logarithmic and dilogarithmic functions are annihilated by the weight-shifting operator in the trispectrum, and we find the IR-finite shape function expressed by rational polynomials. Unlike the dS-invariant result in \eqref{trispec-dS}, here we find both total-energy and partial-energy poles, which means that this result cannot be given by a field redefinition.
Meanwhile, as the coupling constant $g_{\rm I}$ is less constrained in the EFT, this massless-exchange trispectrum provides potentially large signals that can be tested in the observational surveys.

\subsubsection{Multi-Speed Non-Gaussianity}
\label{sec:multispeed}

Now let's consider one particularly interesting case for the massless-exchange bispectrum with a new class of phenomenology. 
In the above analysis, we have focused on the three-point function with the $\dot\phi\s$ linear mixing, which leads to the logarithmic IR-divergent terms and consequently local-like shape functions.
In both dS-invariant and boost-breaking scenarios, the appearance of the IR divergence can be seen as a consequence of the fact that there is only one derivative in this quadratic interaction of two massless scalars.
Next, we extend the analysis to two-point vertices with higher derivatives. The most generic form of the boost-breaking quadratic interaction can be written as
\be
\mathcal{L}^{\phi\s} = a^{-m} \partial_{i}^{m} \big( \partial_t^{n_1} \phi \partial_t^{n_2}\s \big)~,
\ee
where $m$ is the number of spatial derivatives, and $n_1$ and $n_2$ are the numbers of time derivatives on $\phi$ and $\s$ respectively. For interactions with at least two derivatives, i.e. $m+n_1+n_2\geq2$, the mixed propagators become rational functions of $k$, the logarithms and IR-singular terms disappear \cite{Goodhew:2022ayb}.
Next, we shall see that the multiple sound speeds of the scalars become important for the non-Gaussianity signal.

\begin{framed}
\noindent{\underline{\it Convention}}~~~ Previously, ``$c_s$'' is used as the ratio of the sound speeds of the two fields. For the convenience of analysis in this subsection, we reintroduce the individual sound speeds for each field: $c_s$ for the inflaton $\phi$ and $c_\s$ for the massless field $\s$.  
\end{framed}
We first take a close look at the $\dot\phi\dot\s$ coupling as the simplest example. This quadratic interaction arises in theories with more than one derivative per field, and was also identified in the EFT of multifield inflation \cite{Senatore:2010wk}. One way to generate this coupling is to consider the following interacting operator of the inflaton $\Phi$ and $\s$, and then take one inflaton leg to the background
\be
\partial_{\mu\nu}\Phi \partial^{\mu\nu}\Phi   \s \rightarrow -2\ddot\Phi \dot\phi\dot\s+...
\ee
where we have used integration by parts, and dots represent terms with $\dot\phi\s$ couplings. In the following, we will keep agnostic about the model realization, but focus on the new effects on massless exchange correlators from this higher-derivative linear mixing.
First, we construct the mixed propagator from the $\dot\phi\dot\s$ coupling.
The bulk-to-bulk propagator for $\partial_\eta\s$ from one bulk time to another is given by
\bea
G_{++}^{\partial_\eta\s}(c_\s k,\eta, \eta') &=& \partial_\eta \sigma_{k}(\eta) \partial_{\eta'} \sigma^*_{k}(\eta') \theta(\eta - \eta') +
\partial_\eta \sigma^*_{k}(\eta) \partial_{\eta'} \sigma_{k}(\eta') \theta(\eta' - \eta) \nn\\ 
G_{+-}^{\partial_\eta\s}(c_\s k,\eta, \eta') &=& 
\partial_\eta \sigma^*_{k}(\eta) \partial_{\eta'} \sigma_{k}(\eta') ~,
\eea
where $G_{--}^{\partial_\eta\s}$ and $G_{-+}^{\partial_\eta\s}$ are their complex conjugates respectively. The new mixed propagator from $\s$ to $\phi$ is given by
\begin{small}
\be
\cG_\pm (k, \eta , \eta_0) = \pm i  \int_{-\infty}^{\eta_0} \frac{d\eta'}{H^2\eta'^2} \[ G_{\pm\pm}^{\partial_\eta\s}(c_\s k,\eta, \eta') \partial_{\eta'} K_{\pm}(c_s k, \eta')
- G_{\pm\mp}^{\partial_\eta\s}(c_\s k,\eta, \eta') \partial_{\eta'} K_{\mp}(c_s k, \eta')
\] .
\ee
\end{small}
This integration can be easily solved, and we find a simple analytical expression for $\cG_+$
\begin{align}
\cG_+ (k, \eta  ) = 
        \begin{dcases}
            \frac{H^2\eta}{4c_s k}(1-ic_sk\eta) e^{ic_s k\eta}~,\quad & c_s = c_\s\\
              \frac{{c_\s} }{(c_\s^2-c_s^2)} \frac{H^2\eta}{2 {c_s}k} \(c_\s e^{ic_sk\eta \mathfrak{a}}-c_s e^{ic_\s k\eta}\)  ~,\quad & c_s \neq c_\s
        \end{dcases} \qquad .
\end{align}
The parameter $\mathfrak{a}=1-i\epsilon$ is introduced to take care of the $i\epsilon$-prescription such that $\cG_+ \propto e^{ic_\s k\eta}$ in the early-time limit $\eta\rightarrow -\infty$.
We first notice that there is no logarithmic functions or $\eta_0$-related singular behaviour for this linear mixing with two derivatives. 
When $c_s=c_\s$, the mixed propagator is similar with one free bulk-to-boundary propogator of the external field with sound speed $c_s$. 
The $c_s \neq c_\s$ case is more interesting, where $\cG_+$ becomes a combination of two free bulk-to-boundary propogators with   different sound speeds.
The $e^{ic_s k\eta }$ piece can always be mimicked by the inflaton propagator, and thus its contribution to correlators is present already within single field inflation. The novel effect of the additional light field $\s$ is given by the $e^{ic_\s k\eta }$ piece, which propagates to the boundary with the $\s$ sound speed and thus cannot be mimicked by $K$. This is due to the fact that the mixed propagator encodes some behaviour of the intermediate field. While for the $\dot\phi\s$ interaction the IR-divergent terms always dominate, for $\dot\phi\dot\s$ and other higher derivative couplings, the sound speed effect of the exchanged field gets manifested in a novel way.

\vskip4pt

\begin{figure}[t!]
   \centering
      \includegraphics[height =6cm]{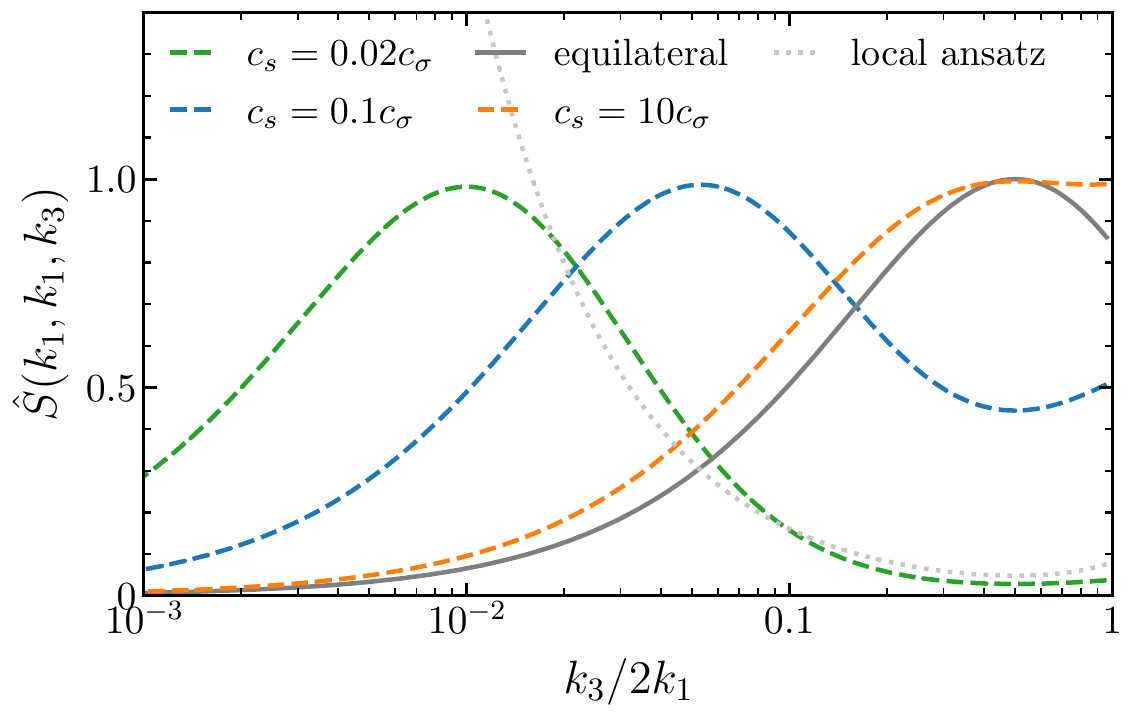}   
      \caption{ The dimensionless shape function of the bispectrum \eqref{multiS} with $k_1=k_2$ and various sound speed ratios. We have also include the standard equilateral and local shapes for comparison. The peaks of the shapes are normalized to be 1 regardless of its location.} 
      \label{fig:multiS}
\end{figure}

Next, to investigate new signatures in correlation functions, we separate out the $c_\s$-related term in $\cG_+$, and introduce the following form of the bulk-to-boundary propagator
\be \label{mix-new}
\cG_+^{c_\s} ( c_\s k, \eta  ) =  \frac{H^2\eta}{2 {c_\s}k} e^{ic_\s k\eta } ~.
\ee
For demonstration, we assume the cubic interaction is given by $\dot\phi^2\dot\s$, and then the single-exchange contribution to the inflaton correlator from the above mixed propagator  becomes
\bea  \label{multiS}
\langle \phi_{\bf k_1} \phi_{\bf k_2} \phi_{\bf k_3} \rangle' &=& i \int_{-\infty}^{\eta_0} d\eta a(\eta) \[ \partial_\eta K_+(c_s k_1,\eta) \partial_\eta K_+ (c_s k_2,\eta) \cG_+^{c_\s} ( c_\s k_3, \eta  )  -c.c. \] + {\rm perm.} \nn\\
&\propto & \frac{1}{k_1^3k_2^3k_3^3} \frac{e_3^2}{(c_s k_{12}+c_\s k_3)^3}  + {\rm perm.}~.
\eea
The bispectrum shape is similar with the equilateral shape from the $\dot\phi^3$ self-interaction, however because of the two sound speeds, the shape function may not be peaked at the equilateral configuration.
Figure \ref{fig:multiS} shows how the shape changes with different choices of the two sound speeds.
For $c_s>c_\s$, as $k_3 \leq k_{12}$, the bispectrum is still close to the equilateral shape.
For $c_s<c_\s$, 
the location of the peak is shifted to $k_3/k_{12} = c_s/c_\s$. For the extremal situation $c_s\ll c_\s$, we find the shape function reaches its maximum around the squeezed limit $k_3 \ll k_{12}$, which is similar with the local shape.
Thus by tuning the sound speed ratio, we find a parameter class of shape functions that can have arbitrary peak location, and approximately interpolate the local shape and the equilateral shape.

\vskip4pt
The shifted location of the peak in the shape function has a simple and intuitive explanation. 
Recall that in single field inflation with a nontrivial sound speed, the bispectrum is generated by contact interactions, and its shape is peaked at the equilateral limit due to the enhancement of resonance when three modes exit the sound horizon at the same time $k_1=k_2=k_3 = a(t_*)H/c_s$.
In the massless exchange here with \eqref{mix-new}, the mixed inflaton leg propagates to the boundary with the sound speed of the exchanged scalar $c_\s$. Thus, at some time $t_*$ when $k_1$ and $k_2$ modes of the free inflaton leg exit the $c_s$ sound horizon with $k_1=k_2 = a(t_*)H/c_s$, the  mixed $k_3$-leg has a different sound horizon crossing with $k_3 = a(t_*)H/c_\s$. As a consequence, the resonance enhancement happens for $c_s k_{12} =c_\s k_3$, and thus the sound speed ratio determines where the bispectrum peak is located. Similar phenomenon was recently reported as the low-speed resonance signal in massive exchanges \cite{Jazayeri:2022kjy}. Our focus here is the massless exchange bispectra, and the shape functions take simpler forms.

\begin{figure} 
   \centering
   \begin{subfigure}{0.4\textwidth}
      \includegraphics[height =4.6cm]{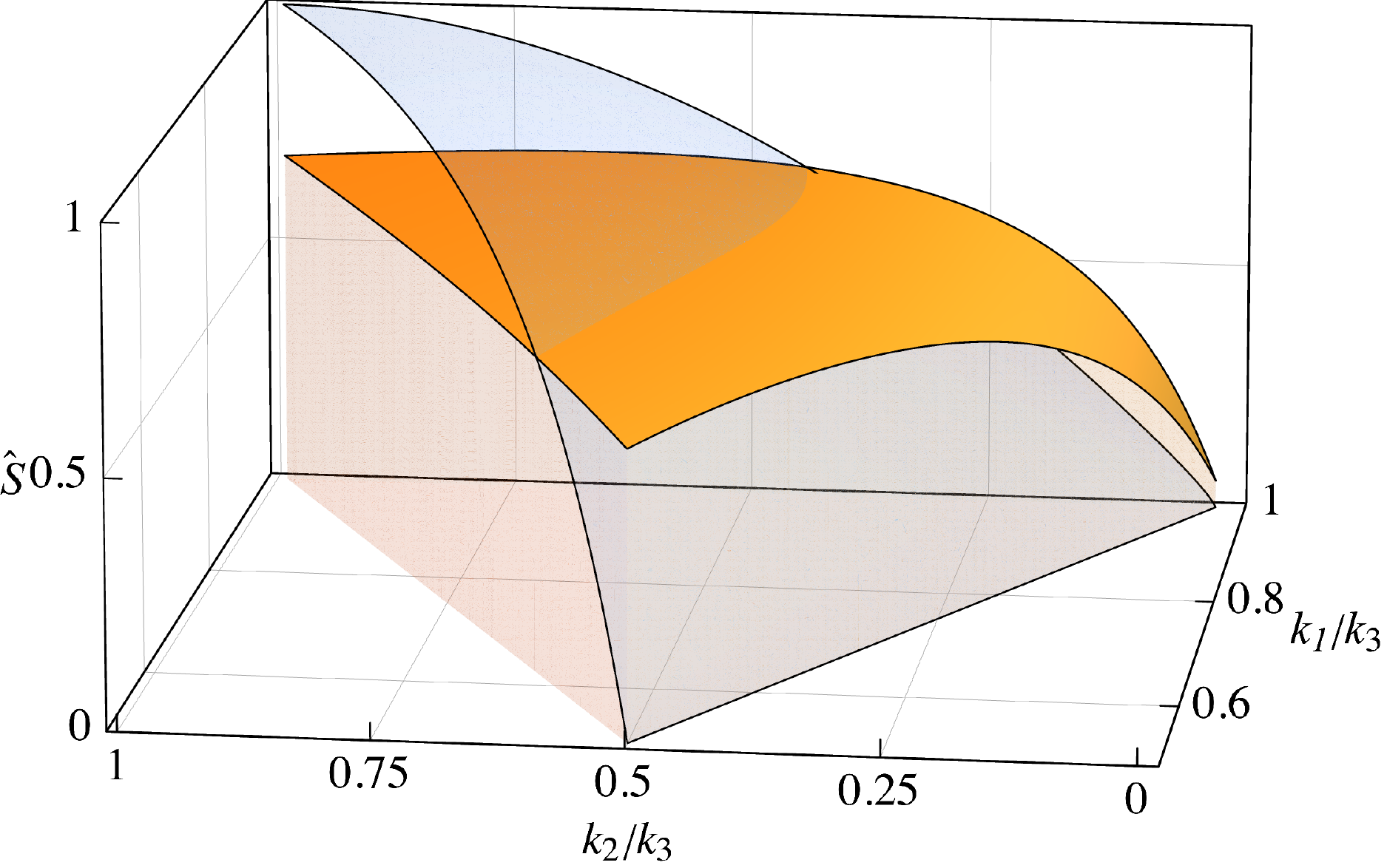} \caption{  $c_1=c_2=0.8$, $c_3=1$}
      \end{subfigure} \hspace{1cm}
   \begin{subfigure}{0.4\textwidth}
      \includegraphics[height =4.6cm]{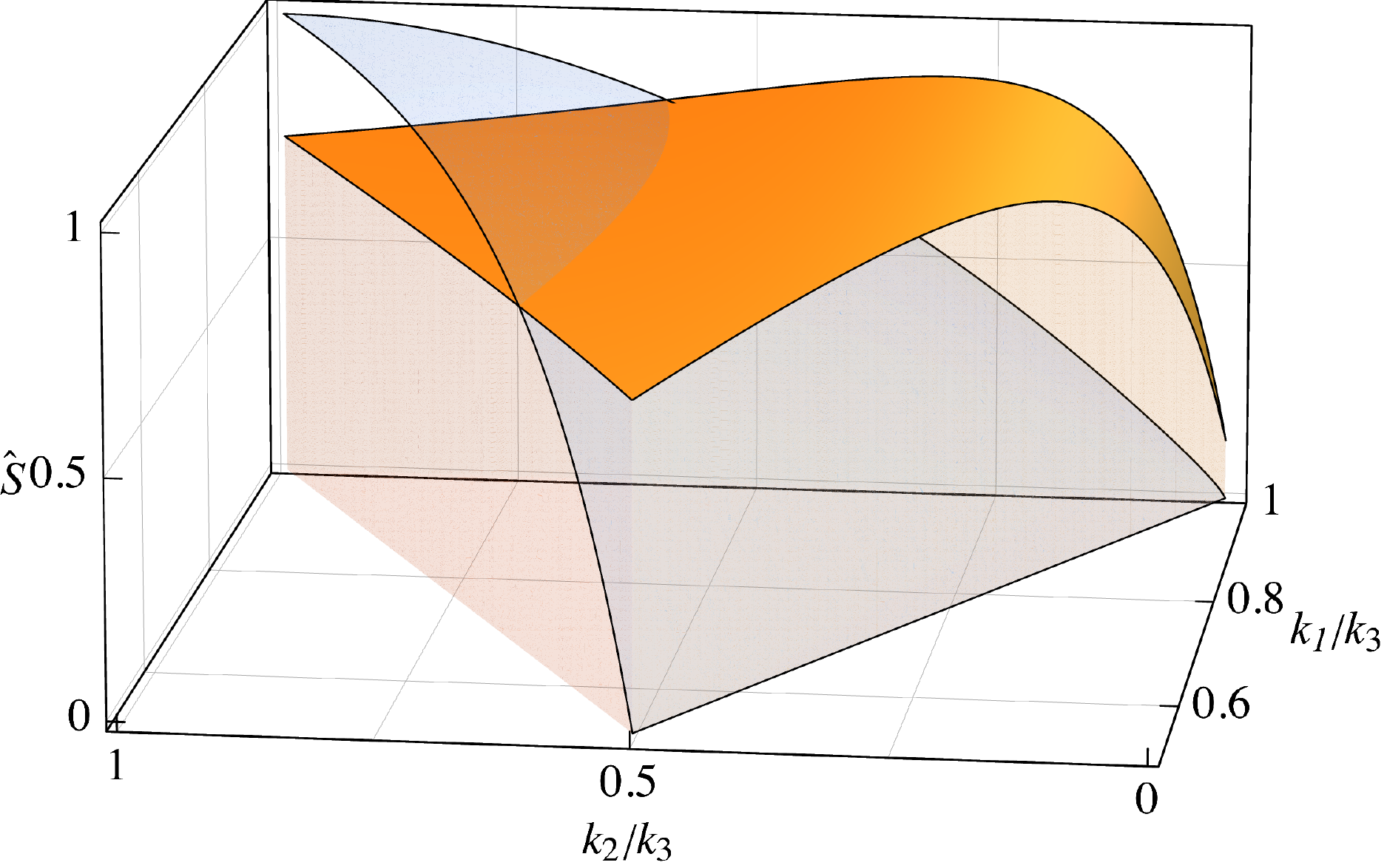} \caption{  $c_1=0.3$, $c_2=0.4$, $c_3=1$}
      \end{subfigure}  \\ \vspace{0.5cm}
       \begin{subfigure}{0.4\textwidth}
      \includegraphics[height =4.6cm]{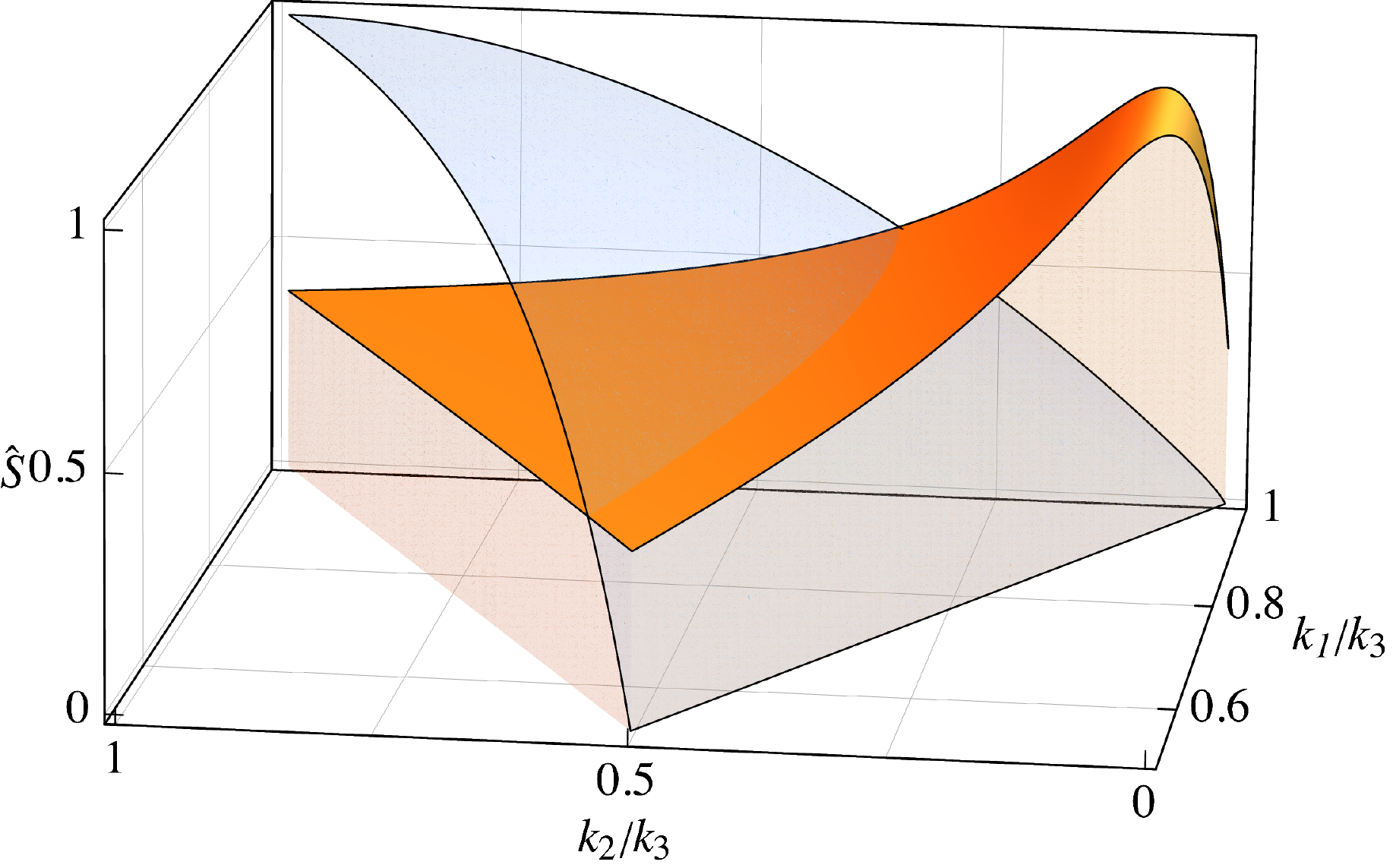} \caption{  $c_1=0.5$, $c_2=0.2$, $c_3=1$}
      \end{subfigure}  \hspace{1cm}
   \begin{subfigure}{0.4\textwidth}
      \includegraphics[height =4.6cm]{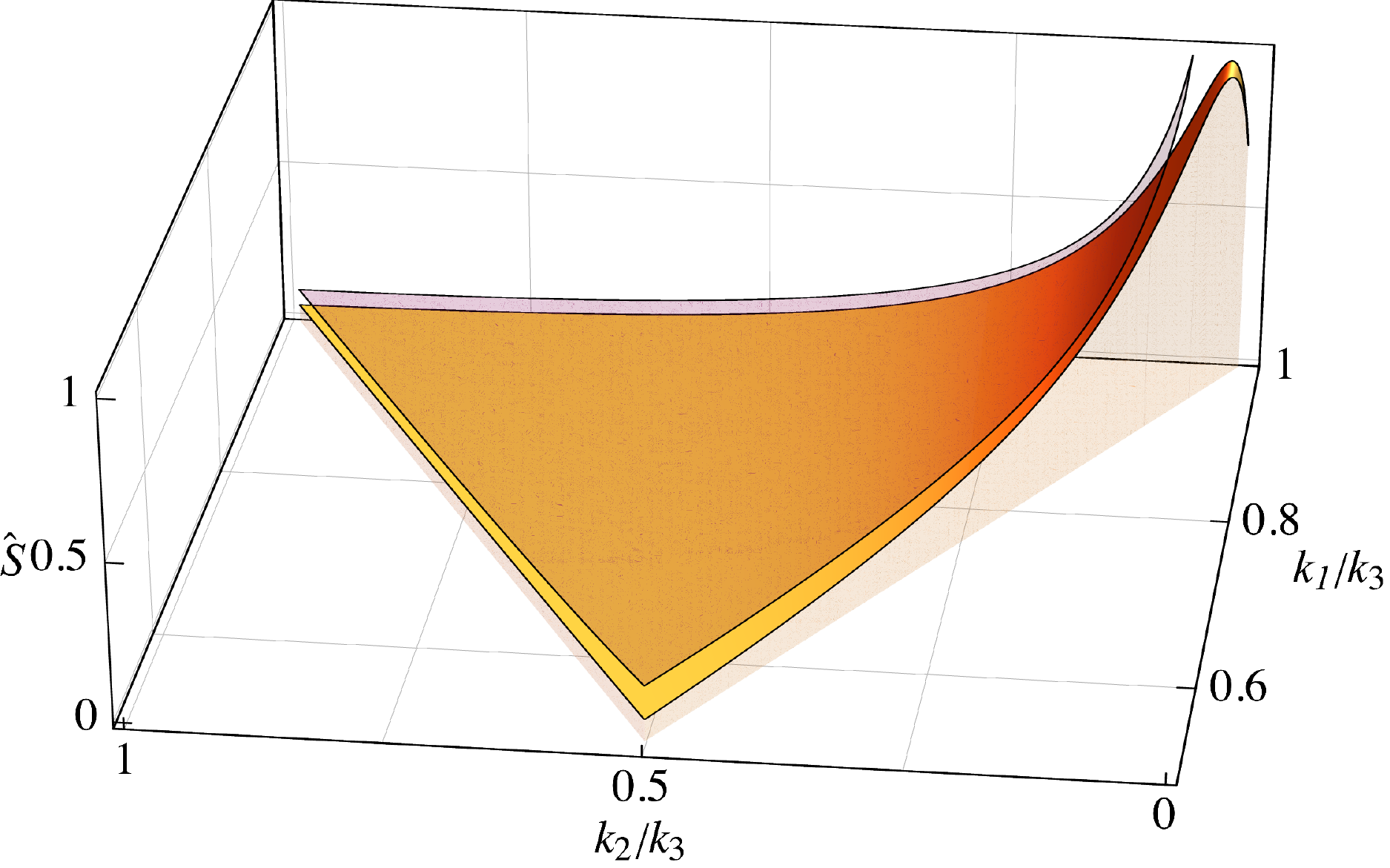} \caption{  $c_1=c_2=0.05$, $c_3=1$}
      \end{subfigure} 
      \caption{The multi-speed shapes $ \hat{S}^{{\rm multi-}c_s}(k_1,k_2,k_3)$ with four different choices of sound speed parameters. For comparison, we have also plotted the standard equilateral shape ({\it blue transparent surfaces})  in figures a), b), c), and the local shape ({\it purple transparent surface})  in figure d). From figure a) to d) we see the peak of shape is shifted from the equilateral configuration $k_1=k_2=k_3=1$ to the squeezed limit $k_2\ll k_1\simeq k_3$.} 
      \label{fig:multics}
\end{figure}

\vskip4pt

Furthermore, we can extend our analysis to exchange processes with two or three mixed propagators. 
Let's consider a double-exchange diagram that has two intermediate fields $\s_{\rm I}$ and $\s_{\rm II}$ with sound speeds $c_{\rm I}$ and $c_{\rm II}$ respectively.
The two mixed propagator $\cG^{\rm I}_+$ and $\cG^{\rm II}_+$ can be expressed by \eqref{mix-new} with their own sound speed  being the one of the exchanged field. Then for the cubic interaction $\dot\phi\dot\s_{\rm I}\dot\s_{\rm II}$, the bispectrum becomes
\bea 
\langle \phi_{\bf k_1} \phi_{\bf k_2} \phi_{\bf k_3} \rangle' &=& i \int_{-\infty}^{\eta_0} d\eta a(\eta) \[ \partial_\eta K_+(c_s k_1,\eta)  \cG_+^{\rm I} ( c_{\rm I} k_2, \eta  )  \cG_+^{\rm II} ( c_{\rm II} k_3, \eta  )  -c.c. \] + {\rm perm.} \nn\\
&\propto & \frac{1}{k_1^3k_2^3k_3^3} \frac{e_3^2}{(c_s k_{1}+c_{\rm I} k_2+c_{\rm II} k_3)^3}  + {\rm perm.}~.
\eea
Now we have three massless fields involved, and each of them has its own sound speed. The resulting bispectrum demonstrates richer structure in the shape function, with each momentum $k$ associated with a sound speed parameter. Intuitively, the location of its peak is determined by two sound speed ratios: $c_s/c_{\rm I}$ and $c_s/c_{\rm II}$. 
Similarly, we can also consider the triple-exchange bispectrum, where all the three sound speeds are the ones of the mediator fields.

\vskip4pt

Let's summarize the major messages from the above example. In the massless exchanges   with higher-derivative quadratic interactions, the IR divergences vanish and we find the bispectrum in terms of rational polynomials of $k$. More importantly, as the mixed propagators inherit the sound speeds of the exchanged fields, it becomes possible for each momentum to be associated with a different sound speed. As a result, the exchange processes can probe multiple sound horizon crossings, which leads to the new phenomenology in the scalar bispectrum. In addition to the simplest $\dot\phi\dot\s$ mixing we considered here, the analysis can be easily extended to other quadratic interactions, such as $\ddot\phi\s$,  $\dot\phi\partial_i^2\s$, etc., and cubic ones, though the expressions of  mixed propagators and the final bispectra become more complicated.

\vskip4pt
We dub this new class of bispectrum shapes the {\it multi-speed non-Gaussianity}, which is a distinctive signature of additional light degree of freedom during inflation. For the convenience of data analysis,   a simple ansatz of the dimensionless shape function is given by
\begin{eBox}
\be \label{multics}
\hat{S}^{{\rm multi}-c_s} (k_1,k_2,k_3)= \frac{k_1k_2k_3}{(c_1 k_1+ c_2 k_2 +c_3 k_3)^3} +  5~{\rm perms}~,
\ee
\end{eBox}
where $0<c_{1,2,3}\leq 1$ are three different sound speed parameters.
When they have the same value, we return to the equilateral shape. But as we are free to tune their ratios, this shape function is able to capture various possibilities of the scalar bispectra that are rational functions of three momenta. Examples are shown in Figure \ref{fig:multics}.
In particular, the location of the peak is  sensitive to the sound speed ratios. 
When the sizes of three sound speed are comparable, we return to the equilateral shape.
While for large hierarchies among sound speeds, the shape function  can even be peaked around the squeezed configuration which mimics the local shape.
For the general parameter choices,  \eqref{multics} interpolates the two standard non-Gaussian shapes.
Therefore, this simple but general template can be of particular interest for the observational search of primordial non-Gaussianity.

\section{Revisiting Multi-Field Non-Gaussianities}
\label{sec:revisit}

The generation of nonlinearities in primordial perturbations has been extensively investigated in the context of multi-field inflation.
The standard approach to compute primordial non-Gaussianities in this class of models is the $\delta N$ formalism, which uses the separate universe assumption, and mainly captures the nonlinearity of curvature perturbations outside of the Hubble radius \cite{Salopek:1990jq, Sasaki:1995aw, Starobinsky:1986fxa, Sasaki:1998ug, Lyth:2004gb, Lee:2005bb, Lyth:2005fi}. 
Meanwhile, as shown in the previous sections, the bootstrap approach takes another perspective, where all the nonlinearities are generated as consequence of interactions among quantum fields in a (quasi-)dS background. 
Thus an interesting question is how these two different approaches are related to each other. 
In this section, we shall review the analysis of multi-field inflation using $\delta N$, and then compare with the bootstrap result explicitly. 

\vskip4pt
First, in Section \ref{sec:deltaN}, we will give a brief review of the $\delta N$ formalism, and in particular the standard computation of the primordial bispectrum and trispectrum in multi-field inflation. 
Then we consider a specific two-field model of inflation with exactly solvable background dynamics in Section \ref{sec:ssoi}. From this concrete example, we shall compare the non-Gaussianity results from the $\delta N$ computation and the dS bootstrap. In Section \ref{sec:higher}, we comment on multi-field models with higher derivatives.

\subsection{The $\delta N$ Formalism}
\label{sec:deltaN}

The  starting point of the $\delta N$ formalism is the observation that the curvature perturbation $\zeta$ outside of the Hubble radius is related to the perturbed scale factor  for a local patch of the Universe.
The super-horizon curvature perturbations can be analyzed by considering  a local FLRW spacetime which evolves like a separate homogeneous universe.
Explicitly, the scale factor of a local universe is given by the following form
\be
a(t,{\bf x}) = a(t) e^{\zeta(t,{\bf x})}~.
\ee
Now consider the number of e-folds of this local FLRW universe $N(t_*,{\bf x})$ from an initial time $t_*$ to the end of inflation at $t_0$:
 we pick a spatially flat slice at time $t_*$, with $\zeta(t_*,{\bf x})=0$; and then consider the comoving curvature slice at time $t_0$. For this local patch of the Universe, the number of e-folds (and the corresponding curvature perturbation $\zeta( {\bf x})$ at  $t_0$) is given by
\be
N(t_*,{\bf x}) = \log\( \frac{a(t_0,{\bf x})}{a(t_*)} \) =N(t_*) +\zeta( {\bf x})  ~~~  \Rightarrow ~~~ \zeta( {\bf x}) = \delta N( {\bf x})  \equiv   N(t_*,{\bf x}) -N(t_*)  ~.
\ee
On the spatially flat slice at $t_*$, though the curvature perturbation vanishes, we are left with fluctuations of scalar fields
$
\Phi^a (t_*, {\bf x}) = \Phi^a_0 (t_*)+  \phi^a (t_*, {\bf x})
$.
In multi-field inflation, these scalar fields constitute the ``multiple inflatons" which control the expansion history of the Universe. Therefore, for a set of initial field values of $\Phi^a(t_*, {\bf x})$, the background evolution of this local universe leads to the corresponding number of e-folds $N(t_*,{\bf x}) = N(\Phi^a)$.
As a result, we are able to expand the $\delta N$ formula  in terms of initial field fluctuations $\phi^a (t_*, {\bf x})$ at time $t_*$
\be \label{deltaN}
\zeta( {\bf x}) = N(\Phi^a_0+\phi^a) -N(\Phi^a_0) = N_a  \phi^a   + \frac{1}{2} N_{ab}   \phi^a    \phi^b  + \frac{1}{6} N_{abc}  \phi^a    \phi^b  \phi^c  + ...~,
\ee
where $N_a \equiv \partial N  (t_*, {\bf x}) / \partial \Phi^a$, and other higher order ones $N_{ab} $, $N_{abc} $ are the derivatives of $N$ defined on the initial slice at $t_*$. 
Typically we work in Fourier space, and 
the initial slice is taken at the time $t_k$ when the Fourier mode $k$ exits the horizon $k=a(t_k)H(t_k)$. Thus  these derivatives of $N$ may acquire some mild  $k$-dependence when background parameters are not exactly time-independent during inflation. To avoid clutter, we will not  explicitly write them as $N_a(k) $, $N_{ab} (k) $, ..., but shall keep in mind that  these e-folds derivatives are associated with the corresponding $k$-mode of the curvature perturbation.

\vskip4pt

One key assumption of the $\delta N$ formalism is the separation of the sub-horizon (quantum) physics from the super-horizon (classical) effects \cite{Dias:2012qy}. 
This can be seen from the above derivation of the $\delta N$  formula \eqref{deltaN}:
 field fluctuations at $t_*$ provide initial conditions  around horizon crossing; the $\delta N$ expansion captures the subsequent evolution on super-horizon scales.
In the conventional approach, these two contributions are supposed to take factorized forms as shown in each term of \eqref{deltaN}, where initial field fluctuations are normally given by the mode functions of {\it free} massless scalars.
For multi-field inflation, as light scalar fluctuations freeze after horizon exit, one expects the major contribution to the nonlinearity of $\zeta$ comes from the super-horizon conversion process. In this sense, 
the $\delta N$ formalism  provides a simple and intuitive method for computing the dominant contribution to  multi-field non-Gaussianities.

\vskip4pt

Now let's briefly review the correlation functions of $\zeta$ in Fourier space using the $\delta N$ formula.
The two-point correlator  in this approach  is simply given by
\be \label{deltaN-2pt}
\langle \zeta_{\bf k} \zeta_{\bf -k}  \rangle' =  N_a  N_b  \langle \phi^a_{\bf k} \phi^b_{\bf -k}  \rangle' = G^{ab}N_a N_b  \mathcal{P}_\phi(k)~,~~~~{\rm with} ~~ \mathcal{P}_\phi(k)=\frac{H^2}{2k^3}~,
\ee
where $G_{ab}$ is the field space metric and $P_\phi$ is the two-point function of a canonically normalized massless scalar. Then the scale-invariant primordial power spectrum is defined as $P_\zeta \equiv  k^3 \langle \zeta_{\bf k} \zeta_{\bf -k}  \rangle' $. Similarly, using \eqref{deltaN} up to the second order, we can find the bispectrum of the primordial curvature perturbation
\bea \label{bispec-deltaN}
\langle \zeta_{{\bf k}_1} \zeta_{{\bf k}_2} \zeta_{{\bf k}_3}  \rangle' 
 &=& N_a N_b N_c \langle \phi^a_{{\bf k}_1}  \phi^b_{{\bf k}_2}  \phi^c_{{\bf k}_3 }   \rangle'_* + \frac{1}{2} N_a N_b N_{cd} ~ \langle \phi^a_{{\bf k}_1}  \phi^b_{{\bf k}_2} \int_{\bf p} \phi^c_{{\bf k}_3-{\bf p}} \phi^d_{{\bf p}} \rangle' + 2 ~{\rm perms.}\nn\\
& = & G^{ac} G^{bd} N_a N_b N_{cd} \big[ \mathcal{P}_\phi(k_1) \mathcal{P}_\phi (k_2) + \mathcal{P}_\phi(k_1) \mathcal{P}_\phi (k_3) +\mathcal{P}_\phi(k_2) \mathcal{P}_\phi (k_3) \big]~,
\eea
where in the second line we have neglected the mild $k$-dependence in the derivatives of $N$.
Meanwhile, the three-point function $\langle \phi\phi\phi \rangle'_*$ is a correlator evaluated at the time $t_*$, and is generated by interactions of scalar fields around horizon crossing. In the second step, this contribution is taken to be very small and subdominant, as shown in \cite{Seery:2005gb} for slow-roll models.
The above formula provides the standard derivation of local non-Gausianity in multi-field inflation. If we compare with the notation in \eqref{local}, we find the above bispectrum shape is simply $S_{\rm local}$, and the size of the non-Gaussianity  is given by
\be
\fnl = \frac{5}{6} \frac{G^{ac} G^{bd} N_a N_b N_{cd}}{\(G^{ab}N_a N_b\)^2}~.
\ee 
The primordial trispectrum can be derived by considering the $\delta N$ expansion \eqref{deltaN} up to the third order. 
We can also neglect the slow-roll suppressed terms from contact interactions, and then the trispectrum has two types of contributions \cite{Seery:2006js, Byrnes:2006vq}
\bea
\langle \zeta_{{\bf k}_1} \zeta_{{\bf k}_2} \zeta_{{\bf k}_3} \zeta_{{\bf k}_4} \rangle' 
& = & G^{ad} G^{be} G^{cf} N_a N_b N_c N_{def} \big[ \mathcal{P}_\phi(k_1) \mathcal{P}_\phi (k_2)\mathcal{P}_\phi (k_3) +  3~ {\rm perms.} \big] \nn\\ [6pt]
&& + G^{ac} G^{be} G^{df} N_a N_b N_{cd} N_{ef} \big[ \mathcal{P}_\phi(k_1) \mathcal{P}_\phi (k_2)\mathcal{P}_\phi (|{\bf k}_1+{\bf k}_3|) +  11~ {\rm perms.} \big] ~.
\eea
Again   we have ignored the $k$-dependence in the derivatives of $N$.
This shape is given by the two local ans\"atze in \eqref{tri-local}, and we find the corresponding size parameters given by
\be
\tau_{\rm NL} =  \frac{ G^{ac} G^{be} G^{df} N_a N_b N_{cd} N_{ef} }{\(G^{ab}N_a N_b\)^3}  ~,~~~~~~   g_{\rm NL}= \frac{25}{54} \frac{G^{ad} G^{be} G^{cf} N_a N_b N_c N_{def}}{\(G^{ab}N_a N_b\)^3} ~.
\ee

\vskip4pt

As we can see, for both the bispectrum and trispectrum, the local non-Gaussianities are generated as a result of the $\delta N$ expansion \eqref{deltaN}, which is a  field redefinition at some local point $\bf x$ in coordinate space that converts fluctuations  of multiple free scalars to the final curvature perturbation on super-horizon scales. 
Again, we notice that in the above computation of correlation functions, the factorization of sub-horizon and super-horizon physics plays an important role. 
In particular, the sub-horizon information is simply given by correlators  at $t_*$, which are the ones of non-interacting scalars.
Next, we are interested in understanding the $\delta N$ analysis from the perspective of field interactions.
In other words, can we justify this treatment in first-principle computations?
If we see it as an approximation, how good is the $\delta N$ formalism, and what are the differences with the exact results?
With these questions in mind, we will explicitly compare the $\delta N$ analysis with the bootstrap by examining one specific two-field example.

\subsection{A Concrete Case Study}
\label{sec:ssoi}

Consider the following inflation model with two scalar fields $\Phi$ and $\Sigma$
\be \label{ssoi}
 \mathcal{L}= - \frac{1}{2} (\partial_\mu\Phi)^2 -\frac{1}{2}(\partial_\mu\Sigma)^2 - \frac{1}{2} \frac{\Sigma}{\Lambda}(\partial_\mu\Phi)^2 - V_{\rm sr}(\Phi) - V(\Sigma) ~.
\ee
The coupling between two field comes from a dimension-five operator in the kinetic term, which can be seen as a field space with the metric $G_{ab} = {\rm diag}\{1+\Sigma / \Lambda, 1\}$. Meanwhile, by requiring that the inflaton rolls in the $\Phi$ direction with any constant $\Sigma$, the  form of the  potential here is given by the Hamilton-Jacobi formalism 
\be
V_{\rm sr} (\Phi) = \frac{1}{2} m^2 \Phi^2~,~~~~~~ 
V(\Sigma) = -\frac{m^2\mpl^2}{3\(1+\Sigma/\Lambda\)} ~.
\ee
This is one version of the shift-symmetric orbital inflation proposed in \cite{Achucarro:2019pux}. Then in the FLRW spacetime, we find the two-field system and the inflationary background can be exactly solved
\be  \label{background}
\dot\Phi = - \sqrt{\frac{2}{3}} \frac{m \mpl}{\(1+\Sigma/\Lambda\)} ~,~~~~~~ \Sigma = \Sigma_0 ~,~~~~~~ H^2 = \frac{m^2}{6\mpl^2} \Phi^2 ~.
\ee
This exact solution ensured by the Hamilton-Jacobi construction shows a constant turning trajectory of the inflaton in the internal field manifold. It is helpful to apply the covariant formalism here and introduce the tangent and normal vectors to the trajectory $T^a =(-1/\sqrt{1+\Sigma_0/\Lambda},~0 )$ and $N^a =(0,~1)$.  Then the slow-roll and turning parameters are given by
\bea
\epsilon  &\equiv &  \frac{G_{ab}\dot\Phi^a \dot\Phi^b}{2H^2\mpl^2}
= \frac{2\mpl^2}{(1+\Sigma_0/\Lambda )\Phi^2 }  ~,
\\ \label{turning}
\Omega &\equiv & -N_a D_t T^a = -\frac{1}{2\sqrt{1+\Sigma_0/\Lambda}}\frac{\dot\Phi}{\Lambda}~.
\eea
As we will see soon, this is also the coupling constant for the linear mixing between the curvature and isocurvature perturbations. With the background solution \eqref{background} we can also solve for the analytical expression of $N$ in this two-field model. From an initial time $t_*$ to the end of inflation, the number of e-folds is given by
\be \label{efolds}
N_* = \frac{1}{4\mpl^2}\( 1+\frac{\Sigma }{\Lambda} \) \Phi_*^2 -\frac{1}{2}~,
\ee
with $\Phi_* = \Phi(t_*)$. As a last comment on the background dynamics, we take a look at the dimension-five operator from the EFT perspective. This coupling can be seen as a leading order expansion which respects the (approximate) shift symmetry of the inflaton field $\Phi$, and thus the validity of the EFT requires that $\Sigma_0 \ll \Lambda$.  

\vskip4pt
Next, we shall look into the generation of non-Gaussianity in this toy model using  the $\delta N$ formalism, and then compare with the bootstrap results.
Before going into the particulars, we notice that the similar analysis can be applied in  other multi-field inflation models, as long as the {\it Elpis} conditions in Section \ref{sec:pandora} are satisfied. Meanwhile, there are several reasons that we choose this specific model for demonstration:
\begin{itemize}
\item First, we have exact solutions for the background dynamics, with no need of slow-roll or other approximations. This makes it possible to apply the $\delta N$ formalism in a controllable fashion where  contributions to non-Gaussianities can be accurately computed. Thus, it becomes much easier to trace the conversion process in this class of models.\footnote{In this sense, the shift-symmetric orbital inflation can be seen as an analogy of power-law inflation where the exact solutions helped to justify the slow-roll approximations in single-field inflation. From this toy model of multi-field inflation, we are able to perform the exact examination of the conversion effects when additional light scalars are present.}
\item Second, since the inflaton moves in the $\Phi$ direction only, the curvature and isocurvature perturbations here are automatically associated with the $\Phi$ and $\Sigma$ field fluctuations. As we discussed in Section \ref{sec:pandora}, in many multi-field inflation models this decomposition is less manifest, and one needs to apply the field-space tangent and normal vectors at each point of the inflaton trajectory to identify the two corresponding perturbations.
\item Third, it is also convenient to  track interactions of field fluctuations in this model.  Perturbing the background solution $\Phi=\Phi_0(t)+\phi$ and $\Sigma=\Sigma_0 +\sigma$, we find the Lagrangian of fluctuations with interaction terms 
\be \label{interac}
\mathcal{L}_{\rm int}  = -  \lambda \dot\phi \sigma   -  g (\partial_\mu\phi)^2 \sigma ~,
\ee
where the coupling constants are given by $\lambda=2\Omega \simeq - {\dot\Phi}/{\Lambda}$ and $g= 1/{2\Lambda}$. 
These are the first two terms in the general form \eqref{int3} of the interacting Lagrangian which are responsible for the conversion.
As shown in \eqref{source}, due to the $\dot\phi \sigma$ linear mixing, the additional light scalar $\s$ can source the growth of the curvature perturbation after horizon-exit.
The  two coupling coefficients $\lambda$ and $g$ are associated with the turning rate \eqref{turning}, which is an important indicator of the multi-field effects. Thus this model provides the simplest setup for the conversion mechanism.
In addition, we notice that the  interactions in \eqref{interac} are the same as the one we discussed in Section \ref{sec:dS4pt3pt} for the bootstrap of the (approximately) dS-invariant correlators. 
\end{itemize}
Therefore, the predictions here can be analytically derived via both the bootstrap and $\delta N$ methods, and thus
this toy model serves as a link between these two approaches. 
It becomes a convenient choice for the purpose of comparison of different computations.

\paragraph{$\delta N$ analysis} Now we perform the computation for perturbations by using the $\delta N$ formalism. For the number of e-folds in \eqref{efolds}, we consider the initial time to be the horizon-exit time $t_k$ of the perturbation mode $k$. Then, the $\delta N$ formula can be written as
\be \label{deltaNk}
\delta N_k = \frac{1}{\sqrt{2\epsilon }\mpl} \phi  + \frac{N_k}{\Lambda} \s  + \frac{1}{\sqrt{2\epsilon }\Lambda\mpl } \phi \s  + \frac{1}{4\mpl^2} \phi^2 +\frac{1}{4\Lambda\mpl^2} \phi^2\s + ... ~,
\ee
where  the  background quantities in the $\delta N$ formula are from the derivatives of $N$  and defined at time $t_k$.
In particular, the $N$-dependent prefactor of the second term demonstrates the growth of the curvature perturbation with the isocurvature source. 
We keep the mild $k$-dependence of $N_k$ explicit here, and later will show its role in the exact $\delta N$ calculation of non-Gaussianities. 

\vskip4pt
We first take a look at the two-point correlator of the curvature perturbation using the $\delta N$ formula
\be  \label{deltaNs}
\mathcal{P}_\zeta(k) \equiv  \langle \zeta_{\bf k} \zeta_{\bf -k} \rangle' =  \frac{1}{2\epsilon \mpl^2}\mathcal{P}_\phi(k)\( 1 + \tilde{\lambda}^2 N_k^2 \)~,
\ee
where $\mathcal{P}_\phi$ is the power spectrum introduced in \eqref{deltaN-2pt}, and $\tilde\lambda = 2\Omega / H$ is a dimensionless coupling. 
This result agrees with what we find for the power spectrum correction \eqref{2pt-corr} from the bootstrap analysis, and matches the one from in-in formalism in Ref.~\cite{Achucarro:2016fby}.
We explicitly see that the IR-divergent logarithmic functions are expressed as the number of e-folds in the $\delta N$ calculation. 
This correction is simply due to the conversion from $\s$ to $\zeta$ on super-horizon scales.
Here we also take the slow-turn approximation, and thus the theory is under perturbative control, which requires $\tilde{\lambda} N_k < 1$. As a result, the single field power spectrum $\mathcal{P}_\zeta(k) \simeq \mathcal{P}_\phi(k)/({2\epsilon \mpl^2})$ remains a good approximation. 
For the primordial bispectrum, the $\delta N$ formula leads to 
\bea \label{bisp-deltaN}
\langle \zeta_{{\bf k}_1} \zeta_{{\bf k}_2} \zeta_{{\bf k}_3}  \rangle'_{\delta N} 
&\simeq & \frac{\tilde{\lambda}^2}{k_1^3k_2^3k_3^3} \Big[ N_{k_1} (k_2^3+k_3^3) +N_{k_2} (k_1^3+k_3^3)+N_{k_3} (k_1^3+k_2^3) \Big] P_\zeta^2 ~,
\eea
where  we have neglected slow-roll suppressed terms.
The shape function matches the second line of the bootstrap result \eqref{true-local} by taking $N_k\simeq -\log(-2k\eta_0)$.
As we noticed in the bootstrap analysis, this contribution can be seen as the consequence of a field redefinition with time-dependent coefficients. 
The $\delta N$ formula \eqref{deltaNk} here provides its explicit form. 
If we ignore the mild $k$-dependence in the number of e-folds and take $N_{k} =N_*$, this bispectrum reproduces the local shape in \eqref{local}.
From perturbativity, the size of non-Gaussianity here is small, as in many other multi-field models with nearly scale-invariant perturbations.

\vskip4pt

\paragraph{Comparison with bootstrap}
Next, we would like to understand the agreements and differences with the bootstrap result in \eqref{true-local}.\footnote{Another approach for computing the bispectrum is the in-in formalism, which tracks the full field interactions during inflation. For this model, the in-in integral is given by the three-point diagram in Figure \ref{fig:inflaton} with interaction vertices in \eqref{interac}. From this computation we found agreement with the bootstrap result.}
First, we notice that, to have a complete result in the $\delta N$ analysis, we also need to 
include the cubic interaction term that was discarded in \eqref{bispec-deltaN}.
Instead of being related to slow-roll suppressed couplings, here this contribution is given by the $(\partial_\mu\phi)^2\s$  vertex, and we find
\be \label{bisp-int}
\langle \zeta_{{\bf k}_1} \zeta_{{\bf k}_2} \zeta_{{\bf k}_3}  \rangle'_{\rm int}   = N_\phi N_\phi N_\s \langle \phi_{{\bf k}_1} \phi_{{\bf k}_2} \s_{{\bf k}_3}  \rangle'_* + {\rm perms.} 
= -\frac{\tilde{\lambda}^2}{2} N_{k_3} \frac{k_1^3 + k_2^3 - k_3^3}{k_1^3 k_2^3 k_3^3}  P_{\zeta}^2 + {\rm perms.} ~,
\ee
where $\langle \phi  \phi \s \rangle'_*$ is evaluated right after horizon crossing.\footnote{This three-point function is of the local form, which can be explained by the observation in \eqref{IbP}.}
In other words, the $\langle \phi  \phi \s \rangle'_*$ correlator provides the non-Gaussian initial condition at $t_*$ that is transferred by the $\delta N$ formula to the correlation of $\zeta$ at the end of inflation. 
While we are looking at the toy model in \eqref{ssoi}, the key point here is that, 
the presence of the $(\partial_\mu\phi)^2\s$ vertex is independent of particular models as long as the multi-field conversion happens (see Section \ref{sec:convert}), and its contribution has comparable size with the result \eqref{bisp-deltaN} from the $\delta N$ formula. 
Adding \eqref{bisp-int} into the final bispectrum, we find the squeezed limit matches the bootstrap result in \eqref{soft-dSbi}.

\vskip4pt
On the other hand, {\it away from the squeezed limit we find a  mismatch between the $\delta N$ and bootstrap results}. 
To understand the difference, we first recall that one primary assumption of the $\delta N$ analysis is the factorization of the super-horizon and sub-horizon physics, as we discussed in Section \ref{sec:deltaN}. Especially, in this approach the initial field fluctuations on the spatially flat slice are simply taken as the free mode function of massless scalars.
Precisely speaking, however, this treatment is not exact if we track the full field interactions during inflation.
For explicit demonstration, let's take a close look at the  $\delta N$ formula \eqref{deltaNk} and the mixed propagator \eqref{mixKp} with $c_s=1$. 
In this specific model, the second term of the  $\delta N$ expansion captures the super-horizon growth of $\zeta$ sourced by the isocurvature field $\s$, and its sub-horizon behaviour is simply given by the $\s$ mode function. 
Meanwhile, in the field-theoretic computation, the same physics process is described by the mixed propagator $\cG$ from the $\dot\phi\s$ linear mixing. We see that the late-time limit of $\cG$ in \eqref{tau0} corresponds to the super-horizon growth from the conversion, which agrees with the second term in the  $\delta N$ formula. 
Nevertheless, the conversion is already turned on {\it before} horizon exit, and field fluctuations are also affected by the $\dot\phi\s$ quadratic interaction  in early times.
As a result, the mixed propagator has a more complicated form, which cannot be simply captured by the free $\s$ mode function. 
This subtle difference leaves nontrivial imprints in the field-theoretic computation, and is responsible for the mismatch away from the squeezed limit.
Or, we may put it another way: to  implement the $\delta N$ computation more accurately, one needs to use the mixed propagator $\cG$ in \eqref{mixKp} as  the initial field fluctuations on the spatially flat slice, instead of the free $\s$ mode function.

\vskip4pt
As a consequence, one novelty from the bootstrap is the logarithmic $k_t$-pole in \eqref{true-local}. This type of singularity at $k_t\rightarrow 0$ is a ramification of field interactions at very early times $\eta\rightarrow -\infty$.
In general, the residue of  correlators in this limit is  associated with the scattering amplitudes in flat spacetime \cite{Maldacena:2011nz, Raju:2012zr}. 
Thus the $k_t$-pole terms arise only in field-theoretic computations, and cannot be mimicked by field redefinitions.
More specifically, the logarithmic $k_t$-pole of the massless-exchange bispectrum is generated by the early-time limit of the mixed propagator in \eqref{earlycG}.
This is a feature of the $\dot\phi\s$ field interaction deep inside the horizon. Thus it is absent when we separate the super-horizon and sub-horizon effects in a factorized form like in \eqref{deltaNk}.

\vskip4pt
The primordial 
trispectrum can also be derived from the $\delta N$ formula
\be \label{tri-deltaN}
\langle \zeta_{{\bf k}_1} \zeta_{{\bf k}_2} \zeta_{{\bf k}_3} \zeta_{{\bf k}_4} \rangle' = \tilde{\lambda}^2 \[   T_{\rm local1} 
   + \frac{3}{2}  T_{\rm local2} \] P_\zeta^3 ~,
\ee
with the two local ansatzs given in \eqref{tlocal1} and \eqref{tlocal2}. 
Compared with the bootstrap result  \eqref{trispec-dS}, the $\delta N$ computation gives a different combination of these two shapes, for which the soft limit of \eqref{tri-deltaN} does not vanish. Like in the bispectrum, the difference here is due to field interactions from cubic vertices, which is missed by just using the $\delta N$ formula.

\vskip4pt
To summarize, from this toy model we see that the $\delta N$ formalism provides a good approximation for the multi-field non-Gaussianities in the squeezed limit, and we need to be careful to incorporate
 the full nonlinearities of the conversion process: one is given by the $\delta N$ formula; another is from the  $(\partial_\mu\phi)^2\s$ cubic vertex.
In the bootstrap analysis, the super-horizon contribution to the  multi-field non-Gaussianity is associated with the long-wavelength behaviour of the mixed propagator \eqref{tau0} with the IR-divergent term. 
Meanwhile, away from the squeezed limit, there are differences in these two approaches. In particular, the total-energy singularity 
due to sub-horizon quantum interactions is missed in the previous  $\delta N$ computation. 
As the multi-field conversion is in general due to the $\dot\phi\s$ and $(\partial_\mu\phi)^2\s$ field interactions,
the specific model discussed here provides the minimal setup  for demonstrating the agreement and difference in the bootstrap and $\delta N$ results. Similar analysis can be  extended to other multi-field models with the conversion mechanism, which may have more complicated background dynamics and nontrivial decomposition of the curvature and isocurvature modes.

\subsection{Models with Higher Derivatives}
\label{sec:higher}

While the dS bootstrap corresponds to multi-field inflation with two-derivative kinetic terms,
the boost-breaking scenarios we analyzed in Section \ref{sec:boostless} can be compared with models with higher derivative interactions.
One example is the multi-field version of $P(X)$-type theories, 
where both curvature and isocurvature perturbations may have reduced sound speeds \cite{Langlois:2008mn, Langlois:2008wt, Langlois:2008qf, Arroja:2008yy, Cai:2008if, Cai:2009hw}. In these previous studies, it was shown that this class of models can generate both local- and equilateral-type primordial non-Gaussianities. 
Comparing with the results in Section \ref{sec:cs-IR}, we find the boostless bootstrap  provides the full  expressions  \eqref{dotphi2sigma} and \eqref{diphi2sigma}  for the local-like shapes from the first-principle computation, while the equilateral shape is generated through the inflaton self-interaction as in single field inflation. 
Although the squeezed-limit behaviour there remains similar with the dS-invariant case, we identify richer singularity structures in these analytical results, which are consequences of nontrivial sound speeds and boost-breaking interactions.

\vskip4pt

Multi-field inflation has also been studied by using the EFT of fluctuations \cite{Senatore:2010wk}.
There the authors  chose to be agnostic about the conversion mechanism, and proposed a parametrization for the relation between additional light scalars and the final curvature perturbation. In that approach, apart from the conversion, the higher-derivative EFT operators with light scalars were constructed in a systematic way.
New shapes of non-Gaussianities were identified as a result of the nontrivial self-interactions of additional fields. 
In our work, the bootstrap analysis has been focused on the single exchange processes with an intermediate massless state. In Section \ref{sec:multispeed}, we see that a careful investigation of the higher derivative interactions, especially the quadratic ones such as $\dot\phi\dot\s$,  leads to a novel class of non-Gaussian signals. These multi-speed shapes has not been reported in previous literatures.

\vskip4pt
As a final remark, as we have discussed in Section \ref{sec:convert}, the conversion mechanism can be more accurately described by field interactions in a model-independent fashion.
The coupling forms are constrained by the nonlinearly realized spacetime symmetry and the orthogonality condition between curvature and isocurvature modes.
Therefore, it remains a very interesting question about how to embed these interactions in the EFT, which may require a construction based on internal field spaces. 
We leave this for future work.

\section{Conclusions and Outlook}
\label{sec:concl}

In this paper, we presented a systematic investigation of primordial correlation functions with the presence of intermediate massless scalars, working at leading order in the weak coupling of this scalar to curvature perturbations. In inflationary cosmology, the signatures correspond to primordial non-Gaussianity from multi-field models. The resulting non-Gaussianities couple short and long distances,   which make them a suitable target for cosmological observations. Despite having been extensively studied in the literature, the recent theoretical advances coming from the bootstrap motivated us to revisit this important topic.

\vskip4pt
We began with the analysis of IR divergences caused by interacting massless scalars in de Sitter space. 
As light fields freeze after horizon crossing, there are cumulative effects due to interactions on super-horizon scales, which typically lead to logarithmic-type singular behaviour in correlators at the future boundary. An explicit late time cutoff $\eta_0$, related to the end of inflation, is necessary to regulate the IR effects.
Focusing on the leading order in perturbation theory, we showed that the boundary differential equations acquire explicit $\eta_0$-dependence, and proceeded to solve the equations in detail. 
Using weight-shifting operators, we obtained a large menu of multi-field inflationary bispectra and trispectra. Our results incorporate both the (approximately) dS-invariant and boost-breaking scenarios:

\begin{itemize}
\item For the (approximately) {\bf dS-invariant} case, we recover much of the previous literature on multi-field inflation.
We show that the super-horizon conversion from isocurvature to curvature perturbations, which is responsible for the generation of local non-Gaussianity, can be reformulated in terms of field interactions and a $\dot\phi\s$ mixed propagator. Using a benchmark example, we compared the bootstrap results with the ones from the $\delta N$ formalism. 
Both of them resemble the local shape, and
we find agreement in the squeezed limit. Meanwhile, contributions from field interactions around and before horizon-exit are more accurately captured using our methods.

\item  For theories with {\bf broken boosts}, the number of possibilities is richer. Having a different dispersion relation for the inflaton boosts the overall signal, and we can now make the mediator subluminal, and also couple it to the inflaton in a variety of ways. We systematically classified the possible signals in this scenario. In particular, IR-divergent terms remain in the three-point function, due to the $\dot\phi\s$ linear mixing. Consequently, the squeezed bispectra are similar to the dS-invariant ones, though the shape functions differ away from that limit. For the case of higher derivative quadratic interactions, the light scalars generate what we called multi-speed non-Gaussianity. The resulting shapes are equilateral-like, but the location of the peak is determined by the sound speed ratios among light scalars. With a very simple ansatz, we capture the various phenomenological possibilities: depending on the sound speed parameters, we interpolate between the standard local and equilateral shapes.

\end{itemize}

We close by listing several interesting directions for future exploration:

\begin{itemize}

\item The analysis in this paper has been restricted within the perturbative regime of tree diagrams with massless scalars. In order to go beyond that, we need to understand IR effects in cosmology better. Thankfully, recent studies on IR divergences in dS suggest that the stochastic formalism provides an effective framework for analyzing correlators beyond perturbation theory \cite{Gorbenko:2019rza, Mirbabayi:2019qtx, Cohen:2020php,Cohen:2021fzf, Green:2022ovz, Cohen:2022clv, Panagopoulos:2019ail, Achucarro:2021pdh}. It would be very interesting to consider the non-perturbative behaviour of inflation with additional light scalars, where ``stochastic" effects are expected to become important.

\item We mainly studied single exchanges of one additional light scalar. In general multi-field models, there can also be double- or triple-exchange contributions to the inflaton bispectrum. We briefly discussed these possible channels in Section \ref{sec:multispeed}. New tools are required to systematically deal with these diagrams. It would be interesting if the resulting shapes have new phenomenology compared to the single-exchange cases discussed here.

\item Another intriguing question is the role of symmetries in cosmological correlators from multi-field scenarios. Like single field inflation, here the interaction forms between the inflaton and additional fields are constrained by nonlinearly realized spacetime symmetry, as we have seen in Section \ref{sec:convert}. Unlike single field inflation, (broken) internal symmetries may also be expected when multiple light scalars are present.
It would be really interesting if one could identify signatures of the symmetry breaking pattern associated with the inflaton field space from correlation functions.

\end{itemize}

\

\vspace{0.5cm}
\paragraph{Acknowledgements} 

We would like to acknowledge many stimulating conversations with Sebastian Cespedes, Anne-Christine Davis, Carlos Duaso Pueyo,  Victor Gorbenko, Scott Melville, Enrico Pajer, Sebastien Renaux-Petel, Misao Sasaki, David Seery, and Xi Tong.  DGW thanks the Scuola Normale Superiore di Pisa for hospitality while parts of this work were being completed. GLP thanks the Universities of Leiden and Amsterdam for their hospitality and support.
DGW would like to thank his previous collaborators on multi-field inflation, especially Ed Copeland, Oksana Iarygina, Renata Kallosh, Andrei Linde, Gonzalo Palma and Yvette Welling, who helped to shape his thinking of this subject and inspired many aspects of this work.

DGW gratefully acknowledges the support from a Rubicon Postdoctoral Fellowship awarded by the Netherlands Organisation for Scientific Research (NWO). GLP is supported by a Rita-Levi Montalcini fellowship from the Italian Ministry of Universities and Research (MUR). 
AA acknowledges the Spanish Ministry MCIU/AEI/FEDER grants (PID2021-123703NB-C21) and the Basque Government grant (IT-1628-22).

\newpage

\appendix

\section{IR Behaviour in the Wavefunction of the Universe}
\label{app:wave}

In this appendix, we briefly review the wavefunction approach to cosmological correlators, and then we shall examine the IR divergences in the wavefunction coefficients for both contact and exchange interactions.  In particular, we provide more details for the discussion in Section \ref{sec:wave}, and show explicitly that for the four-point scalar seed function of massless exchange, the wavefunction coefficient is IR-finite, and the IR divergence of the correlation function comes from the disconneted part.

\subsection{Wavefunction and Correlators}

The wavefunction of the Universe is introduced in the Schr\"{o}dinger-picture approach to QFT (see Refs.~\cite{Anninos:2014lwa,Goon:2018fyu,Baumann:2020dch,Goodhew:2020hob} for more details).
We consider a set of bulk fields $\Sigma(\eta, \bf x)$ in dS spacetime, and then the wavefunctional $\Psi[\Sigma, \eta]$ is expected to contain all the information for spatial field configurations at time $\eta$. 
Within the perturbative regime of the theory, the wavefunctional at the late-time boundary $\eta_0$ is normally expressed as
\begin{equation}
\Psi[\Sigma, \eta_0] = \exp \[ \sum_{n = 2}\frac{1}{n!}\int \frac{{\rm d}^{3}k_1\cdot\cdot\cdot {\rm d}^{3}k_n}{(2\pi)^{3n}}\,\Sigma_{{\bf k}_1}\cdot\cdot\cdot \Sigma_{{\bf k}_n}\,(2\pi)^3  \delta({\bf k}_1+\cdots+{\bf k}_n)\, \psi_n({\bf k}_1,...,{\bf k}_n) \]\,,
\label{eq:wfcoeff}
\end{equation}
where $\psi_n$ are the {\it wavefunction coefficients} in Fourier space. 
Although $\psi_n$ is not physical observable, in many circumstances it is an  object that is more convenient for analysis than the cosmological correlation functions.  
Meanwhile, the equal-time correlators at $\eta_0$ can be derived through the standard quantum mechanics procedure 
\begin{equation} \label{bornrule}
\langle\Sigma({\bf x}_1)\cdots \Sigma({\bf x}_n)\rangle =\frac{ \displaystyle\int{\cal D} \Sigma\,~\Sigma({\bf x}_1)\cdots \Sigma({\bf x}_n) \left\lvert\Psi[\Sigma, \eta_0]\right\rvert^2}{\displaystyle\int{\cal D} \Sigma ~ \left\lvert\Psi[\Sigma,\eta_0]\right\rvert^2}\,.
\end{equation}
In perturbation theory, we can find explicit relations between the
wavefunction coefficients and the corresponding correlators at the end of inflation. 
For illustration, let's consider the $n$-point functions of the conformally coupled scalar $\vp$ with the presence of a general scalar field $\s_\Delta$ as possible intermediate state.
By expanding the exponential in~\eqref{eq:wfcoeff} and performing the Gaussian integrals in \eqref{bornrule}, we find the two-point functions are simply related by
\be
\langle \vp_{\bf k}\vp_{-{\bf k}}  \rangle' = \frac{1}{2\hskip 1pt{\rm Re}\hskip 1pt\psi_2^{\vp\vp}(k)} \,,~~~~~
\langle \s_\Delta({\bf k})\s_\Delta(-{\bf k}) \rangle' = \frac{1}{2\hskip 1pt{\rm Re}\hskip 1pt\psi_2^{\s\s}(k)} \,,
\ee
where $\psi_2$'s on the boundary are fully fixed by the conformal symmetry: $\psi_2^{\vp\vp} =c_\vp k$  and $\psi_2^{\s\s} = c_\Delta k^{2\Delta-3}$, with the normalization $c_\vp =(H\eta_0)^{-2}$ and $ c_\Delta = H^{-2}\eta_0^{2\Delta-6}$.
For the three-point functions, here let's consider the one  from cubic contact interactions\footnote{If there is a mixed propagator due to quadratic interaction, we would also expect contributions to the bispectrum from exchange diagrams. We leave the analysis of this case in Appendix \ref{app:exchange}.} with two $\vp$ fields and one $\s_\Delta$, and then we find the relation
\be \label{equ:3pt}
\langle \vp_{{\bf k}_1}\vp_{{\bf k}_2}  \s_{{\bf k}_3} \rangle' = -\frac{{\rm Re}\hskip 1pt\psi_3^{\vp\vp\s}({\bf k}_1,{\bf k}_2,{\bf k}_3)}{4   {\rm Re}\hskip 1pt\psi_2^{\vp\vp}(k_1) ~ {\rm Re}\hskip 1pt\psi_2^{\vp\vp}(k_2)~{\rm Re}\hskip 1pt\psi_2^{\s\s}(k_3)}\,.
\ee
The four-point function of $\vp$ may have contributions from both contact interactions and the exchange diagrams with an intermediate $\s_\Delta$ field, and the correlator is given by both $\psi_4$ and $\psi_3$ through
\be
\langle \vp_{{\bf k}_1}\vp_{{\bf k}_2}  \vp_{{\bf k}_3}\vp_{{\bf k}_4} \rangle' =- \frac{{\rm Re}\hskip 1pt\psi_4^{\vp\vp\vp\vp}({\bf k}_1,{\bf k}_2,{\bf k}_3,{\bf k}_4) }{8 \prod_{a=1}^4 {\rm Re}\hskip 1pt \psi_2^{\vp\vp}(k_a)} + \langle \vp_{{\bf k}_1}\vp_{{\bf k}_2}  \vp_{{\bf k}_3}\vp_{{\bf k}_4} \rangle'_{\rm d} \, , \label{equ:4pt}
\ee
where the 
{\it disconnected} part has contributions from a product of two $\psi_3^{\vp\vp\s}$
\begin{small}
\be \label{disconn}
\langle \vp_{{\bf k}_1}\vp_{{\bf k}_2}  \vp_{{\bf k}_3}\vp_{{\bf k}_4} \rangle'_{\rm d} =
\frac{1}{{8 \prod_{a=1}^4 {\rm Re} \hskip 1pt\psi_2^{\vp\vp}(k_a)}} \[ \frac{{\rm Re}\hskip 1pt\psi_3^{\vp\vp\s}({\bf k}_1,{\bf k}_2,{\bf s}) \hskip 1pt {\rm Re}\hskip 1pt\psi_3^{\s\vp\vp}(-{\bf s},{\bf k}_3,{\bf k}_4) }{{\rm Re}\hskip 1pt\psi^{\s\s}_2(s)}  + \text{$t$- and $u$-channels}\, \] \, .
\ee \end{small}
While the contact diagrams only contribute to the connected part of the four-point correlator, there are contributions to both parts from exchange processes. As we will show, the disconnected part becomes particularly important for the exchange of massless scalars.

The bulk computation of the wavefunction coefficients in perturbation theory is similar with the in-in formalism for correlators discussed in Section \ref{sec:recap}. 
We also need two types of propagators: the {bulk-to-boundary} propagator ${K}_\Psi 
(k,\eta)$ that connects bulk interaction vertex to the late-time boundary, and the bulk-to-bulk propagator ${G}_\Psi(k,\eta_1,\eta_2)$ which describes the propagation of fields between two bulk insertions. Here we use the lower index $\Psi$ to denote these are propagators in the wavefunction approach, which are distinguished from the in-in propagators $K_\pm$ and $G_{\pm\pm}$ introduced in  Section \ref{sec:recap}. 
These propagators also satisfy the following differential equations 
\be
\(\mathcal{O}_\eta -m^2/H^2 \) {K}_\Psi  (k,\eta) =0 ~,~~~~
\(\mathcal{O}_\eta -m^2/H^2 \) {G}_\Psi  (k,\eta,\eta') =-i H^2 \eta^2\eta'^2\delta(\eta-\eta') ~.
\ee
However, as we have specified a late-time boundary at $\eta_0$ for the wavefunction, these propagators subject to different boundary conditions
\begin{align}
\lim_{\eta\rightarrow\eta_0} {K}_\Psi  (k,\eta) =1~,~~&~~\lim_{\eta\rightarrow-\infty} {K}_\Psi  (k,\eta) =0\\
\lim_{\eta\rightarrow\eta_0} {G}_\Psi  (k,\eta,\eta') =0~,~~&~~\lim_{\eta\rightarrow-\infty}{G}_\Psi  (k,\eta,\eta') =0~.
\end{align}
Let's take a closer look at their explicit forms by considering a general bulk field $\s_\Delta$  with the conformal dimension $\Delta$. Its mode function is given by
\be \label{sigmak}
 \sigma_\Delta(k,\eta) = i \frac{H\sqrt{\pi}}{2} e^{i\pi/4}e^{i\pi\nu/2}
 (-\eta)^{3/2}H^{(1)}_{\nu}(-k\eta) ~ ~ ~ {\rm with} ~ \nu = 
\Delta-\frac{3}{2}~. 
\ee
For the bulk-to-boundary propagator,  ${K}_\Psi$ 
is similar with $K_+$ up to normalizations
\be \label{KPsis}
K^\Delta_\Psi (k,\eta) = \frac{ \sigma^*_\Delta(k,\eta) }{\sigma^*_\Delta(k,\eta_0)}~,
\ee
which goes to unity at late time.
Its expressions  for massless scalar and conformally coupled scalar are given in \eqref{KPsi}  in terms of simple functions. For the bulk-to-bulk propagator, here we have
\begin{align}
    {G}^\Delta_\Psi(k;\eta,\eta') \equiv ~ &  \sigma_\Delta(k,\eta)  \sigma^*_\Delta(k,\eta')  \theta(\eta-\eta') +  \sigma_\Delta^*(k,\eta)  \sigma_\Delta(k,\eta')  \theta(\eta'-\eta) \nn\\ 
    & - \frac{ \sigma_\Delta(k,\eta_0) }{\sigma^*_\Delta(k,\eta_0)} \sigma^*_\Delta(k,\eta)  \sigma^*_\Delta(k,\eta')   
    \,.
\end{align}
Comparing with the in-in propagator $G_{++}$, we notice that the nontrivial difference is given by the last term in ${G}_\Psi$. This term is added to the bulk-to-bulk propagator such that it vanishes in the late-time limit $\eta,\eta'\rightarrow\eta_0$.
One important consequence of this additional term is that, whatever the field mass, the ${G}_\Psi$ propagator always decays after the field fluctuations exit the horizon, which essentially  differs from the super-horizon behaviour of the massless $G_{\pm\pm}$ propagators. 
This observation plays an important role in identifying the IR behaviour of wavefunction coefficients from massless exchanges.

\paragraph{Contact three-point function}
With these wavefunction propagators, one can apply the Feynman rules for computing the wavefunction coeffients. 
As an explicit example, let's again look at the simple contact interaction $\vp^2\phi$, which is also relevant for the later discussion of  massless exchange diagrams.
The bulk computation with an explicit $\eta_0$  is presented in Section \ref{sec:wave}.
Here, instead of using an IR cutoff, we would like to apply another regularization scheme by simutaneously extending the spacetime and conformal dimensions in the following way
\be
d=3 \rightarrow d= 3+2\delta~,~~~~~~ \Delta \rightarrow \Delta + \delta~.
\ee
This type of dimensional regularization is known as the {\it half-integer scheme} \cite{Bzowski:2013sza}, as the indices of Hankel functions for $\Delta=2, 3$ scalars remain to be half-integers.
Thus the advantage of this scheme is that the mode functions of these fields can still be expressed as simple elementary functions.
Explicitly, the index of the Hankel function is given by
\be
\nu = \sqrt{\frac{d^2}{4}-\frac{m^2}{H^2}}=\Delta-\frac{d}{2}~,
\ee
and in ${\rm dS}_{d+1}$ spacetime the mode functions of these two fields become
\begin{align}
\Delta=3+\delta :~~~~ &\phi_k  = \frac{H}{\sqrt{2k^3}}(-\eta)^\delta(1+ik\eta)e^{-ik\eta}
\\
\Delta=2+\delta :~~~~ &\varphi_k = \frac{iH}{\sqrt{2k}}(-\eta)^\delta \eta e^{-ik\eta}~.
\end{align}
Then using the bulk-to-boundary propagator \eqref{KPsis}, we find the wavefunction coefficient 
\bea
\psi_3^{\vp\vp\phi} &=& -2i\int_{-\infty}^{0} d\eta a(\eta)^{d+1} K_\Psi^\vp (k_1,\eta) K_\Psi^\vp (k_2,\eta)   {K}_\Psi(k_3,\eta)\nn\\
&=&  \frac{-2(-i)^{\delta}}{H^{4+2\delta}\eta_0^{2+3\delta}} k_T^{-\delta}\Gamma(\delta)\( k_{12}+\frac{\delta~ k_T}{1-\delta}\) \xrightarrow{\delta\rightarrow 0 } \frac{-2}{H^4	\eta_0^2}\left[ k_T+ k_{12} \(\frac{1}{\delta} -\log k_T-\gamma_E\)\right] ~,
\eea
where in the last step we have absorbed the overall $H$ and $\eta_0$ prefactors in the coupling constant and set them to $1$ for simplicity. We see that the wavefunction coefficient is singular when we take $\delta\rightarrow0$. 
The divergence can be renormalized by adding a counterterm 
with a renormalization scale $\mu$ \cite{Bzowski:2015pba, Bzowski:2018fql}. As a result, we find
\be \label{psi3vvp}
\psi_3^{\vp\vp\phi} = -\frac{2}{H^4	\eta_0^2}\[k_3 - k_{12} \( \log\(\frac{k_T}{\mu}\)+\gamma_E-1 \)\] ,
\ee
which matches the result with an explicit late-time cutoff by choosing $\mu=-1/\eta_0$. 
We refer the reader to Ref. \cite{Bzowski:2015pba, Bzowski:2018fql} for  detailed analysis about the renormalization of the IR-divergent correlators in CFT.
By using relation \eqref{equ:3pt}, we find the corresponding 
correlator at the end of inflation
\be
\langle \vp_{{\bf k}_1}\vp_{{\bf k}_2}  \phi_{{\bf k}_3} \rangle' = -\frac{H^6\eta_0^4}{ 4k_1k_2k_3^3 }\psi_3^{\vp\vp\phi}\,,
\ee
which reproduces the in-in result in \eqref{vvp}.
For this particular contact interaction, the IR divergence of wavefunction coefficient is basically the same with the analysis on correlators in Section \ref{sec:contact}. But next, we will show that the distinction between correlators and wavefunction coefficients becomes nontrivial when we consider massless exchange diagrams.

\subsection{IR Behaviour in Massless Exchange}
\label{app:exchange}

Our main focus here is to examine whether the wavefunction coefficients from massless exchange are IR-divergent or not.
By doing so, we will demonstrate the origin of the IR divergence in the wavefunction approach, and compare results from correlator analysis.

\paragraph{Four-point function} Let us begin with the $\psi_4$ of four conformally coupled scalars with the exchange of a massless field $\s$, for which the corresponding correlator has been discussed in Section \ref{sec:4pt-ex}. Using the wavefunction propagators in \eqref{KPsi} and \eqref{GPsi}, the bulk integral for the $s$-channel contribution is given by 
\be
\psi_4 = 4\int_{-\infty}^{\eta_0} d\eta \int_{-\infty}^{\eta_0}  d\eta' a(\eta)^{4} a(\eta')^{4} K_\Psi^\vp (k_1,\eta) K_\Psi^\vp (k_2,\eta)  G_\Psi^\s(s,\eta,\eta')  K_\Psi^\vp (k_3,\eta') K_\Psi^\vp (k_4,\eta') ~.
\ee
It is more convenient to discuss its dimensionless version
 $\hat{\psi}_4= -\eta_0^4 H^6 s\psi_4/4$, which can be expressed as 
\be
\hat{\psi}_4 = -\frac{s}{H^2}\int_{-\infty}^{\eta_0} \frac{d\eta}{\eta^{2}} \int_{-\infty}^{\eta_0} \frac{d\eta'}{{\eta'}^{2}} e^{ik_{12}\eta +i k_{34}\eta'}  {G}_\Psi^{\s}(s\eta, s\eta')~.
\ee
Here we see that $\hat{\psi}_4$ has a similar form with the four-point scalar seed $\hat{F}$ in \eqref{Fhat}, while the major difference is given by the bulk-to-bulk propagator.
Likewise, we can act the $\eta_0 \partial_{\eta_0}$ operator on $\hat{\psi}_4$ to trace its IR behaviour and find
\be
\lim_{\eta_0\rightarrow0}\eta_0 \partial_{\eta_0} \hat{\psi}_4  = -i\frac{2}{3}  s \eta_0 \rightarrow 0~,
\ee
which shows that $\hat{\psi}_4$ is IR-finite, very different from the behaviour of $\hat{F}$ in \eqref{deta0F}.
This is due to the fact that in the wavefunction approach ${G}_\Psi^{\s}$ decays on super-horizon scales when $\eta,\eta'>-1/s$, and thus that part of the integration only leads to finite result.  
For the correlator seed function $\hat{F}$, the massless bulk-to-bulk propagator $G$ becomes constant outside of the horizon, and as a result of the accumulative effect of the massless field,  the bulk integral of $\hat{F}$ diverges towards the late time.
Therefore, the wavefunction coefficient $\hat{\psi}_4$ is independent of $\eta_0$ at the late-time boundary, and we can apply the standard treatment of the boundary perspective. First, in terms of $u$ and $v$, the conformal symmetry leads to the differential equations 
\be \label{eq:psi4}
\(\Delta_u  -2\) \hat{\psi}_4 (u,v) = \frac{uv}{u+v}~,~~~~
\(\Delta_v  -2\)\hat{\psi}_4 (u,v) = \frac{uv}{u+v}~.
\ee 
Then the analytical solution can be solved by imposing the absence of the folded singularity and the right normalization at the total or partial energy singularity. Explicitly, we find $\hat{\psi}_4=\hat{F}_{\rm fin}$, where $\hat{F}_{\rm fin}$ is the IR-finite part of the scalar seed function in \eqref{eq:Ffin}.

For the exchange process, there are also contributions which are proportional to the product of two three-point functions. 
Thus to compute the four-point correlator, we also need to include the disconnected part in \eqref{disconn}. Again, let's consider the $s$-channel contribution, then the first term in the bracket of \eqref{disconn} becomes
\be
 \frac{H^2}{s^3} \psi^{\vp\vp\s}_3(k_1,k_2,s)\psi_3^{\vp\vp\s}(k_3,k_4,s) = \frac{4}{\eta_0^4H^6 s} \hat{F}_{\rm div} (u,v,\eta_0)~,
\ee
where $\hat{F}_{\rm div}$ is the IR-divergent part of the four-point scalar seed in \eqref{eq:Fdiv}. This confirms that the IR-divergence in the four-point scalar seed of massless exchange comes from the disconnected contribution.
As a result, from \eqref{equ:4pt} the final four-point correlator is given by
\be
\langle \vp_{{\bf k}_1}\vp_{{\bf k}_2}  \vp_{{\bf k}_3}\vp_{{\bf k}_4} \rangle' = 
\frac{H^2\eta_0^4}{2k_1k_2k_3k_4 s} \[    \hat{F}_{\rm fin} (u,v)+ \hat{F}_{\rm div} (u,v,\eta_0) \] + \text{$t$- and $u$-channels} ,
\ee
which precisely matches the result in Section \ref{sec:4pt-ex}. This computation provides supplementary details for our sketchy analysis in Section \ref{sec:wave}.

\paragraph{Three-Point Function} Now we extend our analysis of the wavefunction approach to the boost-breaking scenarios.
We shall focus the three-point exchange diagrams with mixed propagators as we have done in Section \ref{sec:3pt-ex} for correlators.
Our starting point is to consider the quadratic interaction between two massless field $\dot\phi\s$ and introduce the sound speed $c_s$ for $\phi$ field.\footnote{As in Section \ref{sec:3pt-ex}, $c_s$ can be seen as the sound speed ratio between two scalars, and thus is allowed be larger than $1$.} Due to the presence of this linear mixing, we first notice that there is a nonzero  two-point wavefunction coefficient given by
\bea \label{psi2phis}
\psi_2^{\phi\s} &=& -i\int_{-\infty}^{\eta_0} d\eta a^3(\eta) \partial_\eta K^\phi_\Psi(c_s k,\eta) K^\s_\Psi(k,\eta) \nn\\ &=& -\frac{c_s^3 k^3}{H^3}
\[ \frac{i}{c_s k\eta_0} -\frac{i\pi}{2} +  \gamma_E -1-\frac{1}{c_s}+ \log\(-(1+c_s)k\eta_0\) \]~,
\eea
which is singular when we take $\eta_0$ to be $0$. As only the real part of the wavefunction coefficient matters for physical observables, we shall not worry about the divergence in the $\eta_0^{-1}$ term.
Next, the quadratic interaction also leads to a mixed propagator for wavefunction
\be 
\cG_\Psi (k,\eta,\eta_0) = i\int^{\eta_0}_{-\infty} d\eta' a(\eta')^3 \partial_{\eta'} K_{\Psi}(c_s k,\eta')G_\Psi(k,\eta,\eta')~,
\ee
which satisfies the similar differential equation for the in-in mixed propagator $\hat{\cG}_+$
\be \label{eq:curlyKPsi}
\mathcal{O}_\eta \cG_\Psi = -\frac{1}{H} c_s^2k^2\eta^2e^{ic_s k \eta}
\ee
but subjects to a different boundary condition at the late time.
It is convenient to take a look at its analytical expression 
\be
\cG_\Psi (k,\eta,\eta_0)=  \frac{c_s}{2H} e^{-i k \eta} (1+ik \eta)~{\rm Ei}\big[i(1+c_s)k\eta\big] - \frac{c_s}{2H}  e^{i k \eta} (1-ik\eta )~\tilde{\mathcal{D}} - \frac{1}{H}e^{ic_sk \eta}   
\ee
with
\begin{align}
\tilde{\mathcal{D}}= 
        \begin{dcases}
             \gamma_E-2 - \frac{i\pi}{2}+\log \({-2k\eta}\) ~,\quad & c_s = 1\\
        {\rm Ei}\big(i(-1+c_s)k\eta\big)+ \log\(\frac{1+c_s}{1-c_s}\)~,\quad & c_s \neq 1
        \end{dcases} \qquad .
\end{align}
Thus this mixed propagator is IR-finite with no dependence on the late-time cutoff $\eta_0$, and can be seen as a function of the combination $k\eta$. Now we check its behaviour in the late-time limit
\be
\lim_{\eta\rightarrow 0}\mathcal{K}_\Psi(k \eta) = \frac{1}{2}k^2\eta^2 + \mathcal{O}(\eta^3)~,
\ee
which differs from the logartimically divergent one of $\hat{\cG}_+$  in \eqref{Kx0}.
This fall-off  is again a consequence of the decaying $G_\Psi^\s$ on super-horizon scales. 
It is also interesting to notice that this mixed propagator satisfies $\partial_k \cG_\Psi(k\eta)|_{k=0}=0$, thus the manifestly local test \cite{Jazayeri:2021fvk} can also be applied in the wavefunction coefficients with $\cG_\Psi(k\eta)$. This is not valid for the mixed propagator of correlators $\hat{\cG}_\pm$, which again shows the distinction between wavefunction coefficients and correlation functions. 

With the new mixed propagator, the wavefunction coefficient of the three-point exchange diagram with two conformally coupled scalar is simply given by
\be
\psi_3 = -2i \int^{\eta_0}_{-\infty} d\eta a(\eta)^4 K_\Psi^\vp (c_sk_1,\eta) K_\Psi^\vp (c_sk_2,\eta) \mathcal{K}_\Psi(k_3\eta) 
= \frac{2k_3c_s^2}{H^5\eta_0^2} \hat{\psi}_3
\ee
where for the later convenience we have introduced the dimensionless version
\be
\hat{\psi}_3 = -\frac{iH}{c_s^2 k_3}\int_{-\infty}^{\eta_0}\frac{d\eta}{\eta^2} e^{ic_sk_{12}\eta} \mathcal{K}_\Psi(k_3\eta) ~,
\ee
which can be written as a function of the momentum ratio $w\equiv k_3/c_sk_{12}$.
This integral is IR-finite, which can be simply seen by acting the $\eta_0\partial_{\eta_0}$ operator. However, it remains quite difficult to do the integration directly.
Instead, using the equation \eqref{eq:curlyKPsi}, we are able to find the differential equation for $\hat{\psi}_3$ in terms of $w$
\be
(\Delta_w -2) \hat{\psi}_3 = \frac{w}{1+c_s w}~,
\ee
which is the same with the equation for the IR-finite part of the three-point scalar seed $\hat{\mathcal{I}}_{\rm fin}$. Therefore, we identify the solution of $\hat{\psi}_3(w) = \hat{\mathcal{I}}_{\rm fin}(w)$, with the explicit expression given in \eqref{eq:Ifin}.

Next, to compute the corresponding correlator, we notice that it has the following relation with the wavefunction coefficients
\be \label{equ:3pt-ex}
\langle \vp_{{\bf k}_1}\vp_{{\bf k}_2}  \phi_{{\bf k}_3} \rangle' = -\frac{1}{4 {\rm Re}\hskip 1pt\psi_2^{\vp\vp}(k_1) ~ {\rm Re}\hskip 1pt\psi_2^{\vp\vp}(k_2)~{\rm Re}\hskip 1pt\psi_2^{\phi\phi}(k_3)}
\[
{\rm Re}\hskip 1pt\psi_3 - \frac{{\rm Re}\psi_3^{\vp\vp\s}{\rm Re}\psi_2^{\phi\s}}{{\rm Re}\psi_2^{\s\s}}\]\,.
\ee
In the disconnected part, $\psi_2^{\phi\s}$ is given by \eqref{psi2phis} and $\psi_3^{\vp\vp\s}$ is simply \eqref{psi3vvp} with $k_{12}\rightarrow c_s k_{12}$. In the end, the single-exchange three-point correlator becomes
\be \label{equ:3pt-ex}
\langle \vp_{{\bf k}_1}\vp_{{\bf k}_2}  \phi_{{\bf k}_3} \rangle' = -\frac{H\eta_0^2}{2k_1k_2k_3^2c_s^3}
\[
\hat{\psi}_3 +\hat{\psi}_3^{\rm div} \]\,,
\ee
where the IR-divergent term has the same form with $\mathcal{\hat I}_{\rm div} $ in \eqref{eq:Idiv}
\be
\hat{\psi}_3^{\rm div} = \frac{1}{w} \big[c_s(\gamma_E-1)-1+c_s\log\(-(1+c_s) k_3\eta_0\)\big]
\big[ \gamma_E-1-w + \log(-(c_sk_{12}+k_3)\eta_0) \big]~.
\ee
Again, we identify that the late-time singular behaviour of the correlator comes from the disconnected part.
The above result from the wavefunction approach provides a consistency check for our computation in Section \ref{sec:3pt-ex}.

\clearpage
\phantomsection
\addcontentsline{toc}{section}{References}
\bibliographystyle{utphys}
\bibliography{refs}

\providecommand{\href}[2]{#2}\begingroup\raggedright\begin{thebibliography}{100}

\bibitem{Meerburg:2019qqi}
P.~D. Meerburg {\em et~al.}, ``{Primordial Non-Gaussianity},''
  \href{http://arxiv.org/abs/1903.04409}{{\ttfamily arXiv:1903.04409
  [astro-ph.CO]}}.

\bibitem{Achucarro:2022qrl}
A.~Ach\'ucarro {\em et~al.}, ``{Inflation: Theory and Observations},''
  \href{http://arxiv.org/abs/2203.08128}{{\ttfamily arXiv:2203.08128
  [astro-ph.CO]}}.

\bibitem{Lyth:2005fi}
D.~H. Lyth and Y.~Rodriguez, ``{The Inflationary prediction for primordial
  non-Gaussianity},''
  \href{http://dx.doi.org/10.1103/PhysRevLett.95.121302}{{\em Phys. Rev. Lett.}
  {\bfseries 95} (2005) 121302},
  \href{http://arxiv.org/abs/astro-ph/0504045}{{\ttfamily
  arXiv:astro-ph/0504045}}.

\bibitem{Seery:2005gb}
D.~Seery and J.~E. Lidsey, ``{Primordial non-Gaussianities from multiple-field
  inflation},'' \href{http://dx.doi.org/10.1088/1475-7516/2005/09/011}{{\em
  JCAP} {\bfseries 09} (2005) 011},
  \href{http://arxiv.org/abs/astro-ph/0506056}{{\ttfamily
  arXiv:astro-ph/0506056}}.

\bibitem{Bassett:2005xm}
B.~A. Bassett, S.~Tsujikawa, and D.~Wands, ``{Inflation dynamics and
  reheating},'' \href{http://dx.doi.org/10.1103/RevModPhys.78.537}{{\em Rev.
  Mod. Phys.} {\bfseries 78} (2006) 537--589},
  \href{http://arxiv.org/abs/astro-ph/0507632}{{\ttfamily
  arXiv:astro-ph/0507632}}.

\bibitem{Rigopoulos:2005ae}
G.~I. Rigopoulos, E.~P.~S. Shellard, and B.~J.~W. van Tent, ``{Large
  non-Gaussianity in multiple-field inflation},''
  \href{http://dx.doi.org/10.1103/PhysRevD.73.083522}{{\em Phys. Rev.}
  {\bfseries D73} (2006) 083522},
\href{http://arxiv.org/abs/astro-ph/0506704}{{\ttfamily arXiv:astro-ph/0506704
  [astro-ph]}}.

\bibitem{Vernizzi:2006ve}
F.~Vernizzi and D.~Wands, ``{Non-gaussianities in two-field inflation},''
  \href{http://dx.doi.org/10.1088/1475-7516/2006/05/019}{{\em JCAP} {\bfseries
  05} (2006) 019}, \href{http://arxiv.org/abs/astro-ph/0603799}{{\ttfamily
  arXiv:astro-ph/0603799}}.

\bibitem{Seery:2006js}
D.~Seery and J.~E. Lidsey, ``{Non-Gaussianity from the inflationary
  trispectrum},'' \href{http://dx.doi.org/10.1088/1475-7516/2007/01/008}{{\em
  JCAP} {\bfseries 01} (2007) 008},
  \href{http://arxiv.org/abs/astro-ph/0611034}{{\ttfamily
  arXiv:astro-ph/0611034}}.

\bibitem{Byrnes:2006vq}
C.~T. Byrnes, M.~Sasaki, and D.~Wands, ``{The primordial trispectrum from
  inflation},'' \href{http://dx.doi.org/10.1103/PhysRevD.74.123519}{{\em Phys.
  Rev. D} {\bfseries 74} (2006) 123519},
  \href{http://arxiv.org/abs/astro-ph/0611075}{{\ttfamily
  arXiv:astro-ph/0611075}}.

\bibitem{Byrnes:2009qy}
C.~T. Byrnes and G.~Tasinato, ``{Non-Gaussianity beyond slow roll in
  multi-field inflation},''
  \href{http://dx.doi.org/10.1088/1475-7516/2009/08/016}{{\em JCAP} {\bfseries
  08} (2009) 016}, \href{http://arxiv.org/abs/0906.0767}{{\ttfamily
  arXiv:0906.0767 [astro-ph.CO]}}.

\bibitem{Byrnes:2010em}
C.~T. Byrnes and K.-Y. Choi, ``{Review of local non-Gaussianity from
  multi-field inflation},'' \href{http://dx.doi.org/10.1155/2010/724525}{{\em
  Adv. Astron.} {\bfseries 2010} (2010) 724525},
  \href{http://arxiv.org/abs/1002.3110}{{\ttfamily arXiv:1002.3110
  [astro-ph.CO]}}.

\bibitem{Wands:2010af}
D.~Wands, ``{Local non-Gaussianity from inflation},''
  \href{http://dx.doi.org/10.1088/0264-9381/27/12/124002}{{\em Class. Quant.
  Grav.} {\bfseries 27} (2010) 124002},
  \href{http://arxiv.org/abs/1004.0818}{{\ttfamily arXiv:1004.0818
  [astro-ph.CO]}}.

\bibitem{Senatore:2010wk}
L.~Senatore and M.~Zaldarriaga, ``{The Effective Field Theory of Multifield
  Inflation},'' \href{http://dx.doi.org/10.1007/JHEP04(2012)024}{{\em JHEP}
  {\bfseries 04} (2012) 024}, \href{http://arxiv.org/abs/1009.2093}{{\ttfamily
  arXiv:1009.2093 [hep-th]}}.

\bibitem{Peterson:2010mv}
C.~M. Peterson and M.~Tegmark, ``{Non-Gaussianity in Two-Field Inflation},''
  \href{http://dx.doi.org/10.1103/PhysRevD.84.023520}{{\em Phys. Rev. D}
  {\bfseries 84} (2011) 023520},
  \href{http://arxiv.org/abs/1011.6675}{{\ttfamily arXiv:1011.6675
  [astro-ph.CO]}}.

\bibitem{Gong:2016qmq}
J.-O. Gong, ``{Multi-field inflation and cosmological perturbations},''
  \href{http://dx.doi.org/10.1142/S021827181740003X}{{\em Int. J. Mod. Phys.}
  {\bfseries D26} no.~01, (2016) 1740003},
\href{http://arxiv.org/abs/1606.06971}{{\ttfamily arXiv:1606.06971 [gr-qc]}}.

\bibitem{Komatsu:2001rj}
E.~Komatsu and D.~N. Spergel, ``{Acoustic signatures in the primary microwave
  background bispectrum},''
  \href{http://dx.doi.org/10.1103/PhysRevD.63.063002}{{\em Phys. Rev. D}
  {\bfseries 63} (2001) 063002},
  \href{http://arxiv.org/abs/astro-ph/0005036}{{\ttfamily
  arXiv:astro-ph/0005036}}.

\bibitem{Namjoo:2012aa}
M.~H. Namjoo, H.~Firouzjahi, and M.~Sasaki, ``{Violation of non-Gaussianity
  consistency relation in a single field inflationary model},''
  \href{http://dx.doi.org/10.1209/0295-5075/101/39001}{{\em Europhys. Lett.}
  {\bfseries 101} (2013) 39001},
\href{http://arxiv.org/abs/1210.3692}{{\ttfamily arXiv:1210.3692
  [astro-ph.CO]}}.

\bibitem{Chen:2013eea}
X.~Chen, H.~Firouzjahi, E.~Komatsu, M.~H. Namjoo, and M.~Sasaki, ``{In-in and
  $\delta N$ calculations of the bispectrum from non-attractor single-field
  inflation},'' \href{http://dx.doi.org/10.1088/1475-7516/2013/12/039}{{\em
  JCAP} {\bfseries 1312} (2013) 039},
\href{http://arxiv.org/abs/1308.5341}{{\ttfamily arXiv:1308.5341
  [astro-ph.CO]}}.

\bibitem{Chen:2013aj}
X.~Chen, H.~Firouzjahi, M.~H. Namjoo, and M.~Sasaki, ``{A Single Field
  Inflation Model with Large Local Non-Gaussianity},''
  \href{http://dx.doi.org/10.1209/0295-5075/102/59001}{{\em Europhys. Lett.}
  {\bfseries 102} (2013) 59001},
\href{http://arxiv.org/abs/1301.5699}{{\ttfamily arXiv:1301.5699 [hep-th]}}.

\bibitem{Cai:2017bxr}
Y.-F. Cai, X.~Chen, M.~H. Namjoo, M.~Sasaki, D.-G. Wang, and Z.~Wang,
  ``{Revisiting non-Gaussianity from non-attractor inflation models},''
  \href{http://dx.doi.org/10.1088/1475-7516/2018/05/012}{{\em JCAP} {\bfseries
  1805} no.~05, (2018) 012},
\href{http://arxiv.org/abs/1712.09998}{{\ttfamily arXiv:1712.09998
  [astro-ph.CO]}}.

\bibitem{Akrami:2019izv}
{\bfseries Planck} Collaboration, Y.~Akrami {\em et~al.}, ``{Planck 2018
  results. IX. Constraints on primordial non-Gaussianity},''
\href{http://arxiv.org/abs/1905.05697}{{\ttfamily arXiv:1905.05697
  [astro-ph.CO]}}.

\bibitem{Arkani-Hamed:2018kmz}
N.~Arkani-Hamed, D.~Baumann, H.~Lee, and G.~L. Pimentel, ``{The Cosmological
  Bootstrap: Inflationary Correlators from Symmetries and Singularities},''
  \href{http://dx.doi.org/10.1007/JHEP04(2020)105}{{\em JHEP} {\bfseries 04}
  (2020) 105}, \href{http://arxiv.org/abs/1811.00024}{{\ttfamily
  arXiv:1811.00024 [hep-th]}}.

\bibitem{Baumann:2019oyu}
D.~Baumann, C.~Duaso~Pueyo, A.~Joyce, H.~Lee, and G.~L. Pimentel, ``{The
  cosmological bootstrap: weight-shifting operators and scalar seeds},''
  \href{http://dx.doi.org/10.1007/JHEP12(2020)204}{{\em JHEP} {\bfseries 12}
  (2020) 204}, \href{http://arxiv.org/abs/1910.14051}{{\ttfamily
  arXiv:1910.14051 [hep-th]}}.

\bibitem{Baumann:2020dch}
D.~Baumann, C.~Duaso~Pueyo, A.~Joyce, H.~Lee, and G.~L. Pimentel, ``{The
  Cosmological Bootstrap: Spinning Correlators from Symmetries and
  Factorization},'' \href{http://dx.doi.org/10.21468/SciPostPhys.11.3.071}{{\em
  SciPost Phys.} {\bfseries 11} (2021) 071},
  \href{http://arxiv.org/abs/2005.04234}{{\ttfamily arXiv:2005.04234
  [hep-th]}}.

\bibitem{Arkani-Hamed:2017fdk}
N.~Arkani-Hamed, P.~Benincasa, and A.~Postnikov, ``{Cosmological Polytopes and
  the Wavefunction of the Universe},''
  \href{http://arxiv.org/abs/1709.02813}{{\ttfamily arXiv:1709.02813
  [hep-th]}}.

\bibitem{Arkani-Hamed:2018bjr}
N.~Arkani-Hamed and P.~Benincasa, ``{On the Emergence of Lorentz Invariance and
  Unitarity from the Scattering Facet of Cosmological Polytopes},''
  \href{http://arxiv.org/abs/1811.01125}{{\ttfamily arXiv:1811.01125
  [hep-th]}}.

\bibitem{Benincasa:2018ssx}
P.~Benincasa, ``{From the flat-space S-matrix to the Wavefunction of the
  Universe},'' \href{http://arxiv.org/abs/1811.02515}{{\ttfamily
  arXiv:1811.02515 [hep-th]}}.

\bibitem{Sleight:2019mgd}
C.~Sleight, ``{A Mellin Space Approach to Cosmological Correlators},''
  \href{http://dx.doi.org/10.1007/JHEP01(2020)090}{{\em JHEP} {\bfseries 01}
  (2020) 090}, \href{http://arxiv.org/abs/1906.12302}{{\ttfamily
  arXiv:1906.12302 [hep-th]}}.

\bibitem{Sleight:2019hfp}
C.~Sleight and M.~Taronna, ``{Bootstrapping Inflationary Correlators in Mellin
  Space},'' \href{http://dx.doi.org/10.1007/JHEP02(2020)098}{{\em JHEP}
  {\bfseries 02} (2020) 098}, \href{http://arxiv.org/abs/1907.01143}{{\ttfamily
  arXiv:1907.01143 [hep-th]}}.

\bibitem{Sleight:2020obc}
C.~Sleight and M.~Taronna, ``{From AdS to dS Exchanges: Spectral
  Representation, Mellin Amplitudes and Crossing},''
  \href{http://arxiv.org/abs/2007.09993}{{\ttfamily arXiv:2007.09993
  [hep-th]}}.

\bibitem{Pajer:2020wnj}
E.~Pajer, D.~Stefanyszyn, and J.~Supel{}, ``{The Boostless Bootstrap:
  Amplitudes without Lorentz boosts},''
  \href{http://dx.doi.org/10.1007/JHEP12(2020)198}{{\em JHEP} {\bfseries 12}
  (2020) 198}, \href{http://arxiv.org/abs/2007.00027}{{\ttfamily
  arXiv:2007.00027 [hep-th]}}.

\bibitem{Pajer:2020wxk}
E.~Pajer, ``{Building a Boostless Bootstrap for the Bispectrum},''
  \href{http://dx.doi.org/10.1088/1475-7516/2021/01/023}{{\em JCAP} {\bfseries
  01} (2021) 023}, \href{http://arxiv.org/abs/2010.12818}{{\ttfamily
  arXiv:2010.12818 [hep-th]}}.

\bibitem{Jazayeri:2021fvk}
S.~Jazayeri, E.~Pajer, and D.~Stefanyszyn, ``{From locality and unitarity to
  cosmological correlators},''
  \href{http://dx.doi.org/10.1007/JHEP10(2021)065}{{\em JHEP} {\bfseries 10}
  (2021) 065}, \href{http://arxiv.org/abs/2103.08649}{{\ttfamily
  arXiv:2103.08649 [hep-th]}}.

\bibitem{Bonifacio:2021azc}
J.~Bonifacio, E.~Pajer, and D.-G. Wang, ``{From amplitudes to contact
  cosmological correlators},''
  \href{http://dx.doi.org/10.1007/JHEP10(2021)001}{{\em JHEP} {\bfseries 10}
  (2021) 001}, \href{http://arxiv.org/abs/2106.15468}{{\ttfamily
  arXiv:2106.15468 [hep-th]}}.

\bibitem{Cabass:2021fnw}
G.~Cabass, E.~Pajer, D.~Stefanyszyn, and J.~Supel{}, ``{Bootstrapping Large
  Graviton non-Gaussianities},''
  \href{http://arxiv.org/abs/2109.10189}{{\ttfamily arXiv:2109.10189
  [hep-th]}}.

\bibitem{Hillman:2021bnk}
A.~Hillman and E.~Pajer, ``{A Differential Representation of Cosmological
  Wavefunctions},'' \href{http://arxiv.org/abs/2112.01619}{{\ttfamily
  arXiv:2112.01619 [hep-th]}}.

\bibitem{Pimentel:2022fsc}
G.~L. Pimentel and D.-G. Wang, ``{Boostless cosmological collider bootstrap},''
  \href{http://dx.doi.org/10.1007/JHEP10(2022)177}{{\em JHEP} {\bfseries 10}
  (2022) 177}, \href{http://arxiv.org/abs/2205.00013}{{\ttfamily
  arXiv:2205.00013 [hep-th]}}.

\bibitem{Jazayeri:2022kjy}
S.~Jazayeri and S.~Renaux-Petel, ``{Cosmological Bootstrap in Slow Motion},''
  \href{http://arxiv.org/abs/2205.10340}{{\ttfamily arXiv:2205.10340
  [hep-th]}}.

\bibitem{Cabass:2022jda}
G.~Cabass, D.~Stefanyszyn, J.~Supel, and A.~Thavanesan, ``{On graviton
  non-Gaussianities in the Effective Field Theory of Inflation},''
  \href{http://dx.doi.org/10.1007/JHEP10(2022)154}{{\em JHEP} {\bfseries 10}
  (2022) 154}, \href{http://arxiv.org/abs/2209.00677}{{\ttfamily
  arXiv:2209.00677 [hep-th]}}.

\bibitem{Goodhew:2020hob}
H.~Goodhew, S.~Jazayeri, and E.~Pajer, ``{The Cosmological Optical Theorem},''
  \href{http://dx.doi.org/10.1088/1475-7516/2021/04/021}{{\em JCAP} {\bfseries
  04} (2021) 021}, \href{http://arxiv.org/abs/2009.02898}{{\ttfamily
  arXiv:2009.02898 [hep-th]}}.

\bibitem{Cespedes:2020xqq}
S.~C\'espedes, A.-C. Davis, and S.~Melville, ``{On the time evolution of
  cosmological correlators},''
  \href{http://dx.doi.org/10.1007/JHEP02(2021)012}{{\em JHEP} {\bfseries 02}
  (2021) 012}, \href{http://arxiv.org/abs/2009.07874}{{\ttfamily
  arXiv:2009.07874 [hep-th]}}.

\bibitem{Melville:2021lst}
S.~Melville and E.~Pajer, ``{Cosmological Cutting Rules},''
  \href{http://dx.doi.org/10.1007/JHEP05(2021)249}{{\em JHEP} {\bfseries 05}
  (2021) 249}, \href{http://arxiv.org/abs/2103.09832}{{\ttfamily
  arXiv:2103.09832 [hep-th]}}.

\bibitem{Goodhew:2021oqg}
H.~Goodhew, S.~Jazayeri, M.~H. Gordon~Lee, and E.~Pajer, ``{Cutting
  cosmological correlators},''
  \href{http://dx.doi.org/10.1088/1475-7516/2021/08/003}{{\em JCAP} {\bfseries
  08} (2021) 003}, \href{http://arxiv.org/abs/2104.06587}{{\ttfamily
  arXiv:2104.06587 [hep-th]}}.

\bibitem{Baumann:2021fxj}
D.~Baumann, W.-M. Chen, C.~Duaso~Pueyo, A.~Joyce, H.~Lee, and G.~L. Pimentel,
  ``{Linking the Singularities of Cosmological Correlators},''
  \href{http://arxiv.org/abs/2106.05294}{{\ttfamily arXiv:2106.05294
  [hep-th]}}.

\bibitem{Meltzer:2021zin}
D.~Meltzer, ``{The inflationary wavefunction from analyticity and
  factorization},'' \href{http://dx.doi.org/10.1088/1475-7516/2021/12/018}{{\em
  JCAP} {\bfseries 12} no.~12, (2021) 018},
  \href{http://arxiv.org/abs/2107.10266}{{\ttfamily arXiv:2107.10266
  [hep-th]}}.

\bibitem{Hogervorst:2021uvp}
M.~Hogervorst, J.~a. Penedones, and K.~S. Vaziri, ``{Towards the
  non-perturbative cosmological bootstrap},''
  \href{http://arxiv.org/abs/2107.13871}{{\ttfamily arXiv:2107.13871
  [hep-th]}}.

\bibitem{DiPietro:2021sjt}
L.~Di~Pietro, V.~Gorbenko, and S.~Komatsu, ``{Analyticity and unitarity for
  cosmological correlators},''
  \href{http://dx.doi.org/10.1007/JHEP03(2022)023}{{\em JHEP} {\bfseries 03}
  (2022) 023}, \href{http://arxiv.org/abs/2108.01695}{{\ttfamily
  arXiv:2108.01695 [hep-th]}}.

\bibitem{Baumann:2022jpr}
D.~Baumann, D.~Green, A.~Joyce, E.~Pajer, G.~L. Pimentel, C.~Sleight, and
  M.~Taronna, ``{Snowmass White Paper: The Cosmological Bootstrap},'' in {\em
  {2022 Snowmass Summer Study}}.
\newblock 3, 2022.
\newblock \href{http://arxiv.org/abs/2203.08121}{{\ttfamily arXiv:2203.08121
  [hep-th]}}.

\bibitem{Goodhew:2022ayb}
H.~Goodhew, ``{Rational Wavefunctions in de Sitter Spacetime},''
  \href{http://arxiv.org/abs/2210.09977}{{\ttfamily arXiv:2210.09977
  [hep-th]}}.

\bibitem{Qin:2022fbv}
Z.~Qin and Z.-Z. Xianyu, ``{Helical Inflation Correlators: Partial
  Mellin-Barnes and Bootstrap Equations},''
  \href{http://arxiv.org/abs/2208.13790}{{\ttfamily arXiv:2208.13790
  [hep-th]}}.

\bibitem{Bonifacio:2022vwa}
J.~Bonifacio, H.~Goodhew, A.~Joyce, E.~Pajer, and D.~Stefanyszyn, ``{The
  graviton four-point function in de Sitter space},''
  \href{http://arxiv.org/abs/2212.07370}{{\ttfamily arXiv:2212.07370
  [hep-th]}}.

\bibitem{Salcedo:2022aal}
S.~A. Salcedo, M.~H.~G. Lee, S.~Melville, and E.~Pajer, ``{The Analytic
  Wavefunction},'' \href{http://arxiv.org/abs/2212.08009}{{\ttfamily
  arXiv:2212.08009 [hep-th]}}.

\bibitem{Chen:2009zp}
X.~Chen and Y.~Wang, ``{Quasi-Single Field Inflation and Non-Gaussianities},''
  \href{http://dx.doi.org/10.1088/1475-7516/2010/04/027}{{\em JCAP} {\bfseries
  04} (2010) 027}, \href{http://arxiv.org/abs/0911.3380}{{\ttfamily
  arXiv:0911.3380 [hep-th]}}.

\bibitem{Baumann:2011nk}
D.~Baumann and D.~Green, ``{Signatures of Supersymmetry from the Early
  Universe},'' \href{http://dx.doi.org/10.1103/PhysRevD.85.103520}{{\em Phys.
  Rev. D} {\bfseries 85} (2012) 103520},
  \href{http://arxiv.org/abs/1109.0292}{{\ttfamily arXiv:1109.0292 [hep-th]}}.

\bibitem{Assassi:2012zq}
V.~Assassi, D.~Baumann, and D.~Green, ``{On Soft Limits of Inflationary
  Correlation Functions},''
  \href{http://dx.doi.org/10.1088/1475-7516/2012/11/047}{{\em JCAP} {\bfseries
  11} (2012) 047}, \href{http://arxiv.org/abs/1204.4207}{{\ttfamily
  arXiv:1204.4207 [hep-th]}}.

\bibitem{Chen:2012ge}
X.~Chen and Y.~Wang, ``{Quasi-Single Field Inflation with Large Mass},''
  \href{http://dx.doi.org/10.1088/1475-7516/2012/09/021}{{\em JCAP} {\bfseries
  09} (2012) 021}, \href{http://arxiv.org/abs/1205.0160}{{\ttfamily
  arXiv:1205.0160 [hep-th]}}.

\bibitem{Pi:2012gf}
S.~Pi and M.~Sasaki, ``{Curvature Perturbation Spectrum in Two-field Inflation
  with a Turning Trajectory},''
  \href{http://dx.doi.org/10.1088/1475-7516/2012/10/051}{{\em JCAP} {\bfseries
  10} (2012) 051}, \href{http://arxiv.org/abs/1205.0161}{{\ttfamily
  arXiv:1205.0161 [hep-th]}}.

\bibitem{Noumi:2012vr}
T.~Noumi, M.~Yamaguchi, and D.~Yokoyama, ``{Effective field theory approach to
  quasi-single field inflation and effects of heavy fields},''
  \href{http://dx.doi.org/10.1007/JHEP06(2013)051}{{\em JHEP} {\bfseries 06}
  (2013) 051}, \href{http://arxiv.org/abs/1211.1624}{{\ttfamily arXiv:1211.1624
  [hep-th]}}.

\bibitem{Baumann:2012bc}
D.~Baumann, S.~Ferraro, D.~Green, and K.~M. Smith, ``{Stochastic Bias from
  Non-Gaussian Initial Conditions},''
  \href{http://dx.doi.org/10.1088/1475-7516/2013/05/001}{{\em JCAP} {\bfseries
  05} (2013) 001}, \href{http://arxiv.org/abs/1209.2173}{{\ttfamily
  arXiv:1209.2173 [astro-ph.CO]}}.

\bibitem{Assassi:2013gxa}
V.~Assassi, D.~Baumann, D.~Green, and L.~McAllister, ``{Planck-Suppressed
  Operators},'' \href{http://dx.doi.org/10.1088/1475-7516/2014/01/033}{{\em
  JCAP} {\bfseries 01} (2014) 033},
  \href{http://arxiv.org/abs/1304.5226}{{\ttfamily arXiv:1304.5226 [hep-th]}}.

\bibitem{Gong:2013sma}
J.-O. Gong, S.~Pi, and M.~Sasaki, ``{Equilateral non-Gaussianity from heavy
  fields},'' \href{http://dx.doi.org/10.1088/1475-7516/2013/11/043}{{\em JCAP}
  {\bfseries 11} (2013) 043}, \href{http://arxiv.org/abs/1306.3691}{{\ttfamily
  arXiv:1306.3691 [hep-th]}}.

\bibitem{Arkani-Hamed:2015bza}
N.~Arkani-Hamed and J.~Maldacena, ``{Cosmological Collider Physics},''
  \href{http://arxiv.org/abs/1503.08043}{{\ttfamily arXiv:1503.08043
  [hep-th]}}.

\bibitem{Lee:2016vti}
H.~Lee, D.~Baumann, and G.~L. Pimentel, ``{Non-Gaussianity as a Particle
  Detector},'' \href{http://dx.doi.org/10.1007/JHEP12(2016)040}{{\em JHEP}
  {\bfseries 12} (2016) 040}, \href{http://arxiv.org/abs/1607.03735}{{\ttfamily
  arXiv:1607.03735 [hep-th]}}.

\bibitem{Flauger:2016idt}
R.~Flauger, M.~Mirbabayi, L.~Senatore, and E.~Silverstein, ``{Productive
  Interactions: heavy particles and non-Gaussianity},''
  \href{http://dx.doi.org/10.1088/1475-7516/2017/10/058}{{\em JCAP} {\bfseries
  10} (2017) 058}, \href{http://arxiv.org/abs/1606.00513}{{\ttfamily
  arXiv:1606.00513 [hep-th]}}.

\bibitem{Chen:2016uwp}
X.~Chen, Y.~Wang, and Z.-Z. Xianyu, ``{Standard Model Background of the
  Cosmological Collider},''
  \href{http://dx.doi.org/10.1103/PhysRevLett.118.261302}{{\em Phys. Rev.
  Lett.} {\bfseries 118} no.~26, (2017) 261302},
  \href{http://arxiv.org/abs/1610.06597}{{\ttfamily arXiv:1610.06597
  [hep-th]}}.

\bibitem{Chen:2016hrz}
X.~Chen, Y.~Wang, and Z.-Z. Xianyu, ``{Standard Model Mass Spectrum in
  Inflationary Universe},''
  \href{http://dx.doi.org/10.1007/JHEP04(2017)058}{{\em JHEP} {\bfseries 04}
  (2017) 058}, \href{http://arxiv.org/abs/1612.08122}{{\ttfamily
  arXiv:1612.08122 [hep-th]}}.

\bibitem{Kehagias:2017cym}
A.~Kehagias and A.~Riotto, ``{On the Inflationary Perturbations of Massive
  Higher-Spin Fields},''
  \href{http://dx.doi.org/10.1088/1475-7516/2017/07/046}{{\em JCAP} {\bfseries
  07} (2017) 046}, \href{http://arxiv.org/abs/1705.05834}{{\ttfamily
  arXiv:1705.05834 [hep-th]}}.

\bibitem{Kumar:2017ecc}
S.~Kumar and R.~Sundrum, ``{Heavy-Lifting of Gauge Theories By Cosmic
  Inflation},'' \href{http://dx.doi.org/10.1007/JHEP05(2018)011}{{\em JHEP}
  {\bfseries 05} (2018) 011}, \href{http://arxiv.org/abs/1711.03988}{{\ttfamily
  arXiv:1711.03988 [hep-ph]}}.

\bibitem{An:2017hlx}
H.~An, M.~McAneny, A.~K. Ridgway, and M.~B. Wise, ``{Quasi Single Field
  Inflation in the non-perturbative regime},''
  \href{http://dx.doi.org/10.1007/JHEP06(2018)105}{{\em JHEP} {\bfseries 06}
  (2018) 105}, \href{http://arxiv.org/abs/1706.09971}{{\ttfamily
  arXiv:1706.09971 [hep-ph]}}.

\bibitem{An:2017rwo}
H.~An, M.~McAneny, A.~K. Ridgway, and M.~B. Wise, ``{Non-Gaussian Enhancements
  of Galactic Halo Correlations in Quasi-Single Field Inflation},''
  \href{http://dx.doi.org/10.1103/PhysRevD.97.123528}{{\em Phys. Rev. D}
  {\bfseries 97} no.~12, (2018) 123528},
  \href{http://arxiv.org/abs/1711.02667}{{\ttfamily arXiv:1711.02667
  [hep-ph]}}.

\bibitem{Baumann:2017jvh}
D.~Baumann, G.~Goon, H.~Lee, and G.~L. Pimentel, ``{Partially Massless Fields
  During Inflation},'' \href{http://dx.doi.org/10.1007/JHEP04(2018)140}{{\em
  JHEP} {\bfseries 04} (2018) 140},
  \href{http://arxiv.org/abs/1712.06624}{{\ttfamily arXiv:1712.06624
  [hep-th]}}.

\bibitem{Kumar:2018jxz}
S.~Kumar and R.~Sundrum, ``{Seeing Higher-Dimensional Grand Unification In
  Primordial Non-Gaussianities},''
  \href{http://dx.doi.org/10.1007/JHEP04(2019)120}{{\em JHEP} {\bfseries 04}
  (2019) 120}, \href{http://arxiv.org/abs/1811.11200}{{\ttfamily
  arXiv:1811.11200 [hep-ph]}}.

\bibitem{Bordin:2018pca}
L.~Bordin, P.~Creminelli, A.~Khmelnitsky, and L.~Senatore, ``{Light Particles
  with Spin in Inflation},''
  \href{http://dx.doi.org/10.1088/1475-7516/2018/10/013}{{\em JCAP} {\bfseries
  10} (2018) 013}, \href{http://arxiv.org/abs/1806.10587}{{\ttfamily
  arXiv:1806.10587 [hep-th]}}.

\bibitem{Goon:2018fyu}
G.~Goon, K.~Hinterbichler, A.~Joyce, and M.~Trodden, ``{Shapes of gravity:
  Tensor non-Gaussianity and massive spin-2 fields},''
  \href{http://dx.doi.org/10.1007/JHEP10(2019)182}{{\em JHEP} {\bfseries 10}
  (2019) 182}, \href{http://arxiv.org/abs/1812.07571}{{\ttfamily
  arXiv:1812.07571 [hep-th]}}.

\bibitem{Anninos:2019nib}
D.~Anninos, V.~De~Luca, G.~Franciolini, A.~Kehagias, and A.~Riotto,
  ``{Cosmological Shapes of Higher-Spin Gravity},''
  \href{http://dx.doi.org/10.1088/1475-7516/2019/04/045}{{\em JCAP} {\bfseries
  04} (2019) 045}, \href{http://arxiv.org/abs/1902.01251}{{\ttfamily
  arXiv:1902.01251 [hep-th]}}.

\bibitem{Kim:2019wjo}
S.~Kim, T.~Noumi, K.~Takeuchi, and S.~Zhou, ``{Heavy Spinning Particles from
  Signs of Primordial Non-Gaussianities: Beyond the Positivity Bounds},''
  \href{http://dx.doi.org/10.1007/JHEP12(2019)107}{{\em JHEP} {\bfseries 12}
  (2019) 107}, \href{http://arxiv.org/abs/1906.11840}{{\ttfamily
  arXiv:1906.11840 [hep-th]}}.

\bibitem{Alexander:2019vtb}
S.~Alexander, S.~J. Gates, L.~Jenks, K.~Koutrolikos, and E.~McDonough,
  ``{Higher Spin Supersymmetry at the Cosmological Collider: Sculpting SUSY
  Rilles in the CMB},'' \href{http://dx.doi.org/10.1007/JHEP10(2019)156}{{\em
  JHEP} {\bfseries 10} (2019) 156},
  \href{http://arxiv.org/abs/1907.05829}{{\ttfamily arXiv:1907.05829
  [hep-th]}}.

\bibitem{Hook:2019zxa}
A.~Hook, J.~Huang, and D.~Racco, ``{Searches for other vacua. Part II. A new
  Higgstory at the cosmological collider},''
  \href{http://dx.doi.org/10.1007/JHEP01(2020)105}{{\em JHEP} {\bfseries 01}
  (2020) 105}, \href{http://arxiv.org/abs/1907.10624}{{\ttfamily
  arXiv:1907.10624 [hep-ph]}}.

\bibitem{Kumar:2019ebj}
S.~Kumar and R.~Sundrum, ``{Cosmological Collider Physics and the Curvaton},''
  \href{http://dx.doi.org/10.1007/JHEP04(2020)077}{{\em JHEP} {\bfseries 04}
  (2020) 077}, \href{http://arxiv.org/abs/1908.11378}{{\ttfamily
  arXiv:1908.11378 [hep-ph]}}.

\bibitem{Liu:2019fag}
T.~Liu, X.~Tong, Y.~Wang, and Z.-Z. Xianyu, ``{Probing P and CP Violations on
  the Cosmological Collider},''
  \href{http://dx.doi.org/10.1007/JHEP04(2020)189}{{\em JHEP} {\bfseries 04}
  (2020) 189}, \href{http://arxiv.org/abs/1909.01819}{{\ttfamily
  arXiv:1909.01819 [hep-ph]}}.

\bibitem{Wang:2019gbi}
L.-T. Wang and Z.-Z. Xianyu, ``{In Search of Large Signals at the Cosmological
  Collider},'' \href{http://dx.doi.org/10.1007/JHEP02(2020)044}{{\em JHEP}
  {\bfseries 02} (2020) 044}, \href{http://arxiv.org/abs/1910.12876}{{\ttfamily
  arXiv:1910.12876 [hep-ph]}}.

\bibitem{Wang:2019gok}
D.-G. Wang, ``{On the inflationary massive field with a curved field
  manifold},'' \href{http://dx.doi.org/10.1088/1475-7516/2020/01/046}{{\em
  JCAP} {\bfseries 01} (2020) 046},
  \href{http://arxiv.org/abs/1911.04459}{{\ttfamily arXiv:1911.04459
  [astro-ph.CO]}}.

\bibitem{Maru:2021ezc}
N.~Maru and A.~Okawa, ``{Non-Gaussianity from $X, Y$ gauge bosons in
  Cosmological Collider Physics},''
  \href{http://arxiv.org/abs/2101.10634}{{\ttfamily arXiv:2101.10634
  [hep-ph]}}.

\bibitem{Lu:2021wxu}
Q.~Lu, M.~Reece, and Z.-Z. Xianyu, ``{Missing scalars at the cosmological
  collider},'' \href{http://dx.doi.org/10.1007/JHEP12(2021)098}{{\em JHEP}
  {\bfseries 12} (2021) 098}, \href{http://arxiv.org/abs/2108.11385}{{\ttfamily
  arXiv:2108.11385 [hep-ph]}}.

\bibitem{Wang:2021qez}
L.-T. Wang, Z.-Z. Xianyu, and Y.-M. Zhong, ``{Precision calculation of
  inflation correlators at one loop},''
  \href{http://dx.doi.org/10.1007/JHEP02(2022)085}{{\em JHEP} {\bfseries 02}
  (2022) 085}, \href{http://arxiv.org/abs/2109.14635}{{\ttfamily
  arXiv:2109.14635 [hep-ph]}}.

\bibitem{Tong:2021wai}
X.~Tong, Y.~Wang, and Y.~Zhu, ``{Cutting rule for cosmological collider
  signals: a bulk evolution perspective},''
  \href{http://dx.doi.org/10.1007/JHEP03(2022)181}{{\em JHEP} {\bfseries 03}
  (2022) 181}, \href{http://arxiv.org/abs/2112.03448}{{\ttfamily
  arXiv:2112.03448 [hep-th]}}.

\bibitem{Cui:2021iie}
Y.~Cui and Z.-Z. Xianyu, ``{Probing Leptogenesis with the Cosmological
  Collider},'' \href{http://arxiv.org/abs/2112.10793}{{\ttfamily
  arXiv:2112.10793 [hep-ph]}}.

\bibitem{Tong:2022cdz}
X.~Tong and Z.-Z. Xianyu, ``{Large Spin-2 Signals at the Cosmological
  Collider},'' \href{http://arxiv.org/abs/2203.06349}{{\ttfamily
  arXiv:2203.06349 [hep-ph]}}.

\bibitem{Reece:2022soh}
M.~Reece, L.-T. Wang, and Z.-Z. Xianyu, ``{Large-Field Inflation and the
  Cosmological Collider},'' \href{http://arxiv.org/abs/2204.11869}{{\ttfamily
  arXiv:2204.11869 [hep-ph]}}.

\bibitem{Chen:2022vzh}
X.~Chen, R.~Ebadi, and S.~Kumar, ``{Classical cosmological collider physics and
  primordial features},''
  \href{http://dx.doi.org/10.1088/1475-7516/2022/08/083}{{\em JCAP} {\bfseries
  08} (2022) 083}, \href{http://arxiv.org/abs/2205.01107}{{\ttfamily
  arXiv:2205.01107 [hep-ph]}}.

\bibitem{Qin:2022lva}
Z.~Qin and Z.-Z. Xianyu, ``{Phase information in cosmological collider
  signals},'' \href{http://dx.doi.org/10.1007/JHEP10(2022)192}{{\em JHEP}
  {\bfseries 10} (2022) 192}, \href{http://arxiv.org/abs/2205.01692}{{\ttfamily
  arXiv:2205.01692 [hep-th]}}.

\bibitem{Xianyu:2022jwk}
Z.-Z. Xianyu and H.~Zhang, ``{Bootstrapping One-Loop Inflation Correlators with
  the Spectral Decomposition},''
  \href{http://arxiv.org/abs/2211.03810}{{\ttfamily arXiv:2211.03810
  [hep-th]}}.

\bibitem{Niu:2022quw}
X.~Niu, M.~H. Rahat, K.~Srinivasan, and W.~Xue, ``{Gravitational Wave Probes of
  Massive Gauge Bosons at the Cosmological Collider},''
  \href{http://arxiv.org/abs/2211.14331}{{\ttfamily arXiv:2211.14331
  [hep-ph]}}.

\bibitem{Achucarro:2016fby}
A.~Ach\'{u}carro, V.~Atal, C.~Germani, and G.~A. Palma, ``{Cumulative effects
  in inflation with ultra-light entropy modes},''
  \href{http://dx.doi.org/10.1088/1475-7516/2017/02/013}{{\em JCAP} {\bfseries
  1702} no.~02, (2017) 013},
\href{http://arxiv.org/abs/1607.08609}{{\ttfamily arXiv:1607.08609
  [astro-ph.CO]}}.

\bibitem{Achucarro:2019lgo}
A.~Ach\'ucarro, G.~A. Palma, D.-G. Wang, and Y.~Welling, ``{Origin of
  ultra-light fields during inflation and their suppressed non-Gaussianity},''
  \href{http://dx.doi.org/10.1088/1475-7516/2020/10/018}{{\em JCAP} {\bfseries
  10} (2020) 018}, \href{http://arxiv.org/abs/1908.06956}{{\ttfamily
  arXiv:1908.06956 [hep-th]}}.

\bibitem{Achucarro:2021pdh}
A.~Achucarro, S.~Cespedes, A.-C. Davis, and G.~A. Palma, ``{The hand-made tail:
  non-perturbative tails from multifield inflation},''
  \href{http://dx.doi.org/10.1007/JHEP05(2022)052}{{\em JHEP} {\bfseries 05}
  (2022) 052}, \href{http://arxiv.org/abs/2112.14712}{{\ttfamily
  arXiv:2112.14712 [hep-th]}}.

\bibitem{Achucarro:2019pux}
A.~Ach\'ucarro, E.~J. Copeland, O.~Iarygina, G.~A. Palma, D.-G. Wang, and
  Y.~Welling, ``{Shift-Symmetric Orbital Inflation: single field or
  multi-field?},''
\href{http://arxiv.org/abs/1901.03657}{{\ttfamily arXiv:1901.03657
  [astro-ph.CO]}}.

\bibitem{Ford:1984hs}
L.~H. Ford, ``{Quantum Instability of De Sitter Space-time},''
\href{http://dx.doi.org/10.1103/PhysRevD.31.710}{{\em Phys. Rev.} {\bfseries
  D31} (1985) 710}.

\bibitem{Antoniadis:1985pj}
I.~Antoniadis, J.~Iliopoulos, and T.~N. Tomaras, ``{Quantum Instability of De
  Sitter Space},''
\href{http://dx.doi.org/10.1103/PhysRevLett.56.1319}{{\em Phys. Rev. Lett.}
  {\bfseries 56} (1986) 1319}.

\bibitem{Tsamis:1994ca}
N.~C. Tsamis and R.~P. Woodard, ``{Strong infrared effects in quantum
  gravity},''
\href{http://dx.doi.org/10.1006/aphy.1995.1015}{{\em Annals Phys.} {\bfseries
  238} (1995) 1--82}.

\bibitem{Tsamis:1996qm}
N.~C. Tsamis and R.~P. Woodard, ``{The Quantum gravitational back reaction on
  inflation},'' \href{http://dx.doi.org/10.1006/aphy.1997.5613}{{\em Annals
  Phys.} {\bfseries 253} (1997) 1--54},
\href{http://arxiv.org/abs/hep-ph/9602316}{{\ttfamily arXiv:hep-ph/9602316
  [hep-ph]}}.

\bibitem{Seery:2010kh}
D.~Seery, ``{Infrared effects in inflationary correlation functions},''
  \href{http://dx.doi.org/10.1088/0264-9381/27/12/124005}{{\em Class. Quant.
  Grav.} {\bfseries 27} (2010) 124005},
  \href{http://arxiv.org/abs/1005.1649}{{\ttfamily arXiv:1005.1649
  [astro-ph.CO]}}.

\bibitem{Giddings:2010nc}
S.~B. Giddings and M.~S. Sloth, ``{Semiclassical relations and IR effects in de
  Sitter and slow-roll space-times},''
  \href{http://dx.doi.org/10.1088/1475-7516/2011/01/023}{{\em JCAP} {\bfseries
  01} (2011) 023}, \href{http://arxiv.org/abs/1005.1056}{{\ttfamily
  arXiv:1005.1056 [hep-th]}}.

\bibitem{Giddings:2010ui}
S.~B. Giddings and M.~S. Sloth, ``{Cosmological diagrammatic rules},''
  \href{http://dx.doi.org/10.1088/1475-7516/2010/07/015}{{\em JCAP} {\bfseries
  07} (2010) 015}, \href{http://arxiv.org/abs/1005.3287}{{\ttfamily
  arXiv:1005.3287 [hep-th]}}.

\bibitem{Anninos:2014lwa}
D.~Anninos, T.~Anous, D.~Z. Freedman, and G.~Konstantinidis, ``{Late-time
  Structure of the Bunch-Davies De Sitter Wavefunction},''
  \href{http://dx.doi.org/10.1088/1475-7516/2015/11/048}{{\em JCAP} {\bfseries
  11} (2015) 048}, \href{http://arxiv.org/abs/1406.5490}{{\ttfamily
  arXiv:1406.5490 [hep-th]}}.

\bibitem{Hu:2018nxy}
B.-L. Hu, ``{Infrared Behavior of Quantum Fields in Inflationary Cosmology --
  Issues and Approaches: an overview},''
  \href{http://arxiv.org/abs/1812.11851}{{\ttfamily arXiv:1812.11851 [gr-qc]}}.

\bibitem{Gorbenko:2019rza}
V.~Gorbenko and L.~Senatore, ``{$\lambda \phi^4$ in dS},''
\href{http://arxiv.org/abs/1911.00022}{{\ttfamily arXiv:1911.00022 [hep-th]}}.

\bibitem{Mirbabayi:2019qtx}
M.~Mirbabayi, ``{Infrared dynamics of a light scalar field in de Sitter},''
  \href{http://dx.doi.org/10.1088/1475-7516/2020/12/006}{{\em JCAP} {\bfseries
  12} (2020) 006}, \href{http://arxiv.org/abs/1911.00564}{{\ttfamily
  arXiv:1911.00564 [hep-th]}}.

\bibitem{Cohen:2020php}
T.~Cohen and D.~Green, ``{Soft de Sitter Effective Theory},''
  \href{http://dx.doi.org/10.1007/JHEP12(2020)041}{{\em JHEP} {\bfseries 12}
  (2020) 041}, \href{http://arxiv.org/abs/2007.03693}{{\ttfamily
  arXiv:2007.03693 [hep-th]}}.

\bibitem{Cohen:2021fzf}
T.~Cohen, D.~Green, A.~Premkumar, and A.~Ridgway, ``{Stochastic Inflation at
  NNLO},'' \href{http://dx.doi.org/10.1007/JHEP09(2021)159}{{\em JHEP}
  {\bfseries 09} (2021) 159}, \href{http://arxiv.org/abs/2106.09728}{{\ttfamily
  arXiv:2106.09728 [hep-th]}}.

\bibitem{Green:2022ovz}
D.~Green, ``{EFT for de Sitter Space},''
  \href{http://arxiv.org/abs/2210.05820}{{\ttfamily arXiv:2210.05820
  [hep-th]}}.

\bibitem{Cohen:2022clv}
T.~Cohen, D.~Green, and A.~Premkumar, ``{Large Deviations in the Early
  Universe},'' \href{http://arxiv.org/abs/2212.02535}{{\ttfamily
  arXiv:2212.02535 [hep-th]}}.

\bibitem{Vilenkin:1983xq}
A.~Vilenkin, ``{The Birth of Inflationary Universes},''
  \href{http://dx.doi.org/10.1103/PhysRevD.27.2848}{{\em Phys. Rev. D}
  {\bfseries 27} (1983) 2848}.

\bibitem{Starobinsky:1986fx}
A.~A. Starobinsky, ``{STOCHASTIC DE SITTER (INFLATIONARY) STAGE IN THE EARLY
  UNIVERSE},'' \href{http://dx.doi.org/10.1007/3-540-16452-9_6}{{\em Lect.
  Notes Phys.} {\bfseries 246} (1986) 107--126}.

\bibitem{Baumann:2014nda}
D.~Baumann and L.~McAllister,
  \href{http://dx.doi.org/10.1017/CBO9781316105733}{{\em {Inflation and String
  Theory}}}.
\newblock Cambridge Monographs on Mathematical Physics. Cambridge University
  Press, 5, 2015.
\newblock \href{http://arxiv.org/abs/1404.2601}{{\ttfamily arXiv:1404.2601
  [hep-th]}}.

\bibitem{Turzynski:2014tza}
S.~Renaux-Petel and K.~Turzynski, ``{On reaching the adiabatic limit in
  multi-field inflation},''
  \href{http://dx.doi.org/10.1088/1475-7516/2015/06/010}{{\em JCAP} {\bfseries
  1506} no.~06, (2015) 010},
\href{http://arxiv.org/abs/1405.6195}{{\ttfamily arXiv:1405.6195
  [astro-ph.CO]}}.

\bibitem{Renaux-Petel:2015mga}
S.~Renaux-Petel and K.~Turzynski, ``{Geometrical Destabilization of
  Inflation},'' \href{http://dx.doi.org/10.1103/PhysRevLett.117.141301}{{\em
  Phys. Rev. Lett.} {\bfseries 117} no.~14, (2016) 141301},
\href{http://arxiv.org/abs/1510.01281}{{\ttfamily arXiv:1510.01281
  [astro-ph.CO]}}.

\bibitem{Brown:2017osf}
A.~R. Brown, ``{Hyperbolic Inflation},''
  \href{http://dx.doi.org/10.1103/PhysRevLett.121.251601}{{\em Phys. Rev.
  Lett.} {\bfseries 121} no.~25, (2018) 251601},
\href{http://arxiv.org/abs/1705.03023}{{\ttfamily arXiv:1705.03023 [hep-th]}}.

\bibitem{Achucarro:2017ing}
A.~Ach\'{u}carro, R.~Kallosh, A.~Linde, D.-G. Wang, and Y.~Welling,
  ``{Universality of multi-field $\alpha$-attractors},''
  \href{http://dx.doi.org/10.1088/1475-7516/2018/04/028}{{\em JCAP} {\bfseries
  1804} no.~04, (2018) 028},
\href{http://arxiv.org/abs/1711.09478}{{\ttfamily arXiv:1711.09478 [hep-th]}}.

\bibitem{Christodoulidis:2018qdw}
P.~Christodoulidis, D.~Roest, and E.~I. Sfakianakis, ``{Angular inflation in
  multi-field ${\alpha}$-attractors},''
\href{http://arxiv.org/abs/1803.09841}{{\ttfamily arXiv:1803.09841 [hep-th]}}.

\bibitem{Garcia-Saenz:2018ifx}
S.~Garcia-Saenz, S.~Renaux-Petel, and J.~Ronayne, ``{Primordial fluctuations
  and non-Gaussianities in sidetracked inflation},''
  \href{http://dx.doi.org/10.1088/1475-7516/2018/07/057}{{\em JCAP} {\bfseries
  1807} no.~07, (2018) 057},
\href{http://arxiv.org/abs/1804.11279}{{\ttfamily arXiv:1804.11279
  [astro-ph.CO]}}.

\bibitem{Achucarro:2019mea}
A.~Ach\'ucarro and Y.~Welling, ``{Orbital Inflation: inflating along an angular
  isometry of field space},''
\href{http://arxiv.org/abs/1907.02020}{{\ttfamily arXiv:1907.02020 [hep-th]}}.

\bibitem{Bjorkmo:2019qno}
T.~Bjorkmo, R.~Z. Ferreira, and M.~C.~D. Marsh, ``{Mild Non-Gaussianities under
  Perturbative Control from Rapid-Turn Inflation Models},''
\href{http://arxiv.org/abs/1908.11316}{{\ttfamily arXiv:1908.11316 [hep-th]}}.

\bibitem{Romano:2020kmj}
A.~E. Romano, S.~A. Vallejo-Pe\~na, and K.~Turzy\'nski, ``{Model-independent
  approach to effective sound speed in multi-field inflation},''
  \href{http://dx.doi.org/10.1140/epjc/s10052-022-10669-3}{{\em Eur. Phys. J.
  C} {\bfseries 82} no.~8, (2022) 767},
  \href{http://arxiv.org/abs/2006.00969}{{\ttfamily arXiv:2006.00969 [gr-qc]}}.

\bibitem{Gordon:2000hv}
C.~Gordon, D.~Wands, B.~A. Bassett, and R.~Maartens, ``{Adiabatic and entropy
  perturbations from inflation},''
  \href{http://dx.doi.org/10.1103/PhysRevD.63.023506}{{\em Phys. Rev.}
  {\bfseries D63} (2001) 023506},
\href{http://arxiv.org/abs/astro-ph/0009131}{{\ttfamily arXiv:astro-ph/0009131
  [astro-ph]}}.

\bibitem{Enqvist:2001zp}
K.~Enqvist and M.~S. Sloth, ``{Adiabatic CMB perturbations in pre - big bang
  string cosmology},''
  \href{http://dx.doi.org/10.1016/S0550-3213(02)00043-3}{{\em Nucl. Phys. B}
  {\bfseries 626} (2002) 395--409},
  \href{http://arxiv.org/abs/hep-ph/0109214}{{\ttfamily arXiv:hep-ph/0109214}}.

\bibitem{Lyth:2001nq}
D.~H. Lyth and D.~Wands, ``{Generating the curvature perturbation without an
  inflaton},'' \href{http://dx.doi.org/10.1016/S0370-2693(01)01366-1}{{\em
  Phys. Lett. B} {\bfseries 524} (2002) 5--14},
  \href{http://arxiv.org/abs/hep-ph/0110002}{{\ttfamily arXiv:hep-ph/0110002}}.

\bibitem{Moroi:2001ct}
T.~Moroi and T.~Takahashi, ``{Effects of cosmological moduli fields on cosmic
  microwave background},''
  \href{http://dx.doi.org/10.1016/S0370-2693(01)01295-3}{{\em Phys. Lett. B}
  {\bfseries 522} (2001) 215--221},
  \href{http://arxiv.org/abs/hep-ph/0110096}{{\ttfamily arXiv:hep-ph/0110096}}.
  [Erratum: Phys.Lett.B 539, 303--303 (2002)].

\bibitem{Dvali:2003em}
G.~Dvali, A.~Gruzinov, and M.~Zaldarriaga, ``{A new mechanism for generating
  density perturbations from inflation},''
  \href{http://dx.doi.org/10.1103/PhysRevD.69.023505}{{\em Phys. Rev. D}
  {\bfseries 69} (2004) 023505},
  \href{http://arxiv.org/abs/astro-ph/0303591}{{\ttfamily
  arXiv:astro-ph/0303591}}.

\bibitem{Starobinsky:1985ibc}
A.~A. Starobinsky, ``{Multicomponent de Sitter (Inflationary) Stages and the
  Generation of Perturbations},'' {\em JETP Lett.} {\bfseries 42} (1985)
  152--155.

\bibitem{Langlois:1999dw}
D.~Langlois, ``{Correlated adiabatic and isocurvature perturbations from double
  inflation},'' \href{http://dx.doi.org/10.1103/PhysRevD.59.123512}{{\em Phys.
  Rev. D} {\bfseries 59} (1999) 123512},
  \href{http://arxiv.org/abs/astro-ph/9906080}{{\ttfamily
  arXiv:astro-ph/9906080}}.

\bibitem{GrootNibbelink:2000vx}
S.~Groot~Nibbelink and B.~J.~W. van Tent, ``{Density perturbations arising from
  multiple field slow roll inflation},''
\href{http://arxiv.org/abs/hep-ph/0011325}{{\ttfamily arXiv:hep-ph/0011325
  [hep-ph]}}.

\bibitem{GrootNibbelink:2001qt}
S.~Groot~Nibbelink and B.~J.~W. van Tent, ``{Scalar perturbations during
  multiple field slow-roll inflation},''
  \href{http://dx.doi.org/10.1088/0264-9381/19/4/302}{{\em Class. Quant. Grav.}
  {\bfseries 19} (2002) 613--640},
\href{http://arxiv.org/abs/hep-ph/0107272}{{\ttfamily arXiv:hep-ph/0107272
  [hep-ph]}}.

\bibitem{Achucarro:2010da}
A.~Achucarro, J.-O. Gong, S.~Hardeman, G.~A. Palma, and S.~P. Patil,
  ``{Features of heavy physics in the CMB power spectrum},''
  \href{http://dx.doi.org/10.1088/1475-7516/2011/01/030}{{\em JCAP} {\bfseries
  01} (2011) 030}, \href{http://arxiv.org/abs/1010.3693}{{\ttfamily
  arXiv:1010.3693 [hep-ph]}}.

\bibitem{Gong:2011uw}
J.-O. Gong and T.~Tanaka, ``{A covariant approach to general field space metric
  in multi-field inflation},''
  \href{http://dx.doi.org/10.1088/1475-7516/2012/02/E01,
  10.1088/1475-7516/2011/03/015}{{\em JCAP} {\bfseries 1103} (2011) 015},
  \href{http://arxiv.org/abs/1101.4809}{{\ttfamily arXiv:1101.4809
  [astro-ph.CO]}}.
[Erratum: JCAP1202,E01(2012)].

\bibitem{Achucarro:2012sm}
A.~Achucarro, J.-O. Gong, S.~Hardeman, G.~A. Palma, and S.~P. Patil,
  ``{Effective theories of single field inflation when heavy fields matter},''
  \href{http://dx.doi.org/10.1007/JHEP05(2012)066}{{\em JHEP} {\bfseries 05}
  (2012) 066}, \href{http://arxiv.org/abs/1201.6342}{{\ttfamily arXiv:1201.6342
  [hep-th]}}.

\bibitem{Achucarro:2012yr}
A.~Achucarro, V.~Atal, S.~Cespedes, J.-O. Gong, G.~A. Palma, and S.~P. Patil,
  ``{Heavy fields, reduced speeds of sound and decoupling during inflation},''
  \href{http://dx.doi.org/10.1103/PhysRevD.86.121301}{{\em Phys. Rev. D}
  {\bfseries 86} (2012) 121301},
  \href{http://arxiv.org/abs/1205.0710}{{\ttfamily arXiv:1205.0710 [hep-th]}}.

\bibitem{Garcia-Saenz:2019njm}
S.~Garcia-Saenz, L.~Pinol, and S.~Renaux-Petel, ``{Revisiting non-Gaussianity
  in multifield inflation with curved field space},''
\href{http://arxiv.org/abs/1907.10403}{{\ttfamily arXiv:1907.10403 [hep-th]}}.

\bibitem{Cheung:2007st}
C.~Cheung, P.~Creminelli, A.~L. Fitzpatrick, J.~Kaplan, and L.~Senatore, ``{The
  Effective Field Theory of Inflation},''
  \href{http://dx.doi.org/10.1088/1126-6708/2008/03/014}{{\em JHEP} {\bfseries
  03} (2008) 014}, \href{http://arxiv.org/abs/0709.0293}{{\ttfamily
  arXiv:0709.0293 [hep-th]}}.

\bibitem{Hui:2022dnm}
L.~Hui, A.~Joyce, I.~Komissarov, K.~Parmentier, L.~Santoni, and S.~S.~C. Wong,
  ``{Soft theorems for boosts and other time symmetries},''
  \href{http://arxiv.org/abs/2210.16276}{{\ttfamily arXiv:2210.16276
  [hep-th]}}.

\bibitem{Maldacena:2002vr}
J.~M. Maldacena, ``{Non-Gaussian features of primordial fluctuations in single
  field inflationary models},''
  \href{http://dx.doi.org/10.1088/1126-6708/2003/05/013}{{\em JHEP} {\bfseries
  05} (2003) 013}, \href{http://arxiv.org/abs/astro-ph/0210603}{{\ttfamily
  arXiv:astro-ph/0210603}}.

\bibitem{Weinberg:2005vy}
S.~Weinberg, ``{Quantum contributions to cosmological correlations},''
  \href{http://dx.doi.org/10.1103/PhysRevD.72.043514}{{\em Phys. Rev. D}
  {\bfseries 72} (2005) 043514},
  \href{http://arxiv.org/abs/hep-th/0506236}{{\ttfamily arXiv:hep-th/0506236}}.

\bibitem{Chen:2010xka}
X.~Chen, ``{Primordial Non-Gaussianities from Inflation Models},''
  \href{http://dx.doi.org/10.1155/2010/638979}{{\em Adv. Astron.} {\bfseries
  2010} (2010) 638979}, \href{http://arxiv.org/abs/1002.1416}{{\ttfamily
  arXiv:1002.1416 [astro-ph.CO]}}.

\bibitem{Chen:2017ryl}
X.~Chen, Y.~Wang, and Z.-Z. Xianyu, ``{Schwinger-Keldysh Diagrammatics for
  Primordial Perturbations},''
  \href{http://dx.doi.org/10.1088/1475-7516/2017/12/006}{{\em JCAP} {\bfseries
  12} (2017) 006}, \href{http://arxiv.org/abs/1703.10166}{{\ttfamily
  arXiv:1703.10166 [hep-th]}}.

\bibitem{Bzowski:2013sza}
A.~Bzowski, P.~McFadden, and K.~Skenderis, ``{Implications of conformal
  invariance in momentum space},''
  \href{http://dx.doi.org/10.1007/JHEP03(2014)111}{{\em JHEP} {\bfseries 03}
  (2014) 111}, \href{http://arxiv.org/abs/1304.7760}{{\ttfamily arXiv:1304.7760
  [hep-th]}}.

\bibitem{Bzowski:2015pba}
A.~Bzowski, P.~McFadden, and K.~Skenderis, ``{Scalar 3-point functions in CFT:
  renormalisation, beta functions and anomalies},''
  \href{http://dx.doi.org/10.1007/JHEP03(2016)066}{{\em JHEP} {\bfseries 03}
  (2016) 066}, \href{http://arxiv.org/abs/1510.08442}{{\ttfamily
  arXiv:1510.08442 [hep-th]}}.

\bibitem{Bzowski:2018fql}
A.~Bzowski, P.~McFadden, and K.~Skenderis, ``{Renormalised CFT 3-point
  functions of scalars, currents and stress tensors},''
  \href{http://dx.doi.org/10.1007/JHEP11(2018)159}{{\em JHEP} {\bfseries 11}
  (2018) 159}, \href{http://arxiv.org/abs/1805.12100}{{\ttfamily
  arXiv:1805.12100 [hep-th]}}.

\bibitem{Bzowski:2019kwd}
A.~Bzowski, P.~McFadden, and K.~Skenderis, ``{Conformal $n$-point functions in
  momentum space},''
  \href{http://dx.doi.org/10.1103/PhysRevLett.124.131602}{{\em Phys. Rev.
  Lett.} {\bfseries 124} no.~13, (2020) 131602},
  \href{http://arxiv.org/abs/1910.10162}{{\ttfamily arXiv:1910.10162
  [hep-th]}}.

\bibitem{Bzowski:2022rlz}
A.~Bzowski, P.~McFadden, and K.~Skenderis, ``{A handbook of holographic 4-point
  functions},'' \href{http://arxiv.org/abs/2207.02872}{{\ttfamily
  arXiv:2207.02872 [hep-th]}}.

\bibitem{Pajer:2016ieg}
E.~Pajer, G.~L. Pimentel, and J.~V.~S. Van~Wijck, ``{The Conformal Limit of
  Inflation in the Era of CMB Polarimetry},''
  \href{http://dx.doi.org/10.1088/1475-7516/2017/06/009}{{\em JCAP} {\bfseries
  06} (2017) 009}, \href{http://arxiv.org/abs/1609.06993}{{\ttfamily
  arXiv:1609.06993 [hep-th]}}.

\bibitem{Salopek:1990jq}
D.~S. Salopek and J.~R. Bond, ``{Nonlinear evolution of long wavelength metric
  fluctuations in inflationary models},''
  \href{http://dx.doi.org/10.1103/PhysRevD.42.3936}{{\em Phys. Rev. D}
  {\bfseries 42} (1990) 3936--3962}.

\bibitem{Sasaki:1995aw}
M.~Sasaki and E.~D. Stewart, ``{A General analytic formula for the spectral
  index of the density perturbations produced during inflation},''
  \href{http://dx.doi.org/10.1143/PTP.95.71}{{\em Prog. Theor. Phys.}
  {\bfseries 95} (1996) 71--78},
  \href{http://arxiv.org/abs/astro-ph/9507001}{{\ttfamily
  arXiv:astro-ph/9507001}}.

\bibitem{Starobinsky:1986fxa}
A.~A. Starobinsky, ``{Multicomponent de Sitter (Inflationary) Stages and the
  Generation of Perturbations},'' {\em JETP Lett.} {\bfseries 42} (1985)
  152--155.
[Pisma Zh. Eksp. Teor. Fiz.42,124(1985)].

\bibitem{Sasaki:1998ug}
M.~Sasaki and T.~Tanaka, ``{Superhorizon scale dynamics of multiscalar
  inflation},'' \href{http://dx.doi.org/10.1143/PTP.99.763}{{\em Prog. Theor.
  Phys.} {\bfseries 99} (1998) 763--782},
\href{http://arxiv.org/abs/gr-qc/9801017}{{\ttfamily arXiv:gr-qc/9801017
  [gr-qc]}}.

\bibitem{Lyth:2004gb}
D.~H. Lyth, K.~A. Malik, and M.~Sasaki, ``{A General proof of the conservation
  of the curvature perturbation},''
  \href{http://dx.doi.org/10.1088/1475-7516/2005/05/004}{{\em JCAP} {\bfseries
  0505} (2005) 004},
\href{http://arxiv.org/abs/astro-ph/0411220}{{\ttfamily arXiv:astro-ph/0411220
  [astro-ph]}}.

\bibitem{Lee:2005bb}
H.-C. Lee, M.~Sasaki, E.~D. Stewart, T.~Tanaka, and S.~Yokoyama, ``{A New delta
  N formalism for multi-component inflation},''
  \href{http://dx.doi.org/10.1088/1475-7516/2005/10/004}{{\em JCAP} {\bfseries
  0510} (2005) 004},
\href{http://arxiv.org/abs/astro-ph/0506262}{{\ttfamily arXiv:astro-ph/0506262
  [astro-ph]}}.

\bibitem{Dias:2012qy}
M.~Dias, R.~H. Ribeiro, and D.~Seery, ``{The \ensuremath{\delta}N formula is
  the dynamical renormalization group},''
  \href{http://dx.doi.org/10.1088/1475-7516/2013/10/062}{{\em JCAP} {\bfseries
  10} (2013) 062}, \href{http://arxiv.org/abs/1210.7800}{{\ttfamily
  arXiv:1210.7800 [astro-ph.CO]}}.

\bibitem{Maldacena:2011nz}
J.~M. Maldacena and G.~L. Pimentel, ``{On graviton non-Gaussianities during
  inflation},'' \href{http://dx.doi.org/10.1007/JHEP09(2011)045}{{\em JHEP}
  {\bfseries 09} (2011) 045}, \href{http://arxiv.org/abs/1104.2846}{{\ttfamily
  arXiv:1104.2846 [hep-th]}}.

\bibitem{Raju:2012zr}
S.~Raju, ``{New Recursion Relations and a Flat Space Limit for AdS/CFT
  Correlators},'' \href{http://dx.doi.org/10.1103/PhysRevD.85.126009}{{\em
  Phys. Rev. D} {\bfseries 85} (2012) 126009},
  \href{http://arxiv.org/abs/1201.6449}{{\ttfamily arXiv:1201.6449 [hep-th]}}.

\bibitem{Langlois:2008mn}
D.~Langlois and S.~Renaux-Petel, ``{Perturbations in generalized multi-field
  inflation},'' \href{http://dx.doi.org/10.1088/1475-7516/2008/04/017}{{\em
  JCAP} {\bfseries 04} (2008) 017},
  \href{http://arxiv.org/abs/0801.1085}{{\ttfamily arXiv:0801.1085 [hep-th]}}.

\bibitem{Langlois:2008wt}
D.~Langlois, S.~Renaux-Petel, D.~A. Steer, and T.~Tanaka, ``{Primordial
  fluctuations and non-Gaussianities in multi-field DBI inflation},''
  \href{http://dx.doi.org/10.1103/PhysRevLett.101.061301}{{\em Phys. Rev.
  Lett.} {\bfseries 101} (2008) 061301},
  \href{http://arxiv.org/abs/0804.3139}{{\ttfamily arXiv:0804.3139 [hep-th]}}.

\bibitem{Langlois:2008qf}
D.~Langlois, S.~Renaux-Petel, D.~A. Steer, and T.~Tanaka, ``{Primordial
  perturbations and non-Gaussianities in DBI and general multi-field
  inflation},'' \href{http://dx.doi.org/10.1103/PhysRevD.78.063523}{{\em Phys.
  Rev. D} {\bfseries 78} (2008) 063523},
  \href{http://arxiv.org/abs/0806.0336}{{\ttfamily arXiv:0806.0336 [hep-th]}}.

\bibitem{Arroja:2008yy}
F.~Arroja, S.~Mizuno, and K.~Koyama, ``{Non-gaussianity from the bispectrum in
  general multiple field inflation},''
  \href{http://dx.doi.org/10.1088/1475-7516/2008/08/015}{{\em JCAP} {\bfseries
  08} (2008) 015}, \href{http://arxiv.org/abs/0806.0619}{{\ttfamily
  arXiv:0806.0619 [astro-ph]}}.

\bibitem{Cai:2008if}
Y.-F. Cai and W.~Xue, ``{N-flation from multiple DBI type actions},''
  \href{http://dx.doi.org/10.1016/j.physletb.2009.09.043}{{\em Phys. Lett. B}
  {\bfseries 680} (2009) 395--398},
  \href{http://arxiv.org/abs/0809.4134}{{\ttfamily arXiv:0809.4134 [hep-th]}}.

\bibitem{Cai:2009hw}
Y.-F. Cai and H.-Y. Xia, ``{Inflation with multiple sound speeds: a model of
  multiple DBI type actions and non-Gaussianities},''
  \href{http://dx.doi.org/10.1016/j.physletb.2009.05.047}{{\em Phys. Lett. B}
  {\bfseries 677} (2009) 226--234},
  \href{http://arxiv.org/abs/0904.0062}{{\ttfamily arXiv:0904.0062 [hep-th]}}.

\bibitem{Panagopoulos:2019ail}
G.~Panagopoulos and E.~Silverstein, ``{Primordial Black Holes from non-Gaussian
  tails},'' \href{http://arxiv.org/abs/1906.02827}{{\ttfamily arXiv:1906.02827
  [hep-th]}}.

\end{thebibliography}\endgroup

\end{document}